\documentclass[twocolumn,superscriptaddress,%
nobibnotes,aps,prd,showpacs,nofootinbib]{revtex4}
\usepackage{eurosym}
\usepackage{graphicx}
\usepackage{epsf}
\usepackage{bm}
\usepackage{amsmath}
\usepackage{amsfonts}
\usepackage{amssymb}
\usepackage{epstopdf}
\usepackage{natbib}
\usepackage{rotating}
\usepackage{pdflscape}
\usepackage{color}
\providecommand{\U}[1]{\protect\rule{.1in}{.1in}}
\newcommand{\be}{\begin{equation}}
\newcommand{\ee}{\end{equation}}

\newcommand{\mincir}{\raise
-3.truept\hbox{\rlap{\hbox{$\sim$}}\raise4.truept\hbox{$<$}\ }}
\newcommand{\magcir}{\raise
-3.truept\hbox{\rlap{\hbox{$\sim$}}\raise4.truept\hbox{$>$}\ }}

\begin{document}

\title{Reconstructing the dark matter and dark energy interaction scenarios
from observations}

\author{Weiqiang Yang}
\email{d11102004@163.com}
\affiliation{Department of Physics, Liaoning Normal University, Dalian, 116029, P. R. China}

\author{Narayan Banerjee}
\email{narayan@iiserkol.ac.in}
\affiliation{Department of Physical Sciences, Indian Institute of Science Education and Research, Kolkata, Mohanpur 741246, West Bengal, India}

\author{Andronikos Paliathanasis}
\email{anpaliat@phys.uoa.gr}
\affiliation{Instituto de Ciencias F\'{\i}sicas y Matem\'{a}ticas, Universidad Austral de
Chile, Valdivia, Chile}
\affiliation{Institute of Systems Science, Durban University of Technology, PO Box 1334,
Durban 4000, Republic of South Africa}

\author{Supriya Pan}
\email{supriya.maths@presiuniv.ac.in}
\affiliation{Department of Mathematics, Presidency University, 86/1 College Street, Kolkata 700073, India}


\begin{abstract}
We consider a class of interacting dark energy models in a  flat and nonflat FLRW universe where the interaction is characterized by the  modified
evolution of the pressureless dark matter as $a^{-3+\delta (a)}$, $a$ being the FLRW scale factor and $\delta (a)$ quantifies the interaction rate. By assuming the most natural and nonsingular parametrization for $\delta (a)$ 
as $\delta \left( a\right) =\sum_{i} \delta_i (1-a)^i $,
where $\delta _{i}$'s ($i=0,1,2,3,..$) are constants, we reconstruct the
expansion history of the universe for three particular choices of the DE sector using different cosmological datasets. Our analyses show that the non-interacting scenario is consistent with the observations while the interaction is not strictly ruled out.  We reconstruct in the following way. We start with the first two terms of $\delta (a)$ above and constrain $\delta_0$, $\delta_1$. Then we consider up to the second order terms in $\delta (a)$ but fix $\delta_0$, $\delta_1$ to their constrained values and constrain $\delta_2$; similarly we constrain $\delta_3$, and finally we constrain $(\delta_0, \delta_1, \delta_1, \delta_3)$ by keeping all of them to be free as a generalized case.   
Our reconstruction technique shows that the constraints on $\delta_2$ (fixing $\delta_0$ and $\delta_1$) and $\delta_3$ (fixing $\delta_0$, $\delta_1$ and $\delta_2$) are almost zero for any interaction model and thus the
effective scenario is well described by the linear parametrization $\delta
(a)\simeq \delta _{0}+\delta _{1}(1-a)$. Additionally, a strong negative correlation between $\delta_0$, $\delta_1$ is observed independently of the dark energy fluid and the curvature of our universe. 
\end{abstract}

\maketitle


\section{Introduction}

Observational evidences from a series of astronomical sources firmly 
point towards an accelerating phase of the universe at the present epoch. Theoretically this accelerating 
phase is realized either by introducing some dark energy fluid in the 
context of Einstein gravity or by introducing new gravitational theories 
different from General Relativity. Concerning the dark energy problem, one
of the simplest dark energy candidate is the cosmological constant $\Lambda$. Along with a cold dark matter (CDM), this model accounts for the current cosmological data quite effectively but it suffers from the
major problem, the so-called \textquotedblleft
Fine-tuning\textquotedblright\ problem \cite{and1,Peebles:2002gy, and2,and3}. 
The Fine-tuning problem has to do with the huge mismatch of the theoretically predicted value of $\Lambda$ which is $\sim 120$ orders of magnitude larger than the value consistent with observations. In order
to overpass this major problem, various cosmological
models has been proposed such as the introduction
of scalar fields, or matter sources with exotic 
equations of state, modification of Einstein's General Relativity and many
others, see \cite{ref1,ref2,ref3,Copeland:2006wr,ref4,ref5, Sotiriou:2008rp,
DeFelice:2010aj,ref6,Cai:2015emx,Nunes:2016aup,Pan:2016jli,Nojiri:2017ncd,Papagiannopoulos:2017whb,Pan:2018ibu} 
and references therein.

In this work we shall deal with cosmological models  where an
interaction exists between the fluid terms in the dark sector; 
that is, the dark energy and dark matter do
not satisfy the conservation equations independently. The
conservation of energy is satisfied only for the combined dark sector. 
The interaction mechanism became focused and popular in the cosmological 
regime when such models were found to explain the cosmic coincidence problem \cite{Amendola-ide1,TocchiniValentini:2001ty,Amendola-ide2,delCampo:2008sr,delCampo:2008jx}, 
another crucial problem of modern cosmology. However, if one looks back into the literature, one can find that the original proposal of an interaction between the matter fields was motivated to 
explain the lowest value of the cosmological constant \cite{Wetterich-ide1}. 
Now,  in presence of an  interaction, the 
conservation laws for the cold dark matter (CDM) and the dark energy\ (DE)
can be recast as $\nabla _{\nu }T_{c}^{\mu \nu }=-Q_{c}$ and $\nabla _{\nu
}T_{x}^{\mu \nu }=Q_{x}$, where $Q_{c}=Q_{x}=Q\,$\ such that $\nabla _{\nu
}\left( T_{c}^{\mu \nu }+T_{x}^{\mu \nu }\right) =0.~$Here, $Q$ is the
energy transfer rate or the interaction function which characterizes the
strength and direction of energy flow between the dark sectors; $c$, $x$ stand for the CDM and DE sectors respectively. However, there is no pressing need to choose a particular form  of the interaction, except perhaps the requirement that the interaction is non-gravitational, but indeed there are ways to choose an ansatz for the interaction rate and estimate the interaction parameters from available data. 

The basic motivation for introducing the interaction is that there is no apriori reason for assuming that there is none. The interaction that is considered is certainly not gravitational, but there is hardly any preferred model for the interaction. However, it has been consitently observed that the interaction in the dark sector might be able to reconcile the tensions in the Hubble constant $H_0$ \cite{Kumar:2017dnp,DiValentino:2017iww,Yang:2018euj,Yang:2018uae} and in the amplitude of the matter power spectrum $\sigma_8$ \cite{vandeBruck:2017idm, Barros:2018efl}. Thus, the mechanism of interaction in the dark sector got massive attention in the cosmological community in the recent time.   
We shall refer to quite a few of the works already there in the literature in relevant places, but for a very thorough review, we refer two works, one by Bolotin, Kostenko, Lemets and Yerokhin \cite{Bolotin:2013jpa} and other work by 
Wang, Abdalla, Atro-Barandela and Pav\'{o}n \cite{fernando}.

In a Friedmann-Lema\^itre-Robertson-Walker universe, 
if the interaction is not present, then one can easily
see that CDM sector, for which the pressure is zero, evolves as $a^{-3}$ where $a$ is the scale factor
of this universe. Certainly, if we allow an interaction in the dark
sector, then the evolution of the CDM  will deviate from this usual
evolution law.

A small deviation from the CDM evolution can be looked as a modification 
of the evolution of the matter density as $a^{-3+\delta }$, where $\delta$ 
could be either constant or variable. In
the next section we shall show that such evolution of the CDM sector
effectively reduces to a particular interaction function. If $\delta$ 
is treated as a variable, then one can explore several
possibilities and moreover one can reconstruct the interaction using the
observational data.

In this work we have considered that dark matter and dark energy
interact with each other where the interaction is characterized by the
evolution of the CDM sector: $\rho_c \propto a^{-3 + \delta (a)}$ 
where $\delta (a)$ is a time varying quantity. 
Since the dark energy could be anything,
hence, we have considered three types of dark energy fluid, namely, the
vacuum energy, dark energy with constant equation of state and the dark energy
fluid with a dynamical equation of state. Finally, by using the observational
data, we have reconstructed all the interacting scenarios. 

The work has been organized in the following way. 
In section \ref{ide-equations} we set up the
basic framework of our study and introduce the models. Section \ref{sec-data}
describes the observational data used to analyze the models. Then in section %
\ref{sec-results} we discuss the observational constraints on the
interacting models. Finally, we close the work with a brief discussion
summarized in section \ref{discussion}.

\section{Interacting dark energy}

\label{ide-equations}

Let us consider the homogeneous and isotropic universe described by the
Friedmann-Lema\^{i}tre-Robertson-Walker line element

\begin{equation}\label{flrw}
\mathrm{d}s^{2}=-\mathrm{d}t^{2}+a^{2}(t)\Bigl[\frac{\mathrm{d}r^{2}}{%
1-Kr^{2}}+r^{2}\left( \mathrm{d}\theta ^{2}+\sin ^{2}\theta \,\mathrm{d}\phi
^{2}\right) \Bigr],
\end{equation}%
in ($t$, $r$, $\theta$, $\phi$) comoving coordinate system 
where $a(t)$ is the expansion scale factor of the universe and 
 $K = 0, \pm 1$ denotes
the constant spatial curvature. In the context of Einstein's General Relativity the
main constituents of the universe are considered to be baryons, radiation,
pressureless dark matter (cold dark matter) responsible for the structure
formation of the universe and finally dark energy fluid that steers the
late time acceleration of the universe. In addition to that, we
consider that there is an interaction in the dark sector of the universe.
In particular, we consider that the cold dark matter (CDM) and dark energy
(DE) are coupled to each other where DE has a barotropic equation of
state parameter, that is, $p_{x}=w_{x}\rho _{x}$,  where $\rho$, $p$ and $w$ are the density, pressure and the equation of state parameter respectively, and the subscript $x$ corresponds to DE.

Due to the coupling between CDM and DE, 
the total conservation equation can be decoupled as follows

\begin{equation}
\dot{\rho}_{c}+3H\rho _{c}=-\dot{\rho}_{x}-3H(1+w_{x})\rho _{x}=Q,
\label{ssd11}
\end{equation}%
where the subscript $c$ corresponds to CDM and 
$Q$ is rate of interaction between the dark sectors already mentioned about
in the introduction. Usually the interacting dark energy models are motivated from phenomenological ground in order to mainly address the cosmic coincidence problem. An action formalism/action integral producing the interacting dark energy models is certainly appealing, however, according to our knowledge there exists Lagrangian for ideal gas. Here, our model is motivated from the phenomenological ground and thus we are not able to find such a mechanics. Of course one could describe the above phenomena by using scalar fields having some action formalism, see for instance \cite{Boehmer:2015kta, Boehmer:2015sha}.

In order to study the interacting dynamics, various phenomenological choices for $Q$ have been widely studied
in the literature \cite{Bolotin:2013jpa,fernando, Boehmer:2015kta, Boehmer:2015sha, He:2008tn, Billyard:2000bh,Amendola:2004ew,Curbelo:2005dh,Gonzalez:2006cj, Amendola:2011ie, Pettorino:2012ts, Salvatelli:2014zta,Yang:2014gza, Wang:2014iua, Yang:2014hea, Pan:2012ki, vandeBruck:2016hpz,
Yang:2016evp, Caprini:2016qxs, Shahalam:2017fqt, Cai:2017yww, Kumar:2017bpv, Pan:2016ngu, Mukherjee:2016shl, Sharov:2017iue, Yang:2017zjs,Yang:2017ccc, Pan:2017ent, Yang:2018ubt, Yang:2018pej, Yang:2018xlt} but no one is universally accepted.

For $Q>0$, the energy flow takes place 
from DE to CDM while for $Q< 0$ the direction of energy flow is reversed. If one pretends that there is no interaction and the DE and CDM behave in a different way, one recasts the conservation equation (\ref{ssd11}) as

\begin{eqnarray}
&&\dot{\rho}_{c}+3H \left(1+ w_c^{\rm eff}\right) \rho _{c} = 0,\\
&&\dot{\rho}_{x}+ 3H (1+w_{x}^{\rm eff}) \rho _{x} = 0
\end{eqnarray}
where $w_c^{\rm eff}$, $w_x^{\rm eff}$ are respectively termed as the effective equation of state for CDM and DE that take the following expressions:

\begin{eqnarray}
&&w_c^{\rm eff} = -\frac{Q}{3H \rho_c}, \label{eff-eos-cdm}\\
&&w_x^{\rm eff} = w_x + \frac{Q}{3 H \rho_x}\label{eff-eos-de}. 
\end{eqnarray}

If the energy flows from DE to CDM ($Q > 0$), the effective state parameter for CDM becomes negative, while the negativity of the effective equation of state for DE decreases. So, if a phantom dark energy interacts with CDM, then the effective dark energy state parameter will be of a quintessence type in this version. For $Q< 0$, one can see that if DE with $w_x > -1$ interacts with CDM, then $w_x$ can cross the phantom divide $w_x = -1$ provided $Q$ is sufficiently negative. However, for the case with interacting vacuum, the effective dark energy equation of state definitely crosses the $w_x =-1$ boundary for a positive $Q$.

In this ``effective equation of state'' formulation, the interaction manifests itself as a deviation
from the standard evolution $a^{-3}$ of CDM, where there is no interaction in the dark sector, as given by  \cite{Amendola:2006dg,Nunes:2014qoa,%
Nunes:2016dlj,Kumar:2016zpg, Yang:2017yme}

\begin{equation}
\rho _{c}=\rho _{c,0}\,a^{-3+\delta (a)}\,,  \label{cdm-evolution}
\end{equation}%
where $\rho _{c,0}$ is the present value of $\rho _{c}$; $\delta (a)$, that
characterizes the effect of the interaction, can in general be varying with evolution
and we assume that it is an analytic function of the scale factor. Here, the present value of the scale factor is scaled to be unity.

The flow of energy between the dark sectors  depends on the sign of $%
\delta (a)$. Also, for $\delta (a)<0$, the evolution of 
the CDM sector is faster, 
while on the other hand, for $\delta (a)>0$, 
the evolution of the CDM factor is slower.

In the special case in which $\delta (a)=0$, the standard CDM evolution is recovered, i.e., 
$\rho _{c}\propto a^{-3}$, which indicates that there is  no coupling or interaction between the CDM and DE.  When $\delta (a)=$ constant,
one recovers the cosmological scenarios discussed in \cite{Nunes:2016dlj,
Kumar:2016zpg, Yang:2017yme}. For a variable $\delta (a)$,  one particular ansatz
was introduced in Ref. \cite{Rosenfeld:2007ri}, namely, $\delta (a) = \delta_0\left( \frac{2a}{1+a^2}\right)$, while we think that a detailed analysis will be worth in this direction. The rate of interaction with a varying $\delta (a)$ can be
written as

\begin{eqnarray}\label{Q}
Q = H \rho_{c} \Bigl(\delta(a)+ a\, \delta^{\prime}(a) \, \ln a \Bigr),
\end{eqnarray}
where $\delta ^{\prime}(a)$ is the differentiation of $\delta (a)$ with
respect to the scale factor $a$. For the above explicit form of the interaction rate, the effective equation of state parameters (\ref{eff-eos-cdm}) and (\ref{eff-eos-de}) become, 
\begin{eqnarray}
&&w_c^{\rm eff} = \frac{1}{3} \Bigl[\delta (a) + \delta^{\prime} (a) \ln (a) \Bigr],\label{eff-eos-cdm1}\\
&&w_x^{\rm eff} = w_x + \frac{r}{3} \Bigl[\delta (a) + \delta^{\prime} (a) \ln (a) \Bigr]\label{eff-eos-de1},
\end{eqnarray}
where $r= \rho_c/\rho_x$ is a positive quantity.
Now, for the evolution of CDM as in eqn. (%
\ref{cdm-evolution}), one can write the DE density evolution as

\begin{eqnarray}
\rho_x = \frac{1}{f(a)} \Bigg[ \rho_{x,0}\, f(1) - \rho_{c,0} \int_{1}^{a}
a^{-4+ \delta (a)} f(a) \Bigl(\delta (a)  \notag \\
+ a\, \delta^{\prime}(a) \, \ln a \Bigr) \, da \Bigg],
\end{eqnarray}
where the new quantities $\rho_{x,0}$, $\rho_{c,0}$ are the current values of $\rho_x$ and $\rho_{c}$ respectively; the functional $f(a)$ is given by

\begin{equation*}
f(a)=\exp \left( 3\int \frac{1+w_{x}}{a}da\right),
\end{equation*}%
and $f(1)$ is the value of $f(a)$ at $a=a_{0}=1$. Here,  $a_0$ is the present value of the scale factor $a (t)$ which is related to the redshift $z$ by the relation, $1+z = a_0/a$. 

We assume a smooth function for $\delta \left( a\right)$, which has a Taylor expansion around the present value of $a = a_0 = 1$ as
\begin{equation}
\delta \left( a\right) =\delta _{0}+\delta _{1}\left( 1-a\right) +\delta
_{2}\left( 1-a\right) ^{2}+...
\end{equation}%
where $\delta _{0}=\delta \left( a\right) |_{a\rightarrow 1}$, $\delta
_{1}=\delta ^{\prime }\left( a\right) |_{a\rightarrow 1}$,~$\delta _{2}=%
\frac{\delta ^{\prime \prime }\left( a\right) |_{a\rightarrow 1}}{2!}$ etc.
Thus, we actually assume that $\delta (a)$ is a differentiable function and a Taylor expansion is possible, rather than any complicated choice or parametrization as in \cite{Rosenfeld:2007ri}, and explore the interacting dynamics. 

Now, within the above parametric choice of $\delta (a)$, we can reconstruct the $\delta _{i}$'s ($i= 0, 1, 2...$) and consequently the function $\delta \left( a\right) $ using the observational data. Such an approach has been applied in other facets  of cosmology such as the
reconstruction of the inflationary model \cite{Copeland:1993jj,
Mangano:1995rh}. 

The evolution of the energy density for the DE fluid, for such a function
given by in eqn. (\ref{xi}) takes the form

\begin{align}
\rho_x = \frac{1}{f(a)} \Bigg[ \rho_{x,0}\, f(1) - \rho_{c,0} \int_{1}^{a}
a^{-4+ \delta_0 + \delta_1 (1-a)} f(a) \Bigl(\delta_0 + \delta_1  \notag \\
- \delta_1 a\, (1 +\ln a) \Bigr) \, da \Bigg].
\end{align}
when the Taylor expansion is taken only up to the first order,

\begin{equation}
\delta (a)=\delta_{0}+ \delta_{1}(1-a). \label{xi}
\end{equation}%

Considering the next higher order terms in $\delta (a)$, in a similar fashion, one can calculate the evolutions of $\rho_x$. In this work we consider upto the second and the third order expansion,

\begin{equation}
\delta (a)=\delta_{0}+ \delta_{1}(1-a)+ \delta_{2}(1-a)^{2},  \label{xi-2}
\end{equation}

\begin{equation}
\delta (a)=\delta_{0}+ \delta_{1}(1-a)+ \delta_{2}(1-a)^{2} +
\delta_{3}(1-a)^{3}.  \label{xi-3}
\end{equation}

Now we introduce three equation of state parameters for the dark energy fluid
which corresponds to three different scenarios as follows:

\begin{enumerate}
\item If CDM interacts with the constant vacuum energy which is characterized by $w_{x}=-1$.

\item When CDM interacts with a dark energy fluid having constant equation
of state $w_{x}$.

\item Finally, we consider a scenario where the interacting dark
energy has a dynamical equation of state as

\begin{equation}
w_{x}(a)=w_{0}+w_{a}(1-a),  \label{cpl}
\end{equation}%
where $w_{0}$, $w_{a}$, are real parameters. This parametrization is
well known in the literature as the  Chevallier--Polarski--Linder 
(CPL) parametrization \cite{Chevallier:2000qy,Linder:2002et}.
\end{enumerate}

\section{Observational data}
\label{sec-data}

 In this section we describe the main observational datasets that are employed to constrain the parameters of models of  interacting dark energy scenarios.  

\begin{enumerate}
\item \textit{Cosmic Microwave Background data (CMB):} Observations from the cosmic microwave  background radiation plays a crucial role in constraining the 
cosmological models. Here, we use the CMB data from
the Planck's 2015 measurements \cite{Adam:2015rua, Aghanim:2015xee}. In particular, we have used the Planck TT, TE, EE likelihoods at multipoles $\ell > 30$ together with low$-\ell$ temperature+polarization likelihood. In literature this dataset is referred to as ``Planck TTTEEE+lowP''.

\item \textit{Supernovae Type Ia (SNIa):} The Supernovae Type Ia (SNIa) 
data were the first observational data that indicated an accelerating phase of the universe at late-time. We take the latest joint light
curves (JLA) sample \cite{Betoule:2014frx} of SnIa containing 740 SNIa in the
redshift span $z\in[0.01, 1.30]$.

\item \textit{Baryon acoustic oscillations (BAO) distance measurements:} For
the BAO data we use the estimated ratio $r_s/D_V$ as a `standard ruler' in
which $r_s$ is the comoving sound horizon at the baryon drag epoch and $D_V$
is the effective distance determined by the angular diameter distance $D_A$
and Hubble parameter $H$ as $D_V(z)=\left[(1+z)^2D_A(a)^2\frac{z}{H(z)}
\right]^{1/3}$. We consider three different measurements, 
$r_s(z_d)/D_V(z=0.106)=0.336\pm0.015$ from 6-degree Field Galaxy Redshift
Survey (6dFGRS) data \cite{Beutler:2011hx}, $r_s(z_d)/D_V(z=0.35)=0.1126
\pm0.0022$ from Sloan Digital Sky Survey Data Release 7 (SDSS DR7) 
data \cite{Padmanabhan:2012hf}, and finally $r_s(z_d)/D_V(z=0.57)=0.0732\pm0.0012$,
from the SDSS DR9 \cite{Manera:2012sc}.

\item \textit{Cosmic chronometers (CC):} 
The cosmic chronometers are some extremely massive and 
passively evolving galaxies in our universe. 
We employ the recent 
cosmic chronometers data comprising $30$ measurements of the Hubble
parameter in the redshift interval $0 < z< 2$ \cite{Moresco:2016mzx}.
Here, we determine the Hubble parameter values through the relation 
$H(z)= -\left(1/1+z \right)dz/dt$ where 
the measurement of $dz$ is obtained through the spectroscopic method with high accuracy. and the precise measurement of $dt$ $-$ the differential age evolution  of such galaxies. As a result, a precise measurement of the Hubble parameter is obtained and thus, these measurements can be taken as model independent for the cosmological studies.  For a detail reading on the cosmic chronometers we refer to \cite{Moresco:2016mzx}. 
\end{enumerate}

In order to constrain the interacting scenarios, we use a fastest cosmological package, namely the Markov chain Monte carlo package \texttt{cosmomc} \cite%
{Lewis:2002ah, Lewis:1999bs} equipped with the Gelman-Rubin statistics $R-1$ \cite{Gelman-Rubin} that determines the convergence of the chains. Moreover, we note that the package \texttt{cosmomc} supports the Planck 2015 Likelihood Code \cite{Aghanim:2015xee} (see \url{http://cosmologist.info/cosmomc/}, a publicly available code). Now, since we consider three different interacting DE scenarios, therefore, we work with different parameter spaces during the statistical analyses. For the interacting vacuum scenario in a spatially flat universe, we work with the following parameter spaces, namely,
\begin{eqnarray}
&&\Theta_{01} \equiv \Bigl\{\Omega_bh^2, \Omega_{c}h^2, 100\theta_{MC}, \tau, n_s, log[10^{10}A_{s}], \delta_{0}, \delta_1 \Bigr\},\nonumber\\
&&\Theta_{11} \equiv \Bigl\{\Omega_bh^2, \Omega_{c}h^2, 100\theta_{MC}, \tau, n_s, log[10^{10}A_{s}], \delta_2 \Bigr\},\nonumber\\
&&\Theta_{21} \equiv \Bigl\{\Omega_bh^2, \Omega_{c}h^2, 100\theta_{MC}, \tau, n_s, log[10^{10}A_{s}], \delta_3 \Bigr\},\nonumber\\
&&\Theta_{31} \equiv \Bigl\{\Omega_bh^2, \Omega_{c}h^2, 100\theta_{MC}, \tau, n_s, log[10^{10}A_{s}], \delta_{0}, \delta_1, \delta_2, \delta_3 \Bigr\},\nonumber
\end{eqnarray}
where $\Theta_{01}$ is eight dimensional; $\Theta_{11}$ and $\Theta_{21}$ are seven dimensional each; $\Theta_{31}$ is ten dimensional. When we include curvature into the above framework as a free parameter, then one extra dimension is added to each parameter space $\Theta_{i1}$ ($i =0, 1, 2, 3$). For the next two interacting DE scenarios, namely, interacting DE with constant EoS, $w_x$, in the spatially flat FLRW universe, we work with the following parameter spaces 
\begin{eqnarray}
&&\Theta_{02} \equiv \Bigl\{\Omega_bh^2, \Omega_{c}h^2, 100\theta_{MC}, \tau, n_s, log[10^{10}A_{s}], w_x, \delta_{0}, \delta_1 \Bigr\},\nonumber\\
&&\Theta_{12} \equiv \Bigl\{\Omega_bh^2, \Omega_{c}h^2, 100\theta_{MC}, \tau, n_s, log[10^{10}A_{s}], w_x, \delta_2 \Bigr\},\nonumber\\
&&\Theta_{22} \equiv \Bigl\{\Omega_bh^2, \Omega_{c}h^2, 100\theta_{MC}, \tau, n_s, log[10^{10}A_{s}], w_x, \delta_3 \Bigr\},\nonumber\\
&&\Theta_{32} \equiv \Bigl\{\Omega_bh^2, \Omega_{c}h^2, 100\theta_{MC}, \tau, n_s, log[10^{10}A_{s}], w_x, \delta_{0}, \delta_1, \delta_2, \delta_3 \Bigr\},\nonumber
\end{eqnarray}
where $\Theta_{02}$ is nine dimensional; $\Theta_{12}$ and $\Theta_{22}$ are ten dimensional each; $\Theta_{32}$ is eleven dimensional. In a similar way, when curvature is included into the above framework as a free parameter, then one extra dimension is added to each parameter space $\Theta_{i2}$ ($i =0, 1, 2, 3$).  For interacting DE with its dynamical EoS [$w_x (a) = w_0 + w_a (1-a)$], the parameter space without curvature is,
\begin{eqnarray}
\Theta_{03} \equiv \Bigl\{\Omega_bh^2, \Omega_{c}h^2, 100\theta_{MC}, \tau, n_s, log[10^{10}A_{s}], \nonumber\\ w_0, w_a,  \delta_{0}, \delta_1 \Bigr\},\nonumber\\
\Theta_{13} \equiv \Bigl\{\Omega_bh^2, \Omega_{c}h^2, 100\theta_{MC}, \tau, n_s, log[10^{10}A_{s}], \nonumber\\ w_0, w_a, \delta_2 \Bigr\},\nonumber\\
\Theta_{23} \equiv \Bigl\{\Omega_bh^2, \Omega_{c}h^2, 100\theta_{MC}, \tau, n_s, log[10^{10}A_{s}], \nonumber\\ w_0, w_a, \delta_3 \Bigr\},\nonumber\\
\Theta_{33} \equiv \Bigl\{\Omega_bh^2, \Omega_{c}h^2, 100\theta_{MC}, \tau, n_s, log[10^{10}A_{s}], \nonumber\\ w_0, w_a,  \delta_{0}, \delta_1, \delta_2,\delta_3 \Bigr\},\nonumber
\end{eqnarray}
where $\Theta_{03}$ is ten dimensional; $\Theta_{13}$ and $\Theta_{23}$ are nine dimensional each; $\Theta_{33}$ is twelve dimensional. In a similar way, when curvature is included into the above framework as a free parameter, then one extra dimension is added to each parameter space $\Theta_{i3}$ ($i =0, 1, 2, 3$). The symbols $\Omega_b h^2$ and $\Omega_c h^2$ respectively denote the physical density of baryons and physical density of cold dark matter; $100 \theta_{MC}$ is the ratio of the sound horizon to the angular diameter distance; $\tau$ is the optical depth; $n_s$ is the scalar spectral index; $A_s$ denotes the amplitude of the primordial scalar power spectrum; and the other parameters, namely, $\delta_i$'s ($i =0, 1, 2, 3$), $w_x$, $w_0$, $w_a$ are already described in section  \ref{ide-equations}. 
In the next section \ref{sec-results} we have described the reconstruction technique and the results on the interacting scenarios. 
Let us note that for each analysis, we kept running the chains until we achieve  $R-1 < 0.1$. In Table \ref{priors} we show the flat priors imposed on various free parameters for the statistical analyses.

\begin{table}
\begin{center}
\renewcommand{\arraystretch}{1.4}
\begin{tabular}{|c@{\hspace{1 cm}}|@{\hspace{1 cm}} c|}
\hline
\textbf{Parameter}                    & \textbf{Prior}\\
\hline\hline
$\Omega_{\rm b} h^2$         & $[0.005\,,\,0.1]$\\
$\Omega_{\rm c} h^2$       & $[0.001\,,\,0.99]$\\
$\theta_{\rm {MC}}$             & $[0.5\,,\,10]$\\
$\tau$                       & $[0.01\,,\,0.8]$\\
$\log_{10}(10^{10}A_{s})$         & $[2\,,\,4]$\\
$n_s$                        & $[0.8\,,\, 1.2]$\\
$w_x$ (constant)                        & $[-2, 0]$\\
$w_{0}$ & $[-2 \,,\,0]$\\ 
$w_a$ & $[-3 \,,\,3]$  \\

$\delta_0$   & $[-1, 1]$\\
$\delta_1$   & $[-1, 1]$\\
$\delta_2$   & $[-1, 1]$\\
$\delta_3$   & $[-1, 1]$\\

$\Omega_{K0}$ & $[-1, 1]$\\

\hline
\end{tabular}
\end{center}
\caption{Flat priors on various free parameters of different interacting scenarios imposed during the statistical analyses.}
\label{priors}
\end{table}

\section{Results of the analysis}
\label{sec-results}

In this section we describe the observational constraints that are imposed on 
different interacting scenarios. In the first half of this section we shall consider the interaction scenarios in a spatially flat FLRW universe whilst in the last half of this section we concentrate on the interaction scenarios in presence of the curvature of the universe.  For all the interacting scenarios we use the same dataset: CMB $+$ JLA $+$ BAO $+$ CC. It should be noted that BAO data break the degeneracies between the parameters and usually the existing degeneracies between the parameters cannot be broken by CMB alone. Thus, the inclusion of  BAO data to CMB dataset is effective to understand the nature of the coupling parameters.

\subsection{Reconstruction of the interaction rate in a spatially flat universe}
\label{results:flat-space}

In this section, we shall describe the reconstruction of the interacting
scenario for three particular types of the dark energy fluids, namely the
vacuum energy, dark energy fluid with constant equation of state and finally
a dark energy fluid with dynamical equation of state for a spatially flat universe.  For the dynamical 
state parameter in DE, we choose the most used and well known state parameter, namely, the CPL parametrization \cite{Chevallier:2000qy,Linder:2002et}. 

The combined observational data for every analysis have been considered to be CMB $+$ JLA $+$ BAO $+$ CC.

\begin{figure*}
\includegraphics[width=0.5\textwidth]{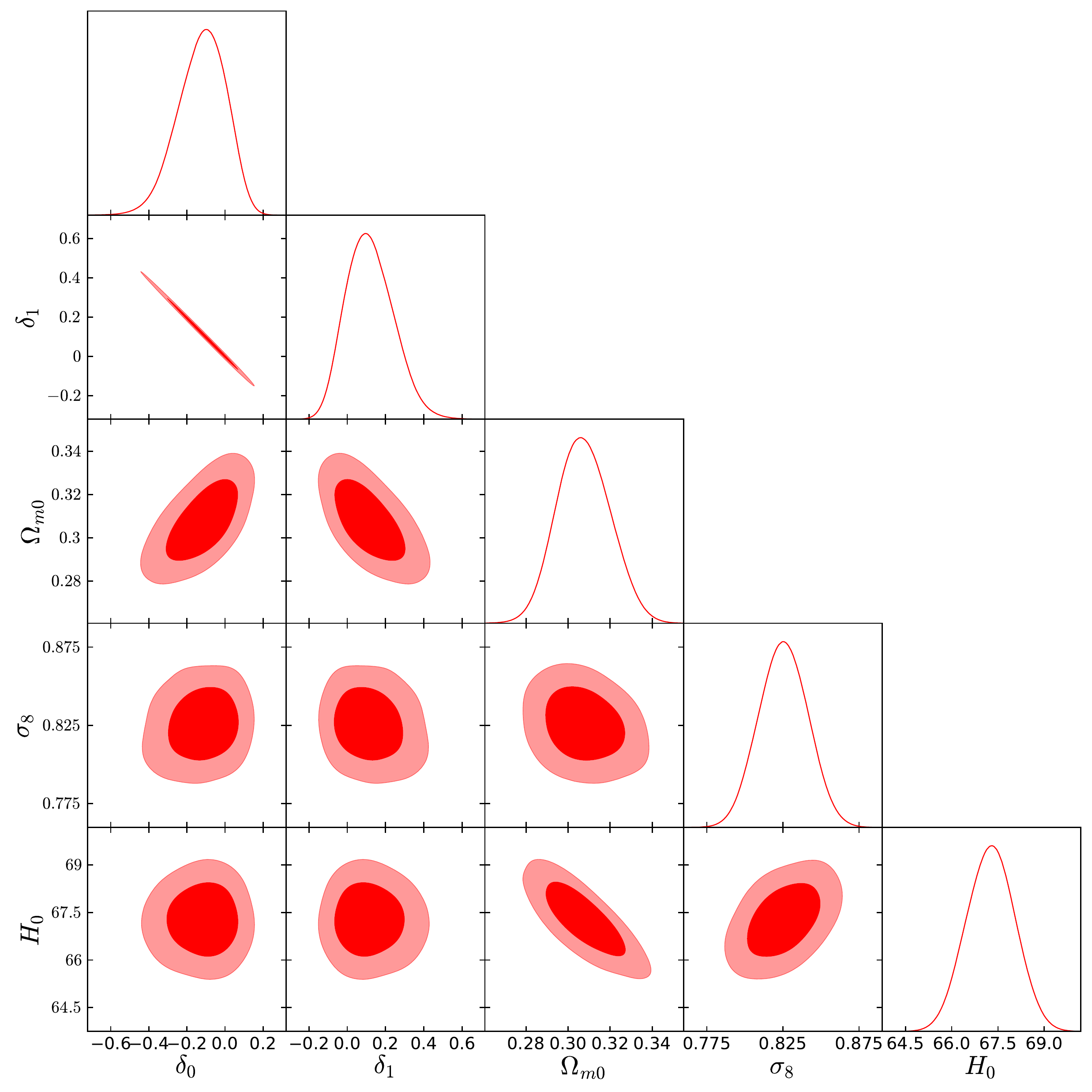}
\caption{68\% and 95\% CL contour plots for different combinations of the model parameters as well as the one dimensional posterior distributions of some selected model parameters of the interacting vacuum
scenario (in a spatially flat universe) where the interaction is parametrized by $\protect\delta (a) =  \protect\delta_0 + \protect\delta_1 (1-a)$.  The combined data for this analysis has been set to be CMB $+$ JLA $+$ BAO $+$ CC. The results are summarized in the second and third columns of Table \ref{tab:Int-vacuum-flat}. }
\label{fig:int-vacuum-flat-1}
\end{figure*}
\begin{figure*}
\includegraphics[width=0.5\textwidth]{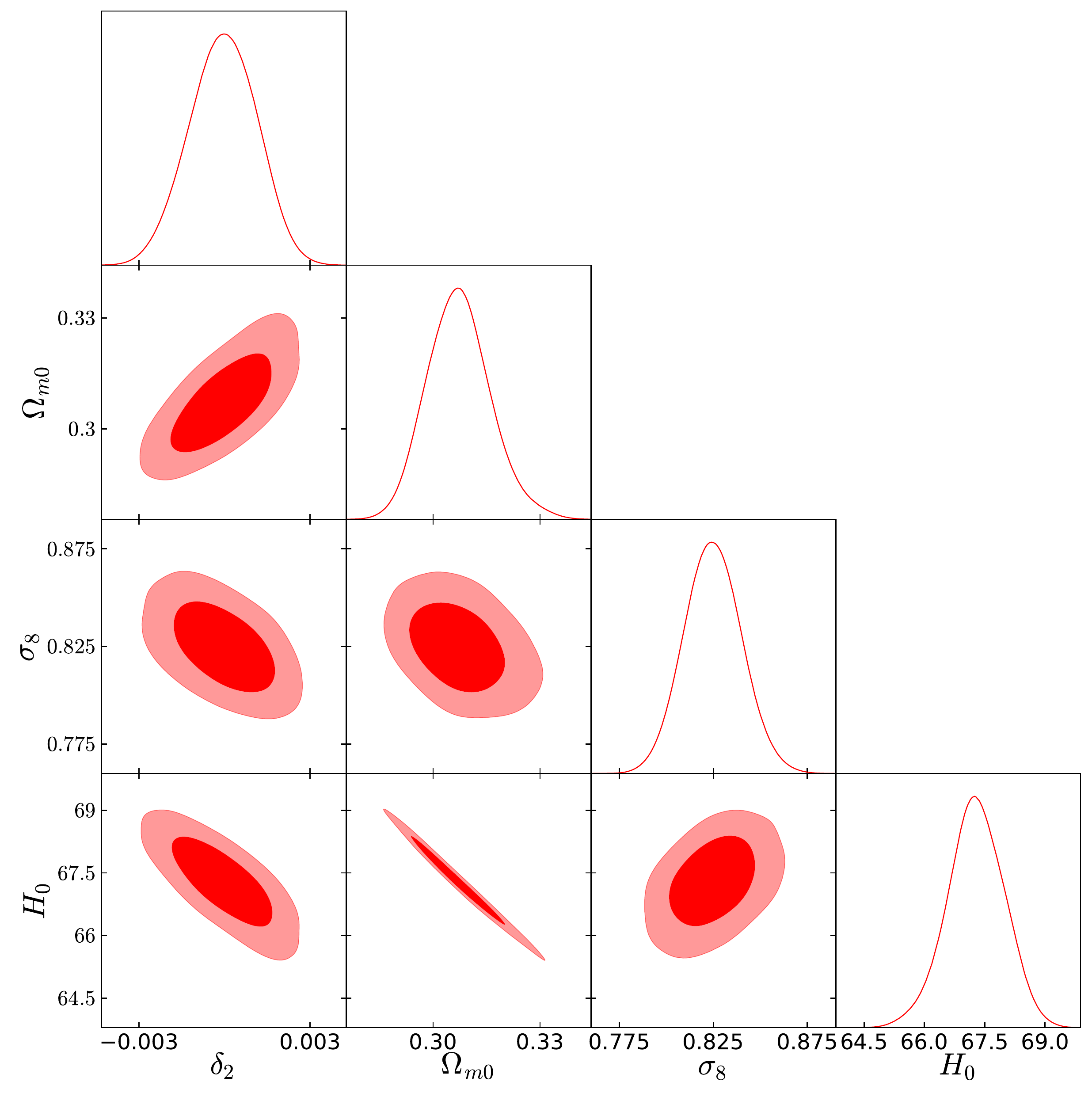}
\caption{68\% and 95\% CL contour plots for
different combinations of the model parameters as well as the one dimensional posterior distributions of some selected model parameters of the interacting vacuum
scenario (in a spatially flat universe) where the interaction is parametrized by $\protect\delta (a) = \protect\delta_0 + \protect\delta_1 (1-a) + \protect\delta_2 (1-a)^2$ in which we fix the mean values of ($\protect\delta_0$, $\protect\delta_1$) from Table \ref{tab:Int-vacuum-flat}. The combined data for this analysis has been set to be CMB $+$ JLA $+$ BAO $+$ CC and the corresponding results are summarized in the fourth and fifth columns of Table \ref{tab:Int-vacuum-flat}.  }
\label{fig:int-vacuum-flat-2}
\end{figure*}
\begin{figure*}
\includegraphics[width=0.5\textwidth]{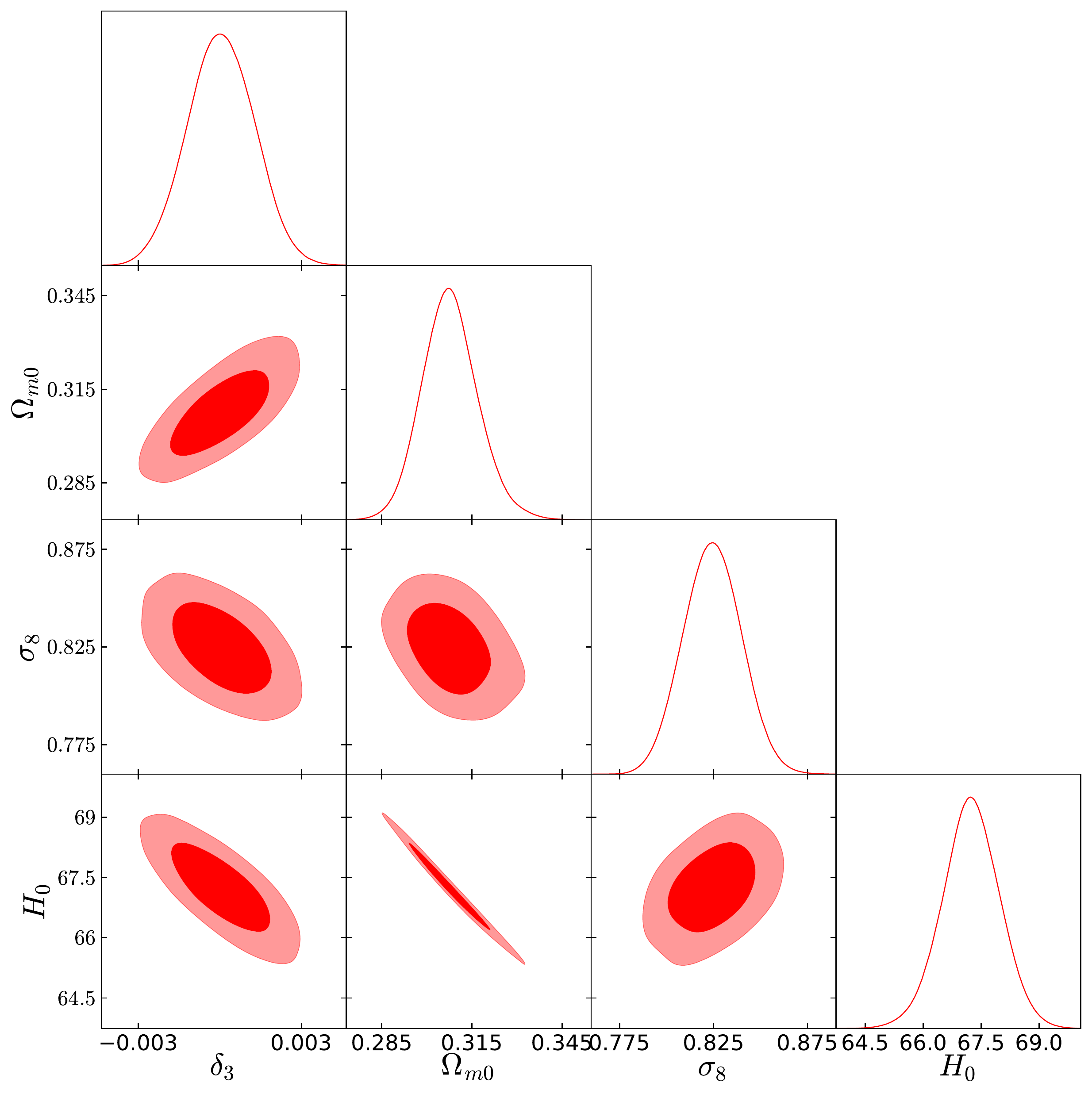}
\caption{68\% and 95\% CL contour plots for
different combinations of the free parameters as well as the one dimensional posterior distributions of some selected model parameters of the interacting vacuum
scenario (in a spatially flat universe) where the interaction is parametrized by $\protect\delta (a) = 
\protect\delta_0 + \protect\delta_1 (1-a) + \protect\delta_2 (1-a)^2+ 
\protect\delta_3 (1-a)^3$ in which the mean values of ($\protect\delta_0$, $%
\protect\delta_1$, $\protect\delta_2$) have been fixed from the previous
analysis summarized in Table \protect\ref{tab:Int-vacuum-flat}. The combined data for this analysis has been set to be CMB $+$ JLA $+$ BAO $+$ CC and the corresponding results are summarized in the sixth and seventh columns of Table \ref{tab:Int-vacuum-flat}. }
\label{fig:int-vacuum-flat-3}
\end{figure*}

\subsubsection{Interacting vacuum}
\label{sec-int-vacuum-flat}

We have reconstructed this interacting model using three steps. We
first constrained the interacting model using $\delta (a)$ given in eqn (\ref%
{xi}), that means for $\delta (a) = \delta_0 + \delta_1 (1-a)$. The results
of this analysis have been summarized in the second and third column of Table %
\ref{tab:Int-vacuum-flat}. From the analysis, we see that  although the mean values of $\delta_0$ and $\delta_1$ are nonzero, but
both $\delta_0$ and $\delta_1$ allow a zero value 
as one can see $\delta_0 = -0.120_{- 0.110}^{+ 0.151}$ (68\% CL)
and $\delta_1 = 0.117_{- 0.147}^{+ 0.107}$ (68\% CL). The results show that the possibility of both interacting and non-interacting scenarios are allowed within this confidence-region, that means within 68\% CL. 
For a better view of the behaviour of the parameters, in Fig. \ref%
{fig:int-vacuum-flat-1}, we show the 68\% and 95\% CL contour plots and
the likelihood analysis for some selected parameters of the model. From Fig. \ref{fig:int-vacuum-flat-1}, we see that the parameters $\delta_0$ and $\delta_1$ have  strong negative correlation between each other. Thus, if one of  $\delta_0$ and $\delta_1$ increases, the other one decreases leading to result that the interaction is hardly effective.

In the next step we extend the  parametrization in (\ref{xi}) to  (\ref{xi-2}) 
with the allowance of one extra parameter $\delta_2$. 
To reconstruct this scenario from the observations, we fix the
values of $\delta_0$, $\delta_1$ to their corresponding mean values
summarized in Table \ref{tab:Int-vacuum-flat} and constrain the new
interaction parameter $\delta_2$. The results of this analysis have been
summarized in the fourth and fifth column of Table \ref{tab:Int-vacuum-flat}. 
From Table \ref{tab:Int-vacuum-flat} one finds that the 
constraint on $\delta_2$ ($=-0.000046_{- 0.001184}^{+ 0.001303}$ at 68\% CL) 
is extremely tiny and close to zero. In addition to that, 
within 68\% CL, $\delta_2$ is 
allowed to have a zero value. In summary, we find that the new interaction
parameter, $\delta_2$  does not really contribute much. In Fig. %
\ref{fig:int-vacuum-flat-2} we show the corresponding contour plots and the
one dimensional posterior distributions for some parameters of this scenario.

After that we consider the parametrization (\ref{xi-3}) and repeat the similar
analysis. We fix the values of $\delta_0$, $\delta_1$, $\delta_2$
to their corresponding mean values obtained in the previous analyses
and perform the fittings with the same set of
observational data. We find that the new interaction
parameter $\delta_3$  is constrained to be extremely small 
($\delta_3 =0.000050_{- 0.001197}^{+0.001247}$ at 68\% CL)
and effectively this does not substantially
contribute to $\delta (a)$. In a similar fashion, 
in Fig. \ref{fig:int-vacuum-flat-3} we display different contour plots of the parameters
and the one-dimensional posterior distributions for different parameters for this particular interaction scenario. Last but not least an important remark is in order: In the above reconstruction technique, for the first step we fix $(\delta_0, \delta_1)$ to their corresponding mean values and obtain $\delta_2$ and in the second step we fix $(\delta_0, \delta_1, \delta_2)$ to their corresponding mean values and obtain $\delta_3$. We note that instead of fixing the interaction parameters, $\delta_i$'s ($i=0, 1$) to their corresponding mean values, if any fiducial values of $\delta_0$, $\delta_1$ are used to constrain $\delta_2$, then the constraint on $\delta_2$ may differ significantly. In a similar fashion, if any fiducial values of $\delta_i$'s ($i =0, 1, 2$) are fixed to constrain $\delta_3$, then its observational constraint may significantly different compared to its constrain obtained by fixing them to their corresponding mean values. We also note that this phenomenon remains true for the next interaction scenarios. 

\begin{figure*}
\includegraphics[width=0.6\textwidth]{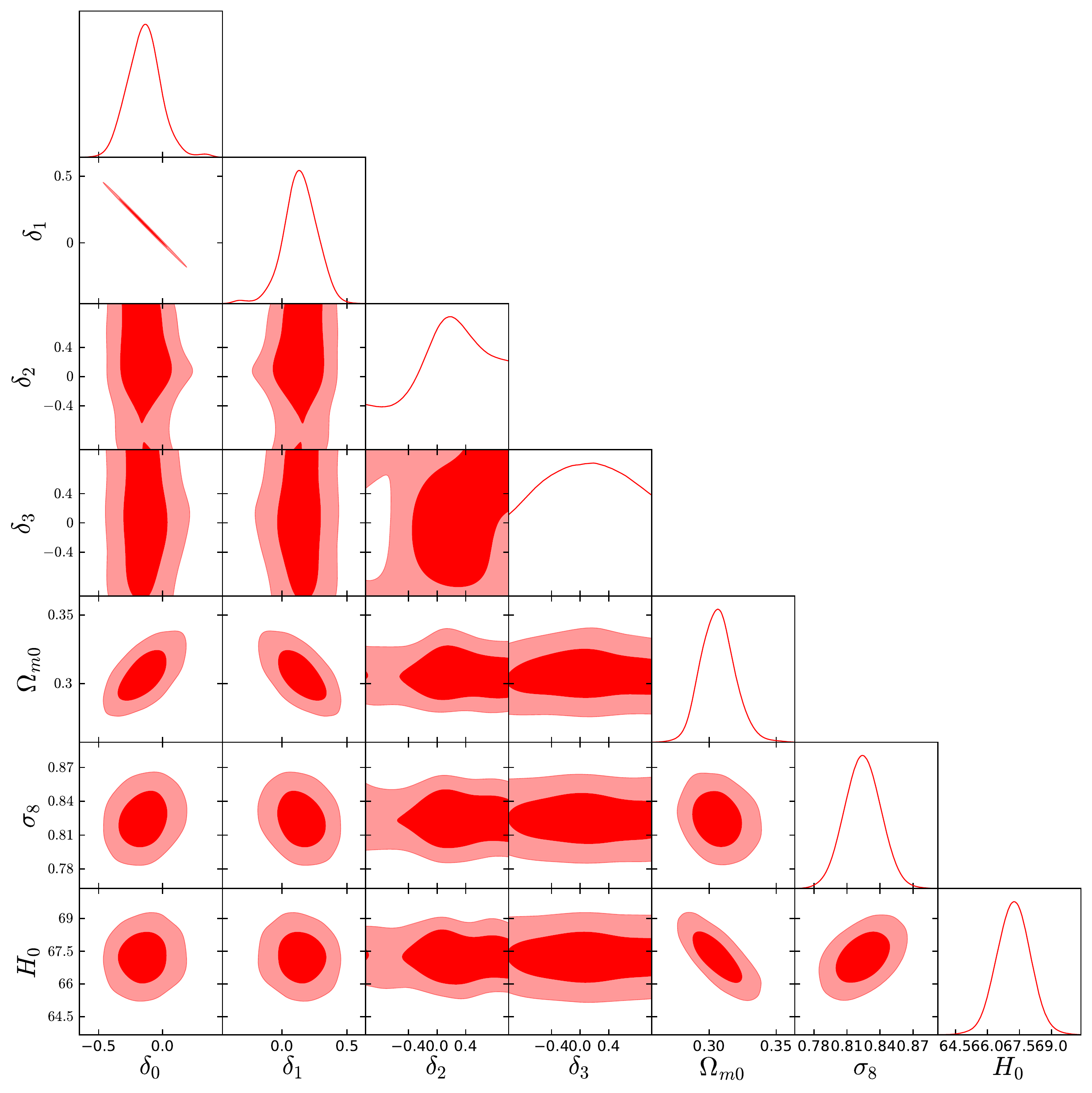}
\caption{68\% and 95\% CL contour plots for
different combinations of the free parameters as well as the one dimensional posterior distributions of some selected model parameters of  the interacting vacuum
 scenario (in a spatially flat FLRW universe) where the interaction is parametrized by the most general 
parametrization $\protect\delta (a) = 
\protect\delta_0 + \protect\delta_1 (1-a) + \protect\delta_2 (1-a)^2+ 
\protect\delta_3 (1-a)^3$. The interaction parameters $\delta_i$'s
are kept free and we use the combined analysis CMB $+$ JLA $+$ BAO $+$ CC. The results are summarized in Table \ref{tab:Int-vacuum-flat-general}. One can clearly notice that the parameters $\delta_2$ and $\delta_3$ are degenerate while other 
parameters are not. }
\label{fig:int-vacuum-flat-4}
\end{figure*}

A natural question that one asks is, what happens if we consider the general parametrization of the interaction function in eqn. (\ref{xi-3}), i.e., $\delta (a) = \delta_{0}+ \delta_{1}(1-a)+ \delta_{2}(1-a)^{2} + \delta_{3}(1-a)^{3}$, and try to constrain $\delta_i$'s. The dimension of the parameters space is big and one apprehends that the degeneracies of the parameters would increase. Anyway, we also try this and fit this more general version with the same observational data CMB $+$ JLA $+$ BAO $+$ CC and show the observational constraints on the model parameters in Table \ref{tab:Int-vacuum-flat-general}. The contour plots for this analysis have been shown in Fig. \ref{fig:int-vacuum-flat-4}. The figure clearly shows that the parameters $\delta_2$ and $\delta_3$ are degenerate. That means, according to the present observational data, this interaction scenario with the general parametrization in eqn. (\ref{xi-3}) cannot be constrained well. That is the reason we made an attempt to constrain these parameters step by step, fixing the lower order parameters to their best fit values.
However, an interesting feature that Fig. \ref{fig:int-vacuum-flat-4} exhibits is that although the parameters $\delta_2$ and $\delta_3$ are found to be degenerate, $\delta_0$ and $\delta_1$ are not and they are still strongly negatively correlated to each other. However, all the attempts indicate quite strongly that the effective $\delta (a)$ is determined by the first two parameters, that means, $\delta (a) \simeq \delta_0 + \delta_1 (1-a)$. Finally, in the last row of Table \ref{tab:Int-vacuum-flat} and Table \ref{tab:Int-vacuum-flat-general} we show the $\chi^2$ values obtained for the best-fit values of different reconstructed interacting scenarios. From Table \ref{tab:Int-vacuum-flat} we observe  a decrease of $\chi^2$ as long as the number of free parameters decrease, however, from Table \ref{tab:Int-vacuum-flat-general} we find reduced $\chi^2$ in contrary to what we expected. The reason might be that some of the parameters are not well constrained in the latter scenario. 

\begingroup                                                                                                                     
\squeezetable                                                                                                                   
\begin{center}
\begin{table*}
\begin{tabular}{|c|c|c|c|c|c|c|c|}
\hline
Parameters & Mean with errors & Best-fit & Mean with errors & Best-fit & 
Mean with errors & Best-fit   \\ \hline
$\Omega_c h^2$ & $0.1161_{- 0.0038- 0.0074}^{+ 0.0041+ 0.0075}$ & $%
0.1194$ & $0.1161_{- 0.0013- 0.0023}^{+ 0.0011+ 0.0024}$ & $%
0.1164$ & $0.1162_{- 0.0013- 0.0022}^{+ 0.0011+ 0.0025}$ & $%
0.1164$  \\ 

$\Omega_b h^2$ & $0.02225_{- 0.00017- 0.00034}^{+ 0.00018+ 0.00033}$ & $%
0.02220$ & $0.02224_{- 0.00015- 0.00033}^{+ 0.00016+ 0.00031}$ & $%
0.02224$ & $0.02225_{- 0.00017- 0.00034}^{+ 0.00017+ 0.00034}$ & $%
0.02224$  \\ 

$100\theta_{MC}$ & $1.0410_{- 0.00043- 0.00081}^{+ 0.00042+ 0.00086}$
& $1.04073$ & $1.04102_{- 0.00033- 0.00057}^{+ 0.00030+ 0.00059}$ & $%
1.04108$ & $1.04102_{- 0.00033- 0.00058}^{+ 0.00030+ 0.00061}$ & $%
1.04108$  \\ 

$\tau$ & $0.084_{- 0.017- 0.035}^{+ 0.017 + 0.034}$ & $%
0.081 $ & $0.082_{- 0.017- 0.033}^{+ 0.017+ 0.032}$ & $%
0.092$ & $0.082_{- 0.017- 0.032}^{+ 0.018+ 0.032}$ & $%
0.092$  \\ 

$n_s$ & $0.9699_{- 0.0061- 0.0117}^{+ 0.0062 + 0.0118}$ & $0.9656$
& $0.9698_{- 0.0040- 0.0077}^{+ 0.0040+ 0.0076}$ & $0.9714$ & $%
0.9697_{- 0.0041- 0.0080}^{+ 0.0041+ 0.0078}$ & $0.9714$  \\ 

$\mathrm{ln}(10^{10} A_s)$ & $3.099_{- 0.033- 0.068}^{+ 0.034+
0.066}$ & $3.097$ & $3.096_{- 0.032- 0.065}^{+ 0.032+
0.062}$ & $3.113$ & $3.097_{- 0.033- 0.064}^{+ 0.032+
0.062}$ & $3.113$ \\ \hline

$\delta_0$ & $-0.120_{- 0.110- 0.247}^{+ 0.151+ 0.237}$ & $%
-0.014$ & $-$ & $-$ & $-$ & $-$ \\ 

$\delta_1$ & $0.117_{- 0.147- 0.231}^{+ 0.107+ 0.241}$ & $%
0.014$ & $-$ & $-$ & $-$ & $-$ \\ 

$\delta_2$ & $\times $ & $\times $ & $-0.000046_{- 0.001184- 0.002420}^{+ 0.001303+
0.002338}$ & $0.000555$ & $-$ & $-$  \\ 

$\delta_3$ & $\times $ & $\times$ & $\times$ & $\times $ & $0.000050_{- 0.001197- 0.002429}^{+
0.001247+ 0.002400}$ & $0.000555$ \\ \hline 

$\Omega_{m0}$ & $0.307_{- 0.014- 0.023}^{+ 0.012+ 0.025}$ & $%
0.316$ & $0.307_{- 0.010- 0.018}^{+ 0.008+ 0.019}$ & $%
0.310$ & $0.308_{- 0.010- 0.019}^{+ 0.009+ 0.019}$ & $%
0.310$  \\ 

$\sigma_8$ & $0.825_{- 0.016- 0.030}^{+ 0.016+ 0.031}$ & $%
0.828$ & $0.825_{- 0.015- 0.029}^{+ 0.015+ 0.031}$ & $%
0.828$ & $0.824_{- 0.015- 0.029}^{+ 0.015+ 0.030}$ & $%
0.827931$ \\
 
$H_0$ & $67.27_{- 0.79- 1.51}^{+ 0.78+ 1.51}$ & $%
67.06$ & $67.28_{- 0.69- 1.47}^{+ 0.73+ 1.47}$ & $%
66.98$ & $67.24_{- 0.72- 1.48}^{+ 0.74+ 1.44}$ & $%
66.98$ \\
\hline   
$\chi^2$ & & 13677.670  &  & 13675.826  & & 13675.826  \\ 
 \hline
\end{tabular}%
\caption{Reconstruction of the interacting vacuum energy scenario for the
spatially flat FLRW universe using the observational data CMB $+$ JLA $+$
BAO $+$ CC. In the columns `$\times$' means that the corresponding parameter has not been considered into the analyses while the sign `$-$' present against any parameter means its mean value has been fixed for the subsequent analyses.   }
\label{tab:Int-vacuum-flat}
\end{table*}
\end{center}
\endgroup                                                                                                                       
\begingroup                                                                                                                     
\squeezetable                                                                                                                   
\begin{center}                                                                                                                  
\begin{table}                                                                                                                   
\begin{tabular}{|c|c|c|c}                                                                                                            
\hline                                                                                                                    
Parameters & Mean with errors & Best fit \\ \hline
$\Omega_c h^2$ & $    0.1156_{-    0.0040-    0.0080}^{+    0.0036+    0.0080}$ & $    0.1176$\\
$\Omega_b h^2$ & $    0.02223_{-    0.00017-    0.00032}^{+    0.00017 +    0.00034}$ & $    0.02218$\\
$100\theta_{MC}$ & $    1.04106_{-    0.00044-    0.00092}^{+    0.00045+    0.00085}$ & $    1.04092700$\\
$\tau$ & $    0.083_{-    0.017-    0.033}^{+    0.017+    0.034}$ & $    0.06773746$\\

$n_s$ & $    0.9705_{-    0.0057-    0.0120}^{+    0.0064+    0.0115}$ & $    0.9670$\\

${\rm{ln}}(10^{10} A_s)$ & $    3.098_{-    0.033-    0.065}^{+    0.034+    0.067}$ & $    3.074$\\

$\delta_0$ & $-0.142_{-    0.133-    0.274}^{+    0.120+    0.270}$ & $   -0.039$\\

$\delta_1$ &  $    0.139_{-    0.116-    0.264}^{+    0.130 +    0.268}$ & $    0.038$\\

$\delta_2$ & $    0.147_{-    0.305-    1.147}^{+    0.732+    0.853}$ & $    0.563$\\

$\delta_3$ & $    0.029_{-    0.631-    1.029}^{+    0.639+    0.971}$ & $    0.622$\\

$\Omega_{m0}$ & $    0.307_{-    0.014-    0.025}^{+    0.011+    0.026}$ & $    0.305$\\

$\sigma_8$ & $    0.824_{-    0.016-    0.032}^{+    0.016+    0.032}$ & $    0.824$\\

$H_0$ & $   67.23_{-    0.79-    1.53}^{+    0.77+    1.52}$ & $   67.81$\\
\hline 
$\chi^2$ & & 13675.468 \\
\hline                                                                                                                 
\end{tabular}                                                                                                                   
\caption{Reconstruction of the interacting vacuum scenario 
for the spatially flat FLRW universe  
and for the general parametrization $\protect\delta (a) = 
\protect\delta_0 + \protect\delta_1 (1-a) + \protect\delta_2 (1-a)^2+ 
\protect\delta_3 (1-a)^3$, using the combined analysis CMB $+$ JLA $+$ BAO $+$ CC.  }\label{tab:Int-vacuum-flat-general}                                                                                                   
\end{table}                                                                                                                     
\end{center}                                                                                                                    
\endgroup         
\begin{figure*}
\includegraphics[width=0.5\textwidth]{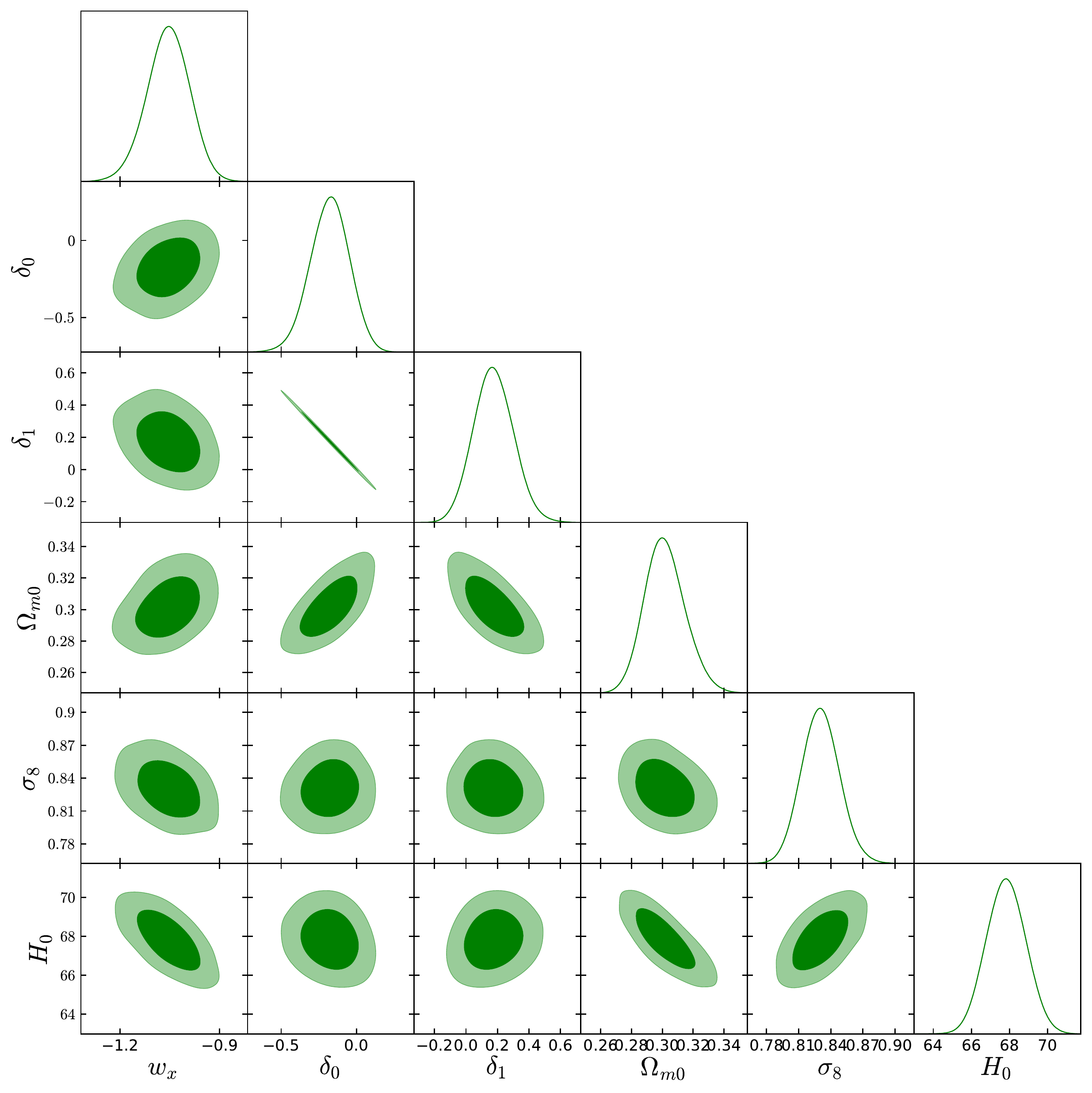}
\caption{68\% and 95\% CL contour plots for
different combinations of the model parameters  as well as the one dimensional posterior distributions of some selected model parameters of the interacting DE 
scenario (for spatially flat case) with constant DE state parameter $w_x$, where the interaction is parametrized by $\protect\delta (a) = \protect\delta_0 + \protect\delta_1
(1-a)$ and the combined observational analysis is CMB $+$ JLA $+$ BAO $+$
CC. The results are summarized in the second and third columns of Table \ref{tab:Int-wCDM-flat}. }
\label{fig:int-wCDM-flat-1}
\end{figure*}
\begin{figure*}
\includegraphics[width=0.5\textwidth]{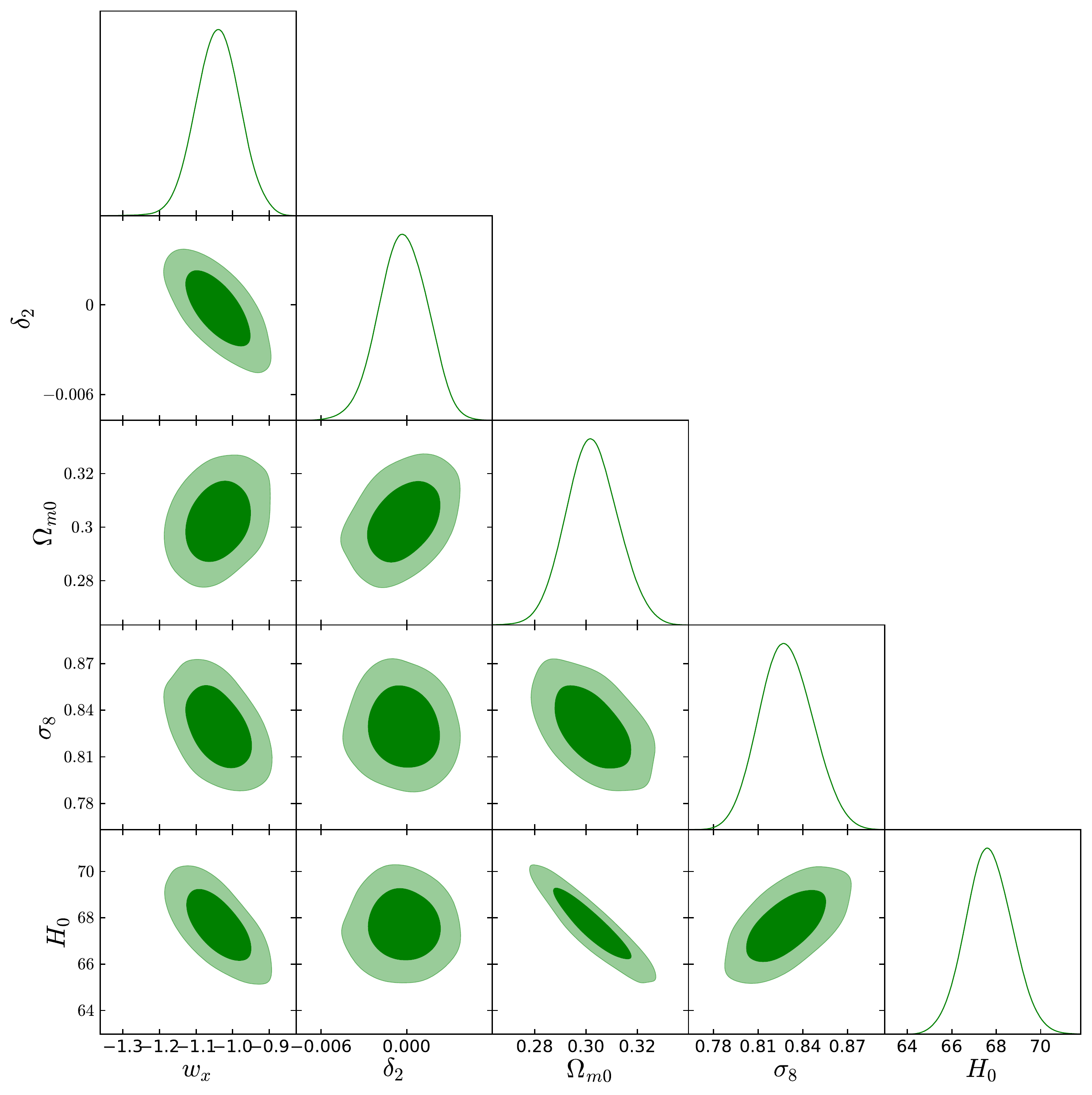}
\caption{68\% and 95\% CL contour plots for
different combinations of the model parameters as well as the one dimensional posterior distributions of some selected model parameters of  the interacting DE
scenario (spatially flat case) with constant DE state parameter $w_x$, where the interaction is  parametrized by $\protect\delta (a) = \protect\delta_0 + \protect\delta_1 (1-a) + \protect\delta_2 (1-a)^2$ in which we fix the mean values of ($\protect\delta_0$, $\protect\delta_1$) from Table \ref{tab:Int-wCDM-flat}. 
The combined observational analysis is CMB $+$ JLA $+$ BAO $+$ CC.  The results are summarized in the fourth and fifth columns of Table \ref{tab:Int-wCDM-flat}.  }
\label{fig:int-wCDM-flat-2}
\end{figure*}
\begin{figure*}
\includegraphics[width=0.5\textwidth]{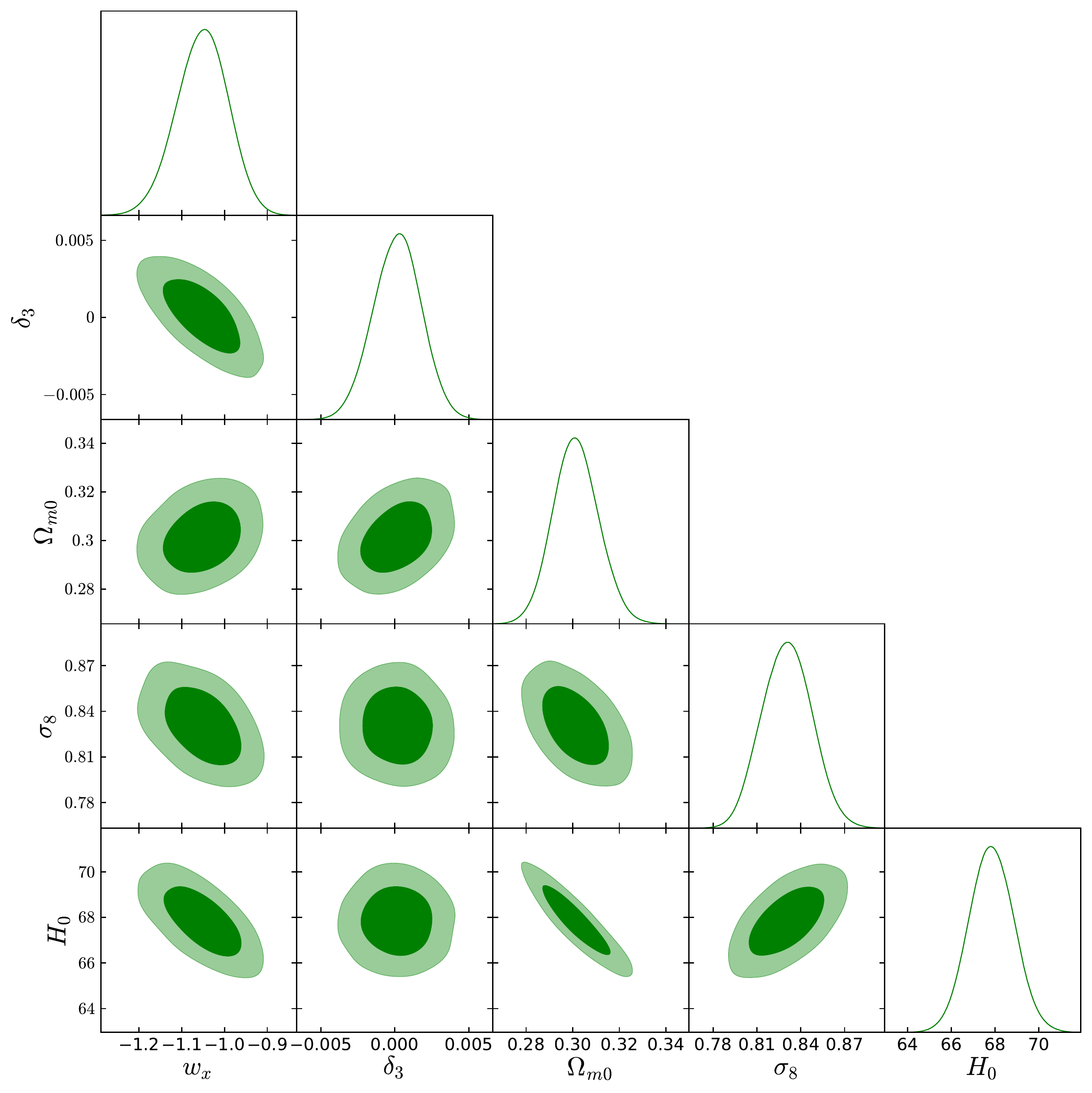}
\caption{68\% and 95\% CL contour plots for
different combinations of the model parameters as well as the one dimensional posterior distributions of some selected model parameters of the interacting DE
scenario (spatially flat case) with constant DE state parameter $w_x$, where the interaction is
parametrized by $\protect\delta (a) = \protect\delta_0 + \protect\delta_1
(1-a) + \protect\delta_2 (1-a)^2 +\protect\delta_3 (1-a)^3$ in which we fix the mean values of ($\protect\delta_0$, $\protect\delta_1$,  $\protect\delta_2$) from Table \ref{tab:Int-wCDM-flat}. The combined
observational analysis is CMB $+$ JLA $+$ BAO $+$ CC and the results are summarized in the last two columns of Table \ref{tab:Int-wCDM-flat}. }
\label{fig:int-wCDM-flat-3}
\end{figure*}

\subsubsection{Interacting DE with constant EoS other than vacuum}
\label{sec-int-cons-eos-flat}

As a second interacting scenario  we pick up a constant EoS in DE, $%
w_x$. Our treatment is similar to the previous case,
$\delta (a)$ given by eqns. (\ref{xi}), (\ref{xi-2})
and (\ref{xi-3}).

We begin the analysis with the simplest case when $\delta (a)$ has the form as in eqn. (\ref{xi}). 
Using the same combined analysis, CMB $+$ JLA $+$ BAO $+$ CC, we constrain the model
parameters that are summarized in Table \ref{tab:Int-wCDM-flat}. From the
analysis, we see that $\delta_0 = -0.178_{- 0.129}^{+ 0.128}$ at 68\% CL ($-0.178_{- 0.260}^{+ 0.247}$ at 95\% CL) and $\delta_1 = 0.175_{- 0.126}^{+ 0.125}$ at 68\% CL ($0.175_{- 0.242}^{+0.254}$ at 95\% CL). Thus, within 68\% CL, the mean values of 
$\delta_0$ and $\delta_1$ are non-zero.
However, within 95\% CL, their zero values are indeed allowed. 
From the observational constraints on the EoS in DE, a
phantom scenario is suggested as one can see, $w_x= -1.055_{- 0.062}^{+ 0.068}$ (68\%
CL). However, in 68\% CL, $w_x$ is allowed its quintessential character as well. 
In Fig.  \ref{fig:int-wCDM-flat-1} we have shown the behaviour of the model
parameters through their two-dimensional contour plots. An worthwhile point
that one must note is that, here too, the parameters $\delta_0$ and $\delta_1$ are strongly negatively correlated. 

Now, we proceed with the next step where we take the second parametrization
of $\delta (a)$ given in eqn. (\ref{xi-2}) containing three parameters 
$(\delta_0, \delta_1, \delta_2)$. We fix the first two parameters, 
$(\delta_0, \delta_1)$ to their mean values from the previous analysis and
constrain the remaining parameter $\delta_2$ in order to reconstruct the
scenario. The constraints on the model parameters have been shown in the
fourth and fifth column of Table \ref{tab:Int-wCDM-flat} and in Fig. 
\ref{fig:int-wCDM-flat-2} we show the 68\% and 95\% confidence-level
contours for various combinations of the model parameters. Our analysis
shows that the new interaction parameter $\delta_2$ is extremely tiny with $%
\delta_2 = -0.000262_{- 0.001665}^{+ 0.001707}$ (68\% CL) and 
within 68\% CL, $\delta_2 = 0$ is allowed. The dark energy 
equation of state parameter may have both its phantom or quintessential behaviour ($w_x = -1.040_{- 0.059}^{+ 0.060}$, 68\% CL).

After that we consider the next parametrization (\ref{xi-3}) having four
parameters ($\delta_0$, $\delta_1$, $\delta_2$, $\delta_3$) following the
same procedure as in section \ref{sec-int-vacuum-flat}, 
that means we fix the first three parameters to their
corresponding mean values summarized in the last two columns of Table 
\ref{tab:Int-wCDM-flat}  and constrain $\delta_3$. Fig. \ref{fig:int-wCDM-flat-3} shows the corresponding graphical variations. 
From the results we find that the parameter $\delta_3$ is not significant in the sense that, it is very small taking $\delta_3= 0.000166_{- 0.001609}^{+ 0.001594}$ (68\% CL) which
also recovers its null value in 68\% CL. Furthermore, the phantom  behaviour in the dark energy equation of state is allowed with $w_x = -1.052_{- 0.059}^{+ 0.064}$ (68\% CL) while its quintessential nature is also allowed in the same confidence-level. 

Thus, following the reconstruction mechanism described in the three analyses above, it is clear that for the interacting
scenario where dark energy state parameter remains constant, the parameters $\delta_2$ and $\delta_3$ in the parametrization (\ref{xi-3}) are insignificant.

\begin{figure*}
\includegraphics[width=0.6\textwidth]{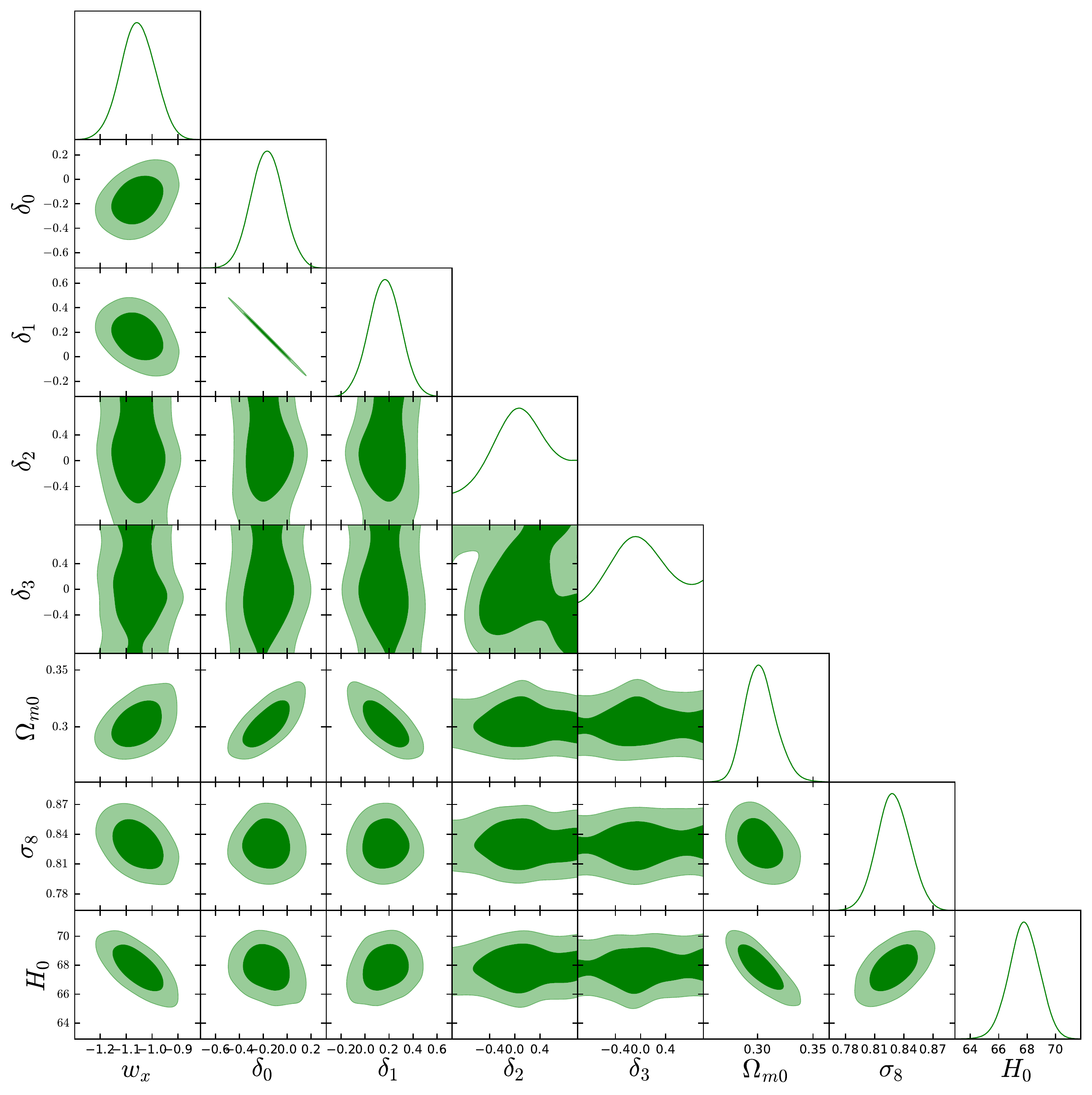}
\caption{68\% and 95\% CL contour plots for
different combinations of the model parameters as well as the one dimensional posterior distributions of some selected parameters of the interacting DE
scenario (spatially flat case) with constant EoS in DE, $w_x$,
for the most general 
parametrization $\protect\delta (a) = 
\protect\delta_0 + \protect\delta_1 (1-a) + \protect\delta_2 (1-a)^2+ 
\protect\delta_3 (1-a)^3$ where  the interaction parameters $\delta_i$'s
are kept free. The analysis is CMB $+$ JLA $+$ BAO $+$ CC and the results are summarized in Table \ref{tab:Int-wCDM-flat-general}. 
One can clearly notice that the parameters $\delta_2$ and $\delta_3$ are degenerate similar to the interaction vacuum scenario. }
\label{fig:int-wCDM-flat-4}
\end{figure*}

We now exercise with the general parametrization 
in eqn. (\ref{xi-3}) and wish to constrain all the four interaction parameters
$\delta_i$ ($i= 0, 1, 2, 3$) by starting with all of them as free. The dimension of the parameters space now becomes eleven. Thus, some of the parameters might be degenerate. We constrained this interaction scenario with the same combined analysis CMB $+$ JLA $+$ BAO $+$ CC. The Table \ref{tab:Int-wCDM-flat-general} shows the constraints on the model parameters and in Fig. \ref{fig:int-wCDM-flat-4} we show the confidence-level contour plots for different combinations of the model parameters. The figure clearly shows that the last two interaction parameters in the general parametrization (\ref{xi-3}), i.e., $\delta_2$ and $\delta_3$ cannot be constrained well as expected. The parameters $\delta_0$ and $\delta_1$ are strongly correlated to one another with negative orientation, that means, although we increase the dimension of the parameters space, the relation between these two parameters seems to be unaltered. Finally, in the last row of Table \ref{tab:Int-wCDM-flat} and Table \ref{tab:Int-wCDM-flat-general} we present the $\chi^2$ values obtained for the best-fit values of different reconstructed interacting scenarios. From Table \ref{tab:Int-wCDM-flat}, looking at the estimated values of $\chi^2$, we have a similar observation to the interacting vacuum scenarios because as we notice from Table \ref{tab:Int-wCDM-flat}, as long as the free parameters decrease in the parameter space, the $\chi^2$ values decrease, however, for the general case (Table \ref{tab:Int-wCDM-flat-general}) as already commented earlier, since still some of the parameters  are unconstrained, thus, no definite conclusion regarding the comparison of $\chi^2$ between Table \ref{tab:Int-wCDM-flat} and Table \ref{tab:Int-wCDM-flat-general} cannot be drawn. 

\begingroup                                                                                                                     
\squeezetable                                                                                                                   

\begin{center}
\begin{table*}
\begin{tabular}{|c|c|c|c|c|c|c|c}
\hline
Parameters & Mean with errors & Best fit & Mean with errors & Best fit & 
Mean with errors & Best fit   \\ \hline
$\Omega_c h^2$ & $0.1158_{- 0.0038- 0.0074}^{+ 0.0038+ 0.0075}$ & $%
0.1157$ & $0.1155_{- 0.0019- 0.0038}^{+ 0.0019+ 0.0037}$ & $%
0.1140$ & $0.1157_{- 0.0019- 0.0037}^{+ 0.0018+ 0.0036}$ & $%
0.1157$   \\ 

$\Omega_b h^2$ & $0.02224_{- 0.00017- 0.00033}^{+ 0.00017+ 0.00032}$ & $%
0.02212$ & $0.02223_{- 0.00018- 0.00033}^{+ 0.00017+ 0.00034}$ & $%
0.02226$ & $0.02224_{- 0.00016- 0.00032}^{+ 0.00016+ 0.00032}$ & $%
0.02235$   \\
 
$100\theta_{MC}$ & $1.04105_{- 0.00042- 0.00083}^{+ 0.00044+ 0.00085}$
& $1.04118$ & $1.04109_{- 0.00033- 0.00063}^{+ 0.00033+ 0.00064}$ & $%
1.04142$ & $1.04108_{- 0.00032- 0.00064}^{+ 0.00032+ 0.00065}$ & $%
1.04137$   \\ 

$\tau$ & $0.083_{- 0.018- 0.033}^{+ 0.016+ 0.034}$ & $%
0.080 $ & $0.083_{- 0.017- 0.033}^{+ 0.017+ 0.034}$ & $%
0.096$ & $0.082_{- 0.016- 0.033}^{+ 0.017+ 0.032}$ & $%
0.087$   \\
 
$n_s$ & $0.9701_{- 0.0059- 0.0115}^{+ 0.0059+ 0.0121}$ & $0.9687$
& $0.9705_{- 0.0046- 0.0087}^{+ 0.0045 + 0.0089}$ & $0.9789$ & $%
0.9699_{- 0.0043- 0.0084}^{+ 0.0043+ 0.0087}$ & $0.9725$  \\ 

$\mathrm{ln}(10^{10} A_s)$ & $3.097_{- 0.033- 0.064}^{+ 0.033+
0.066}$ & $3.097$ & $3.098_{- 0.033- 0.065}^{+ 0.032+
0.066}$ & $3.119$ & $3.097_{- 0.032- 0.064}^{+ 0.032+
0.063}$ & $3.108$   \\  \hline

$\delta_0$ & $-0.178_{- 0.129- 0.260}^{+ 0.128+ 0.247}$ & $%
-0.172$ & $-$ & $-$ & $-$ & $-$   \\ 

$\delta_1$ & $0.175_{- 0.126- 0.242}^{+ 0.125+ 0.254}$ & $%
0.168$ & $-$ & $-$ & $-$ & $-$   \\ 

$\delta_2$ & $\times$ & $\times$ & $-0.00026_{- 0.00167- 0.00332}^{+ 0.00171+
0.00335}$ & $-0.00080$ & $-$ & $-$   \\ 

$\delta_3$ & $\times$ & $\times$ & $\times$ & $\times$ & $0.00017_{- 0.00161- 0.00320}^{+
0.00159+ 0.00311}$ & $0.00027$   \\ \hline

$w_x$ & $-1.055_{- 0.062- 0.130}^{+ 0.068+ 0.121}$ & $%
-1.080$ & $-1.040_{- 0.059- 0.117}^{+ 0.060+ 0.119}$ & $%
-1.011$ & $-1.052_{- 0.059- 0.122}^{+ 0.064+ 0.115}$ & $%
-1.060$  \\ 

$\Omega_{m0}$ & $0.302_{- 0.014- 0.024}^{+ 0.012+ 0.027}$ & $%
0.291$ & $0.302_{- 0.011- 0.019}^{+ 0.010+ 0.020}$ & $%
0.298$ & $0.301_{- 0.010- 0.019}^{+ 0.009+ 0.020}$ & $%
0.297$   \\ 

$\sigma_8$ & $0.830_{- 0.018- 0.034}^{+ 0.017 + 0.036}$ & $%
0.846$ & $0.829_{- 0.019- 0.033}^{+ 0.017+ 0.035}$ & $%
0.833$ & $0.831_{- 0.017- 0.032}^{+ 0.017+ 0.033}$ & $%
0.837$   \\ 

$H_0$ & $67.83_{- 1.05- 1.96}^{+ 1.03+ 1.99}$ & $%
68.95$ & $67.67_{- 1.02- 2.02}^{+ 1.04+ 2.09}$ & $%
67.82$ & $67.84_{- 1.02- 1.97}^{+ 1.02+ 2.03}$ & $%
68.29$  \\ \hline
$\chi^2$ & & 13676.524 & & 13676.058 & & 13676.224 \\
\hline 
\end{tabular}%
\caption{Reconstruction of the interacting DE scenario with constant
equation of state of DE, $w_x$, for the spatially flat FLRW universe using
the observational data CMB $+$ JLA $+$ BAO $+$ CC. In the columns `$\times$' means that the corresponding parameter has not been considered into the analyses while the sign `$-$' present against any parameter means its mean value has been fixed for the subsequent analyses.  }
\label{tab:Int-wCDM-flat}
\end{table*}
\end{center}
\endgroup                                                                                                                       
\begingroup                                                                                                                     
\squeezetable                                                                                                                   
\begin{center}                                                                                                                  
\begin{table}                                                                                                                   
\begin{tabular}{|c|c|c|c|}                                                                                                            
\hline                                                                                                                    
Parameters & Mean with errors & Best fit \\ \hline
$\Omega_c h^2$ & $    0.1161_{-    0.0039-    0.0076}^{+    0.0041+    0.0073}$ & $    0.1136$\\

$\Omega_b h^2$ & $    0.02225_{-    0.00017-    0.00034}^{+    0.00017+    0.00034}$ & $    0.02206$\\

$100\theta_{MC}$ & $    1.04103_{-    0.00048-    0.00081}^{+    0.00041+    0.00090}$ & $    1.04126$\\

$\tau$ & $    0.081_{-    0.018-    0.033}^{+    0.017+    0.035}$ & $    0.072$\\

$n_s$ & $    0.9696_{-    0.0065-    0.0113}^{+    0.0059+    0.0118}$ & $    0.9693$\\

${\rm{ln}}(10^{10} A_s)$ & $    3.095_{-    0.036-    0.065}^{+    0.032+    0.067}$ & $    3.080$\\

$w_x$ & $   -1.055_{-    0.065-    0.130}^{+    0.067+    0.127}$ & $   -1.051$\\

$\delta_0$ & $-0.168_{-    0.133-    0.264}^{+    0.133+    0.274}$ & $   -0.288$\\

$\delta_1$ & $ 0.165_{-    0.130-    0.268}^{+    0.131+    0.258}$ & $    0.283$\\

$\delta_2$ & $    0.082_{-    0.503-    1.082}^{+    0.559+    0.918}$ & $    0.441$\\

$\delta_3$ &  $0.010_{-    0.263-    1.010}^{+    0.990+    0.990}$ & $   -0.639$\\

$\Omega_{m0}$ & $    0.303_{-    0.016-    0.027}^{+    0.012+    0.028}$ & $    0.303$\\

$\sigma_8$ & $    0.830_{- 0.017- 0.033}^{+ 0.017+ 0.033}$ & $    0.817$\\

$H_0$ & $   67.80_{-    1.05-    2.13}^{+    1.07+    2.06}$ & $   67.09$\\
\hline
$\chi^2$ & & 13674.762 \\
\hline                                                                                                                    
\end{tabular}                                                                                                                   
\caption{Reconstruction of the interacting DE scenario with constant
equation of state of DE, $w_x$, for the spatially flat FLRW universe  
and for the general parametrization $\protect\delta (a) = 
\protect\delta_0 + \protect\delta_1 (1-a) + \protect\delta_2 (1-a)^2+ 
\protect\delta_3 (1-a)^3$, using the combined analysis CMB $+$ JLA $+$ BAO $+$ CC. }\label{tab:Int-wCDM-flat-general}                                                                                                   
\end{table}                                                                                                                     
\end{center}                                                                                                                    
\endgroup                                                                                                                       
\begin{figure*}
\includegraphics[width=0.5\textwidth]{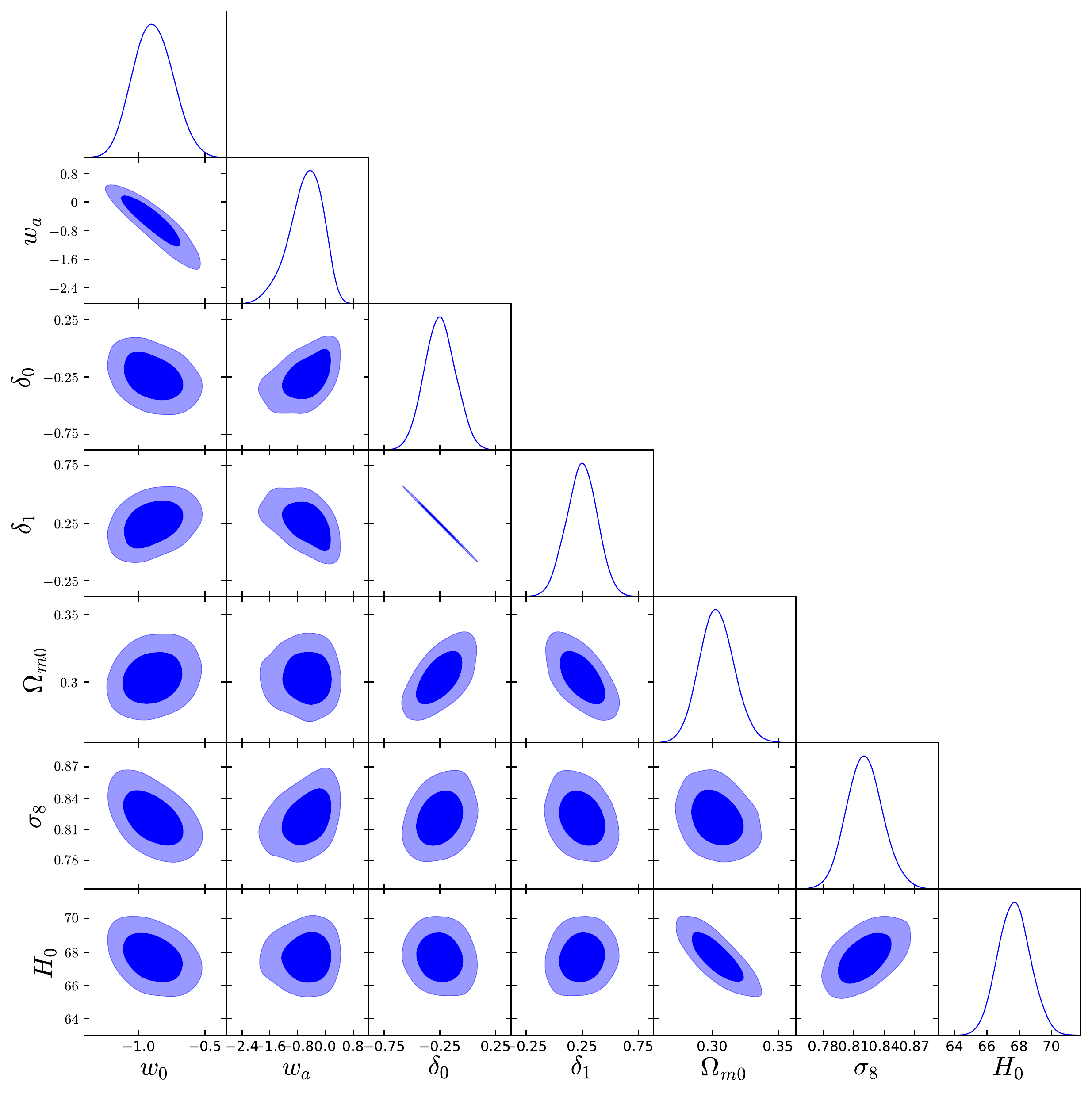}
\caption{68\% and 95\% CL joint contours and one dimensional posterior distributions for the interacting DE scenario (spatially flat case) 
with dynamical DE state parameter $w_x (a) = w_0 + w_a (1-a)$,
where the interaction is parametrized by $\protect\delta (a) = \protect\delta%
_0 + \protect\delta_1 (1-a)$ and the combined observational analysis is CMB $%
+$ JLA $+$ BAO $+$ CC. The results of this analysis are shown in the second and third columns of Table \ref{tab:Int-dyn-flat}. }
\label{fig:int-dyn-flat-1}
\end{figure*}
\begin{figure*}
\includegraphics[width=0.5\textwidth]{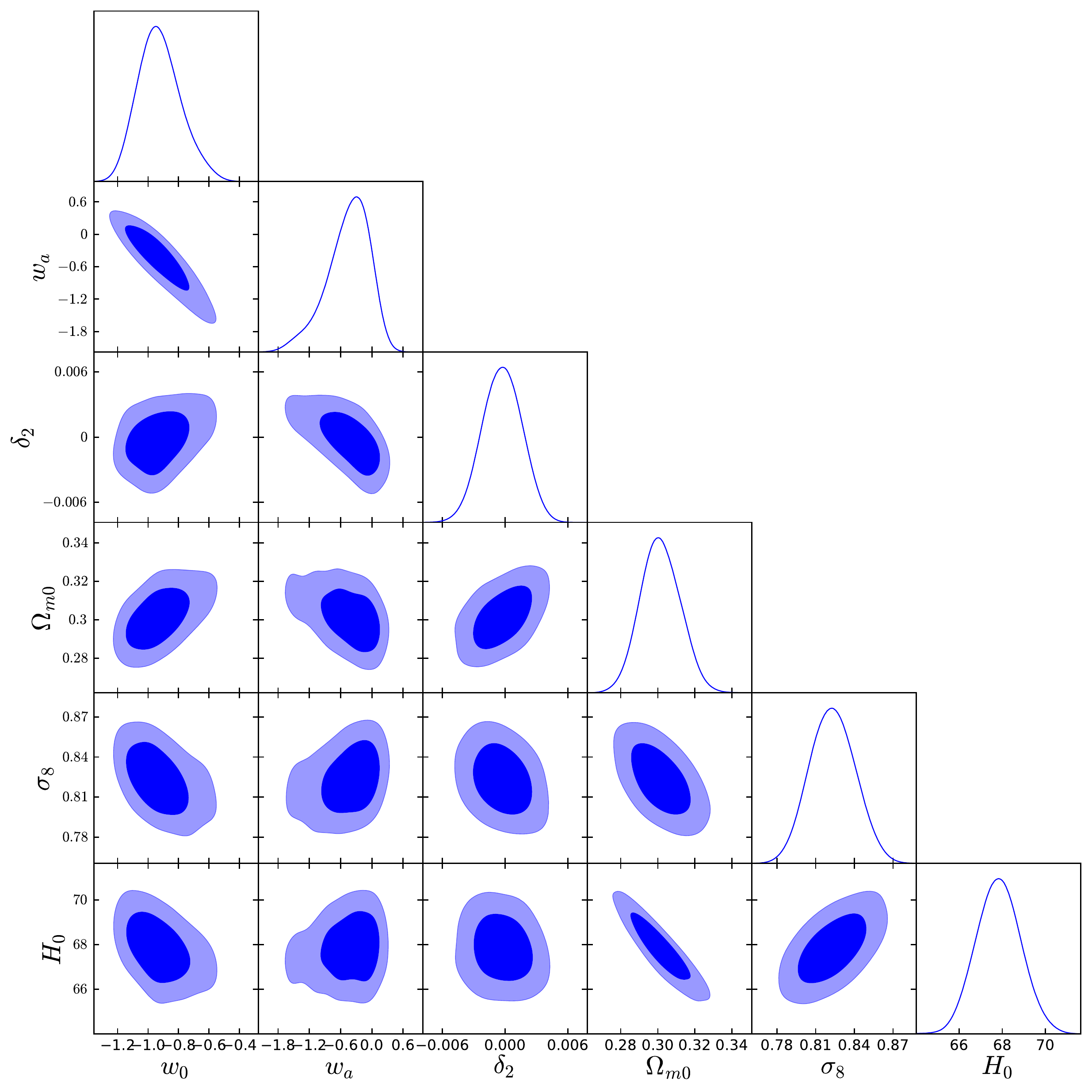}
\caption{68\% and 95\% CL joint contours and one dimensional posterior distributions for the interacting DE scenario (spatially flat case) 
with dynamical DE state parameter $w_x (a) = w_0 + w_a (1-a)$,
where the interaction is parametrized by $\protect\delta (a) = \protect\delta_0 + \protect\delta_1 (1-a) +\protect\delta_2 (1-a)^2$ and the mean values of ($\delta_0$, $\delta_1$) are fixed from Table \ref{tab:Int-dyn-flat}. 
The combined observational analysis is taken to be as usual CMB $+$ JLA $+$ BAO $+$ CC and the results of this analysis are shown in the fourth and fifth columns of Table \ref{tab:Int-dyn-flat}.  }
\label{fig:int-dyn-flat-2}
\end{figure*}
\begin{figure*}
\includegraphics[width=0.5\textwidth]{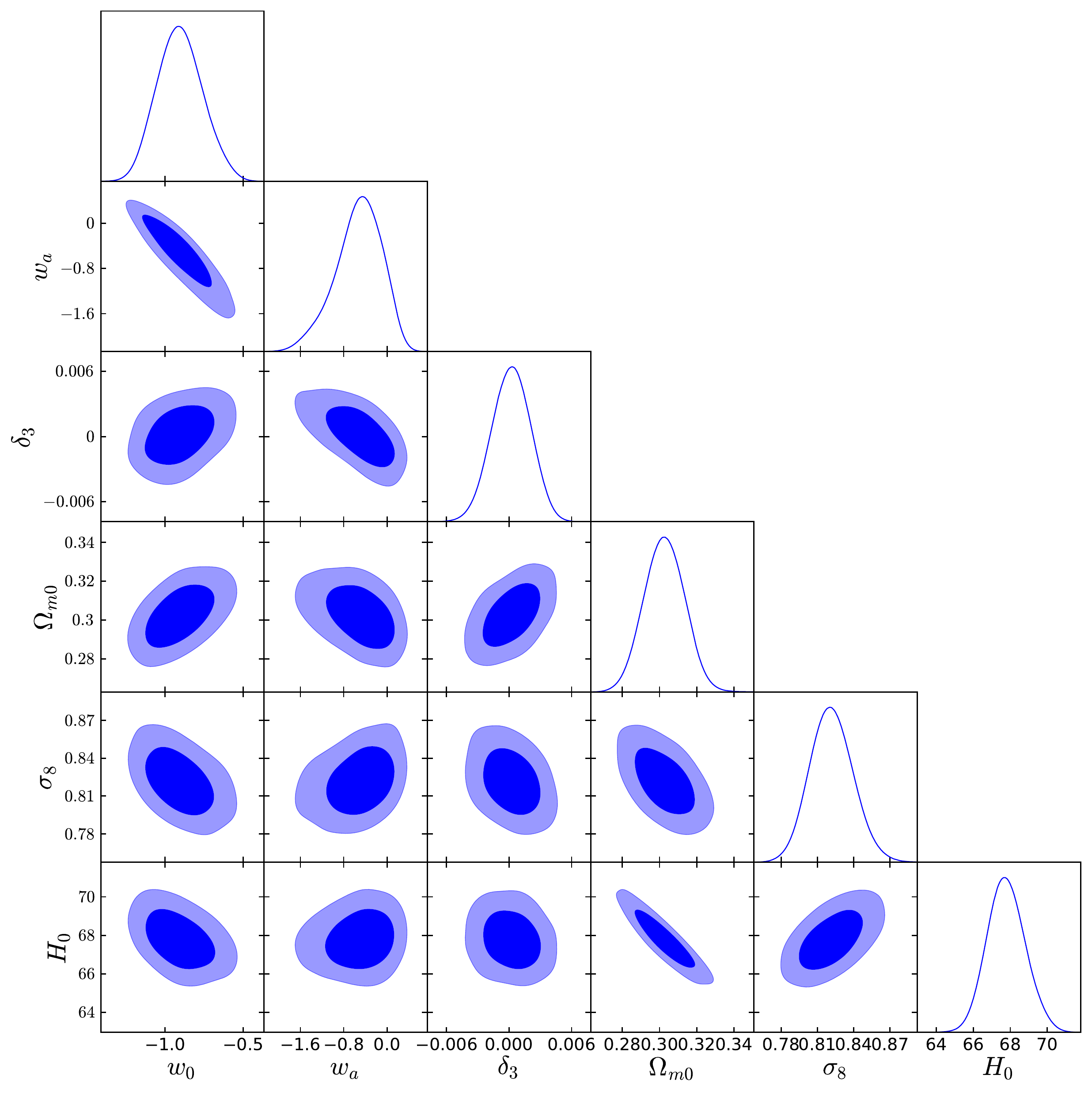}
\caption{68\% and 95\% CL joint contours and one dimensional posterior distributions for the interacting DE scenario (spatially flat case) 
with dynamical DE state parameter $w_x (a) = w_0 + w_a (1-a)$,
where the interaction is parametrized by $\protect\delta (a) = \protect\delta%
_0 + \protect\delta_1 (1-a) +\protect\delta_2 (1-a)^2 +\protect\delta_3
(1-a)^3$ and the parameters ($\delta_0$, $\delta_1$, $\delta_2$) are fixed at their mean values from Table \ref{tab:Int-dyn-flat}. The combined observational analysis is CMB $+$ JLA $+$ BAO $+$ CC and the results of this analysis are shown in the last two columns of Table \ref{tab:Int-dyn-flat}. }
\label{fig:int-dyn-flat-3}
\end{figure*}

\subsubsection{Interacting DE with dynamical EoS}
\label{sec-int-DE-Dyn-flat}

We now consider the interacting scenario between dark matter and dark
energy where the dark energy has a dynamical equation of state.
We choose a very widely used dynamical parametrization for DE namely the
Chevallier--Polarski--Linder (CPL) parametrization \cite{Chevallier:2000qy,Linder:2002et} 
given in eqn. (\ref{cpl}).
One may notice that the inclusion of CPL parametrization increases the
parameters space  compared to the other two interacting scenarios.
However, the procedure is exactly same as  that we adopted in the previous two
analyses.

We first consider the parametrization of $\delta (a) =
\delta_0 + \delta_1 (1-a)$ of eqn. (\ref{xi}) and work out the observational
fittings with the same combined analysis. Table \ref{tab:Int-dyn-flat}
summarizes the mean and the best-fit values of the free and derived
parameters of this model (see the second and third column respectively of
Table \ref{tab:Int-dyn-flat}). 
From Table \ref{tab:Int-dyn-flat} we see that the mean values of both 
$\delta_0$, $\delta_1$ are
nonzero and within 68\% CL, they are strictly nonzero. However, within 95\% CL,
both of them allow their zero values.
Looking at the current value of the dark energy state parameter, $w_0$, a
perfect quintessential scenario is observed ($w_0 =
-0.892_{- 0.152}^{+ 0.151}$ at 68\% CL) while however, 
within 68\% CL, 
$-1.044 <w_0 < -0.741$, implying that the phantom scenario is also equally 
allowed. Fig. \ref{fig:int-dyn-flat-1} reflects the corresponding graphical variations for this scenario.    
Similar to the previous two interaction scenarios, we again observe that the parameters $\delta_0$ and $\delta_1$ are strongly correlated to each other in the negative orientation, and they together yield an insignificantly small value for $\delta$. 

As a second step, we consider the next parametrization (\ref{xi-2}) in this
series having three free parameters $\delta_0$, $\delta_1$ and $\delta_2$.
We apply the same procedure, fix the parameters $\delta_0$, $%
\delta_1$ to their corresponding mean values and constrain the interaction 
scenario to constrain the parameter $\delta_2$. The
results of the analysis have been summarized in Table \ref{tab:Int-dyn-flat}
(fourth and fifth columns) and in Fig. \ref{fig:int-dyn-flat-2} we show the corresponding graphical variations. 
We find that even if we allow the dark energy
state parameter to be dynamical, but the interaction parameter $\delta_2$ is
very small, $\delta_2= -0.000273_{- 0.001874}^{+ 0.001934}$
(68\% CL) and the zero value of $\delta_2$ is allowed in this
CL. The dark energy state parameter (the mean and best-fit values) shows its
quintessential behaviour although within 68\% CL, the phantom crossing is allowed.

Now we do the analysis with the last parametrization of $\delta (a)$
as in equation (\ref{xi-3}) where we fix the interaction parameters $\delta_0$, 
$\delta_1$, $\delta_2$ from their previous analyses and constrain $\delta_3$
using the same observational data. The observational constraints have been
summarized in the last two columns of Table \ref{tab:Int-dyn-flat} where we
find that $\delta_3 = 0.000207_{- 0.001865}^{+ 0.001854}$ at 68\% CL and
one can see that within this CL, $\delta_3 =0$ is possible. The current
value of the dark energy state parameter is found to be $w_0 = -0.905_{-
0.157}^{+ 0.142}$ (68\% CL) which shows its quintessential
character but phantom nature is still allowed. The allowance of the phantom
state parameter in dark energy is further supported from its best fit value
which crosses the phantom boundary.  One may see Fig. \ref{fig:int-dyn-flat-3}
for the graphical variations of this scenario. 

\begin{figure*}
\includegraphics[width=0.6\textwidth]{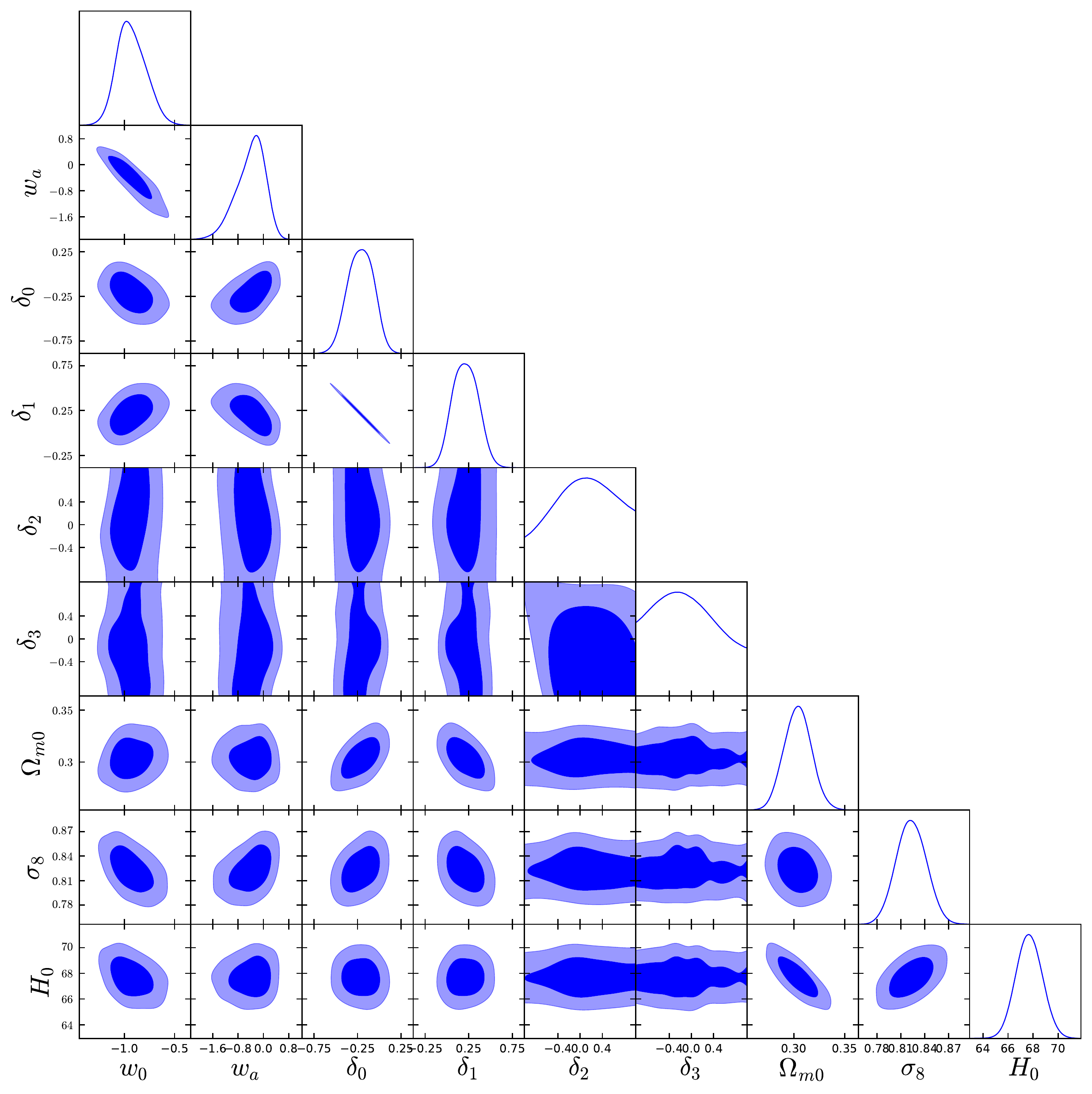}
\caption{68\% and 95\% CL joint contours and one dimensional posterior distributions for the interacting DE scenario (spatially flat case) 
with dynamical DE state parameter $w_x (a) = w_0 + w_a (1-a)$,
for the most general 
parametrization $\protect\delta (a) = 
\protect\delta_0 + \protect\delta_1 (1-a) + \protect\delta_2 (1-a)^2+ 
\protect\delta_3 (1-a)^3$ where  the interaction parameters $\delta_i$'s
are kept free. The combined analysis is CMB $+$ JLA $+$ BAO $+$ CC and the results are summarized in Table \ref{tab:Int-dyn-flat-general}. One can clearly notice that the parameters $\delta_2$ and $\delta_3$ are degenerate similar to the interaction vacuum scenario.}
\label{fig:int-dyn-flat-4}
\end{figure*}

Similar to the previous two interaction models, we constrain the scenario for the general parametrization of eqn. (\ref{xi-3}) taking all the interaction parameters $\delta_i$'s to be free. The dimension of the parameters space for this interaction scenario is twelve which is double of the dimension of the parameters space for the non-interacting $\Lambda$CDM model. We employ the same combined analysis and show the results of the free parameters in Table \ref{tab:Int-dyn-flat-general}. And in Fig. \ref{fig:int-dyn-flat-4} we also show the contour plots for different combinations of the free parameters. From the Fig. \ref{fig:int-dyn-flat-4}, one can clearly see that in a similar fashion, the parameters $\delta_2$, $\delta_3$ are degenerate, at least with the present astronomical data. Finally, we comment on the $\chi^2$ values that we obtain for the best-fit values for this particular scenario. In Tables \ref{tab:Int-dyn-flat} and \ref{tab:Int-dyn-flat-general}, we have shown the values $\chi^2$ for different reconstructed scenario. As already commented, we are not bothered about the $\chi^2$ value for the general scenario corresponding to the results in Table \ref{tab:Int-dyn-flat-general} since some of the parameters are not constrained, so we are inclined towards the $\chi^2$ values shown in Table  \ref{tab:Int-dyn-flat}. From this table we find that the results are not similar to what we find in  the cases with interacting vacuum and interacting DE with $w_x \neq -1$. The dynamical character in $w_x (a)$ works differently in this case, as we observe. 

\begingroup                                                                                                                     
\squeezetable                                                                                                                   
\begin{center}
\begin{table*}
\begin{tabular}{|c|c|c|c|c|c|c|c}
\hline
Parameters & Mean with errors & Best fit & Mean with errors & Best fit & 
Mean with errors & Best fit   \\ \hline
$\Omega_c h^2$ & $0.1160_{- 0.0038- 0.0075}^{+ 0.0038+ 0.0073}$ & $%
0.1193$ & $0.1157_{- 0.0022- 0.0042}^{+ 0.0022+ 0.0041}$ & $%
0.1164$ & $0.1160_{- 0.0021- 0.0041}^{+ 0.0022+ 0.0040}$ & $%
0.1154$  \\ 

$\Omega_b h^2$ & $0.02225_{- 0.00017- 0.00033}^{+ 0.00017+ 0.00034}$ & $%
0.02244$ & $0.02224_{- 0.00017- 0.00034}^{+ 0.00017+ 0.00036}$ & $%
0.02233$ & $0.02225_{- 0.00017- 0.00034}^{+ 0.00017+ 0.00034}$ & $%
0.02219$   \\ 

$100\theta_{MC}$ & $1.04104_{- 0.00042- 0.00084}^{+ 0.00042+ 0.00085}$
& $1.04091$ & $1.04104_{- 0.00033- 0.00067}^{+ 0.00034+ 0.00063}$ & $%
1.04082$ & $1.04102_{- 0.00035- 0.00068}^{+ 0.00035+ 0.00066}$ & $%
1.04113$   \\ 

$\tau$ & $0.080_{- 0.017- 0.034}^{+ 0.017 + 0.034}$ & $%
0.057 $ & $0.079_{- 0.017- 0.032}^{+ 0.016 + 0.032}$ & $%
0.062$ & $0.078_{- 0.017- 0.032}^{+ 0.017+ 0.034}$ & $%
0.076$   \\ 

$n_s$ & $0.9695_{- 0.0061- 0.0115}^{+ 0.0061+ 0.0122}$ & $0.9682$
& $0.9697_{- 0.0047- 0.0091}^{+ 0.0047+ 0.0093}$ & $0.9707$ & $%
0.9693_{- 0.0046- 0.0091}^{+ 0.0046+ 0.0093}$ & $0.9710$   \\ 

$\mathrm{ln}(10^{10} A_s)$ & $3.092_{- 0.034- 0.065}^{+ 0.033+
0.066}$ & $3.047$ & $3.090_{- 0.032- 0.063}^{+ 0.032+
0.064}$ & $3.058$ & $3.089_{- 0.035- 0.063}^{+ 0.032+
0.065}$ & $3.083$  \\ \hline 

$\delta_0$ & $-0.249_{- 0.141- 0.270}^{+ 0.134+ 0.273}$ & $%
-0.193$ & $-$ & $-$ & $-$ & $-$   \\

$\delta_1$ & $0.246_{- 0.131- 0.268}^{+ 0.147+ 0.264}$ & $%
0.194$ & $-$ & $-$ & $-$ & $-$   \\ 

$\delta_2$ & $\times$ & $\times$ & $-0.000273_{- 0.001874- 0.003698}^{+ 0.001934+
0.003600}$ & $0.000609$ & $-$ & $-$   \\ 

$\delta_3$ & $\times$ & $\times$ & $\times$ & $\times$ & $0.000207_{- 0.001865- 0.003653}^{+
0.001854+ 0.003487}$ & $-0.000995$   \\  \hline 

$w_0$ & $-0.892_{- 0.152- 0.293}^{+ 0.151+ 0.297}$ & $%
-0.907$ & $-0.926_{- 0.156- 0.272}^{+ 0.132+ 0.295}$ & $%
-0.942$ & $-0.905_{- 0.157- 0.280}^{+ 0.142+ 0.294}$ & $%
-1.112$   \\ 

$w_a$ & $-0.562_{- 0.380- 1.023}^{+ 0.594+ 0.897}$ & $%
-0.727$ & $-0.450_{- 0.282- 0.906}^{+ 0.517+ 0.736}$ & $%
-0.379$ & $-0.526_{- 0.333- 0.900}^{+ 0.542+ 0.797}$ & $%
0.092$   \\ 

$\Omega_{m0}$ & $0.303_{- 0.013- 0.025}^{+ 0.013+ 0.027}$ & $%
0.303$ & $0.301_{- 0.012- 0.020}^{+ 0.011+ 0.021}$ & $%
0.309$ & $0.303_{- 0.011- 0.021}^{+ 0.011+ 0.021}$ & $%
0.304$   \\
 
$\sigma_8$ & $0.821_{- 0.019- 0.034}^{+ 0.017+ 0.037}$ & $%
0.809$ & $0.823_{- 0.018- 0.033}^{+ 0.018+ 0.035}$ & $%
0.804$ & $0.822_{- 0.019- 0.033}^{+ 0.017+ 0.036}$ & $%
0.829$   \\ 

$H_0$ & $67.68_{- 1.08- 1.93}^{+ 0.97+ 2.03}$ & $%
68.52$ & $67.83_{- 1.03- 1.96}^{+ 1.02+ 2.01}$ & $%
67.18$ & $67.78_{- 1.09- 1.93}^{+ 0.99 + 2.07}$ & $%
67.46$  \\ \hline
$\chi^2$ & & 13674.426 & & 13676.256 & & 13675.758 \\
\hline 
\end{tabular}%
\caption{Reconstruction of the dark matter-dark energy interaction scenario
in a spatially flat universe where dark energy assumes the dynamical state
parameter, namely, the CPL parametrization, $w_x (a) = w_0 + w_a (1-a)$. 
The combined observational data
for all the analyses are, CMB $+$ JLA $+$ BAO $+$ CC. In the columns `$\times$' means that the corresponding parameter has not been considered into the analyses while the sign `$-$' present against any parameter means its mean value has been fixed for the subsequent analyses.  }
\label{tab:Int-dyn-flat}
\end{table*}
\end{center}
\endgroup                                                                                                                       
\begingroup                                                                                                                     
\squeezetable                                                                                                                   
\begin{center}                                                                                                                  
\begin{table}                                                                                                                   
\begin{tabular}{|c|c|c|c|}                                                                                                            
\hline                                                                                                                    
Parameters & Mean with errors & Best fit \\ \hline
$\Omega_c h^2$ & $    0.1162_{-    0.0038-    0.0077}^{+    0.0039+    0.0072}$ & $    0.1203$\\

$\Omega_b h^2$ & $    0.02225_{-    0.00016-    0.00031}^{+    0.00016+    0.00032}$ & $    0.02227$\\

$100\theta_{MC}$ & $    1.04102_{-    0.00043-    0.00086}^{+    0.00043+    0.00087}$ & $    1.04058$\\

$\tau$ & $    0.080_{-    0.017-    0.032}^{+    0.017+    0.032}$ & $    0.067$\\

$n_s$ & $    0.9694_{-    0.0066-    0.0115}^{+    0.0059+    0.0125}$ & $    0.9652$\\

${\rm{ln}}(10^{10} A_s)$ & $    3.092_{-    0.033-    0.063}^{+    0.033+    0.063}$ & $    3.070$\\

$w_0$ & $   -0.930_{-    0.165-    0.266}^{+    0.128+    0.297}$ & $   -1.053$\\

$w_a$ & $   -0.402_{-    0.366-    0.933}^{+    0.540+    0.837}$ & $   -0.198$\\

$\delta_0$ & $   -0.214_{-    0.146-    0.281}^{+    0.163+    0.268}$ & $   -0.097$\\

$\delta_1$ & $    0.211_{-    0.160-    0.262}^{+    0.144+    0.275}$ & $    0.098$\\

$\delta_2$ & $    0.061_{-    0.250-    1.061}^{+    0.9395+    0.940}$ & $    0.047$\\

$\delta_3$ & $   -0.081_{-    0.802-    0.919}^{+    0.347+    1.081}$ & $   -0.347$\\

$\Omega_{m0}$ & $    0.304_{-    0.013-    0.026}^{+    0.013+    0.026}$ & 
$ 0.306$\\

$\sigma_8$ & $    0.823_{-    0.019-    0.036}^{+    0.019+    0.037}$ & $    0.829$\\

$H_0$ & $   67.69_{-    1.03-    1.99}^{+    1.04+    2.07}$ & $   68.45$\\

\hline
$\chi^2$ && 13675.162 \\
\hline                                                                                                                    
\end{tabular}                                                                                                                   
\caption{Reconstruction of the interacting DE scenario with its dynamical EoS   
for the spatially flat FLRW universe  
and for the general parametrization $\protect\delta (a) = 
\protect\delta_0 + \protect\delta_1 (1-a) + \protect\delta_2 (1-a)^2+ 
\protect\delta_3 (1-a)^3$, using the combined analysis CMB $+$ JLA $+$ BAO $+$ CC.}
\label{tab:Int-dyn-flat-general}                                                                                                   
\end{table}                                                                                                                     
\end{center}                                                                                                                    
\endgroup                                                                                                                       
\begin{figure*}
\includegraphics[width=0.5\textwidth]{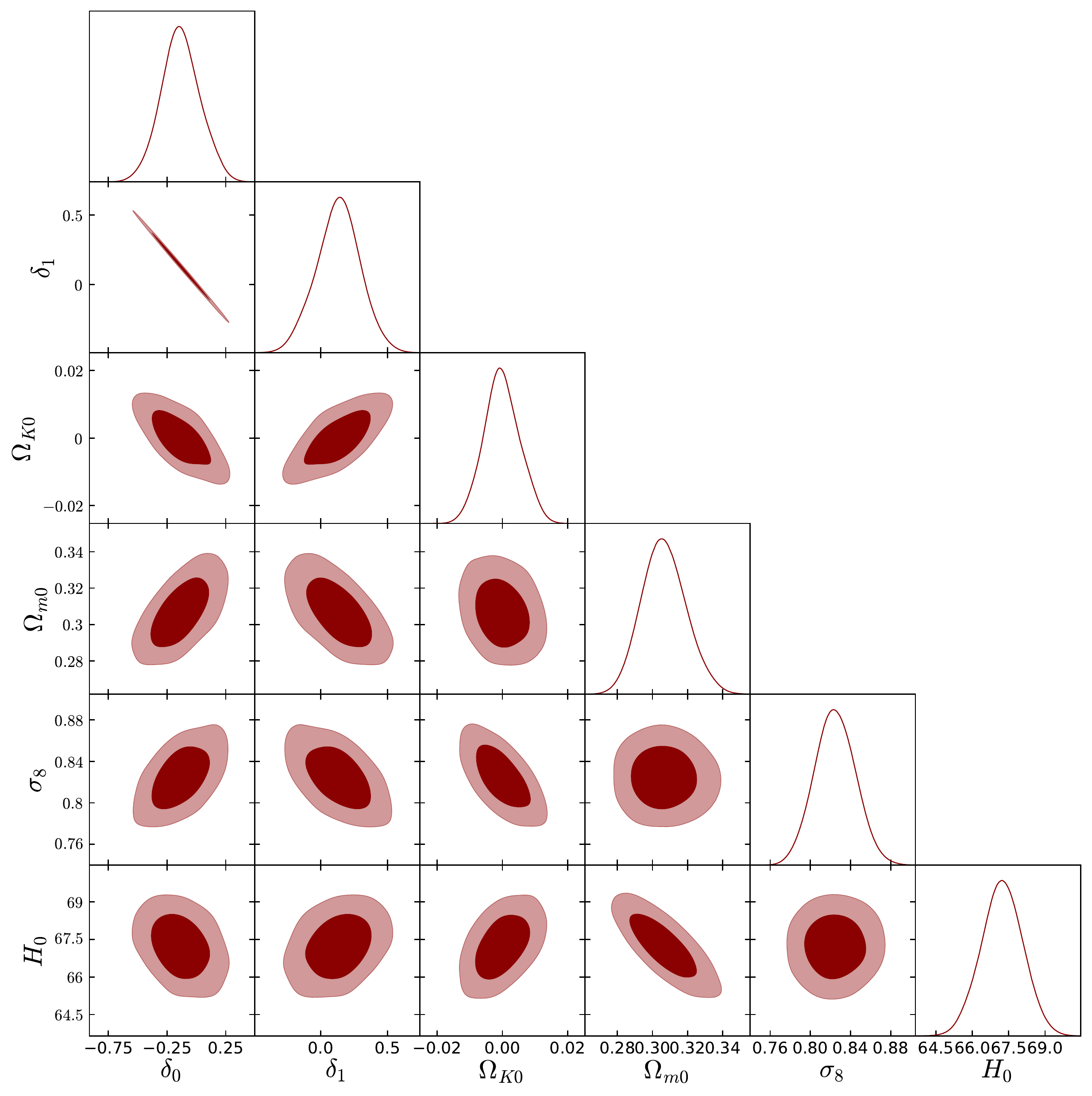}
\caption{68\% and 95\% CL joint contours and one dimensional posterior distributions for the interacting vacuum scenario (nonflat case)
where the interaction is parametrized by $\protect\delta (a) = \protect\delta%
_0 + \protect\delta_1 (1-a)$. The combined data for this analysis have been
set to be CMB $+$ JLA $+$ BAO $+$ CC and the results are displayed in the second and third columns of Table \ref{tab:Int-vacuum-nonflat}. }
\label{fig:int-vacuum-nonflat-1}
\end{figure*}
\begin{figure*}
\includegraphics[width=0.5\textwidth]{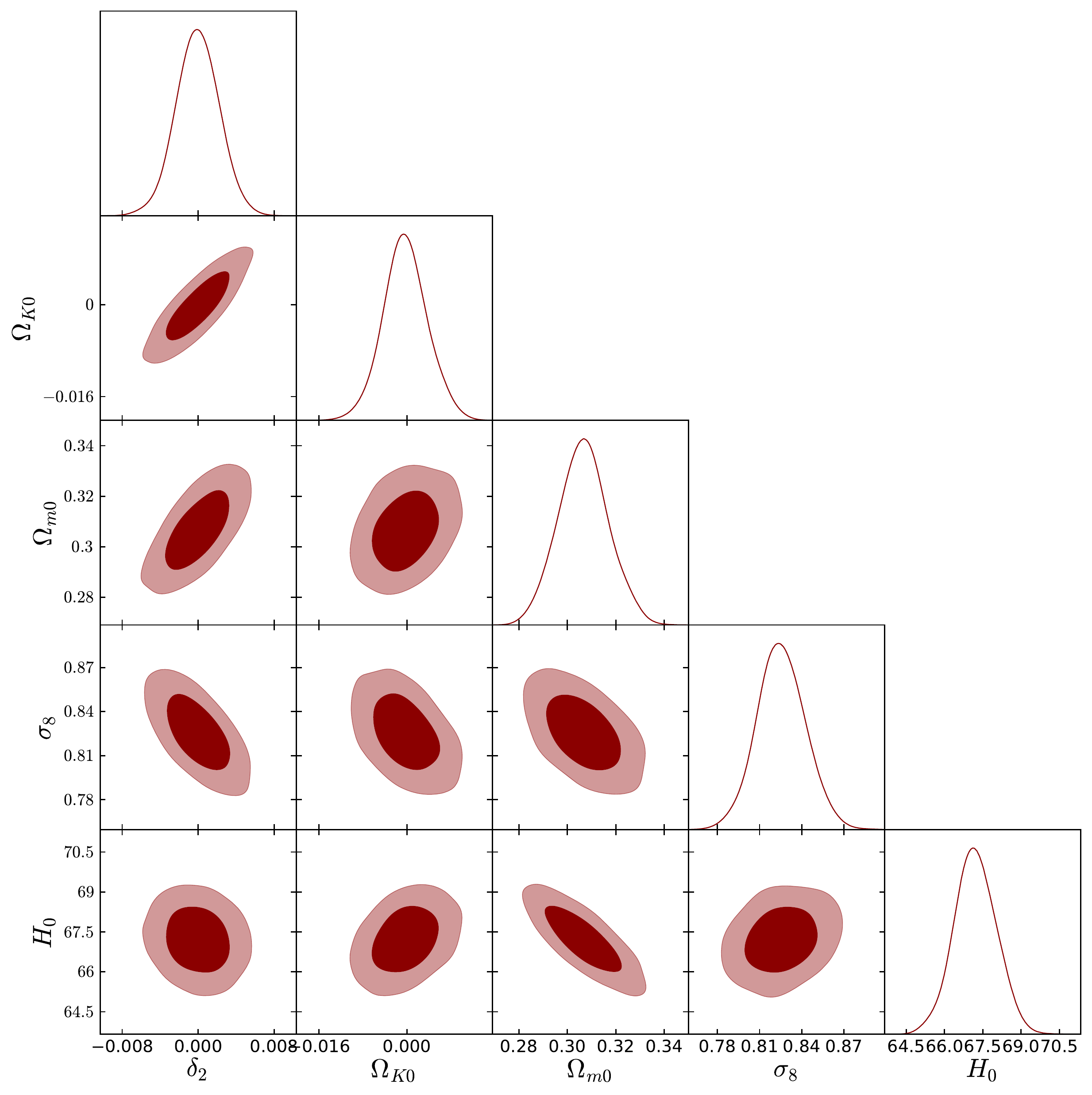}
\caption{68\% and 95\% CL joint contours and one dimensional posterior distributions for the interacting vacuum scenario (nonflat case)
where the interaction is parametrized by $\protect\delta (a) = \protect\delta%
_0 + \protect\delta_1 (1-a) + \protect\delta_2 (1-a)^2$ in which we fix the
values of ($\protect\delta_0$, $\protect\delta_1$) from Table \ref{tab:Int-vacuum-nonflat} but left $\protect\delta_2$ as a free parameter. The combined
observational dataset we fix to be CMB $+$ JLA $+$ BAO $+$ CC and the results are displayed in the fourth and fifth columns of Table \ref{tab:Int-vacuum-nonflat}. }
\label{fig:int-vacuum-nonflat-2}
\end{figure*}
\begin{figure*}
\includegraphics[width=0.5\textwidth]{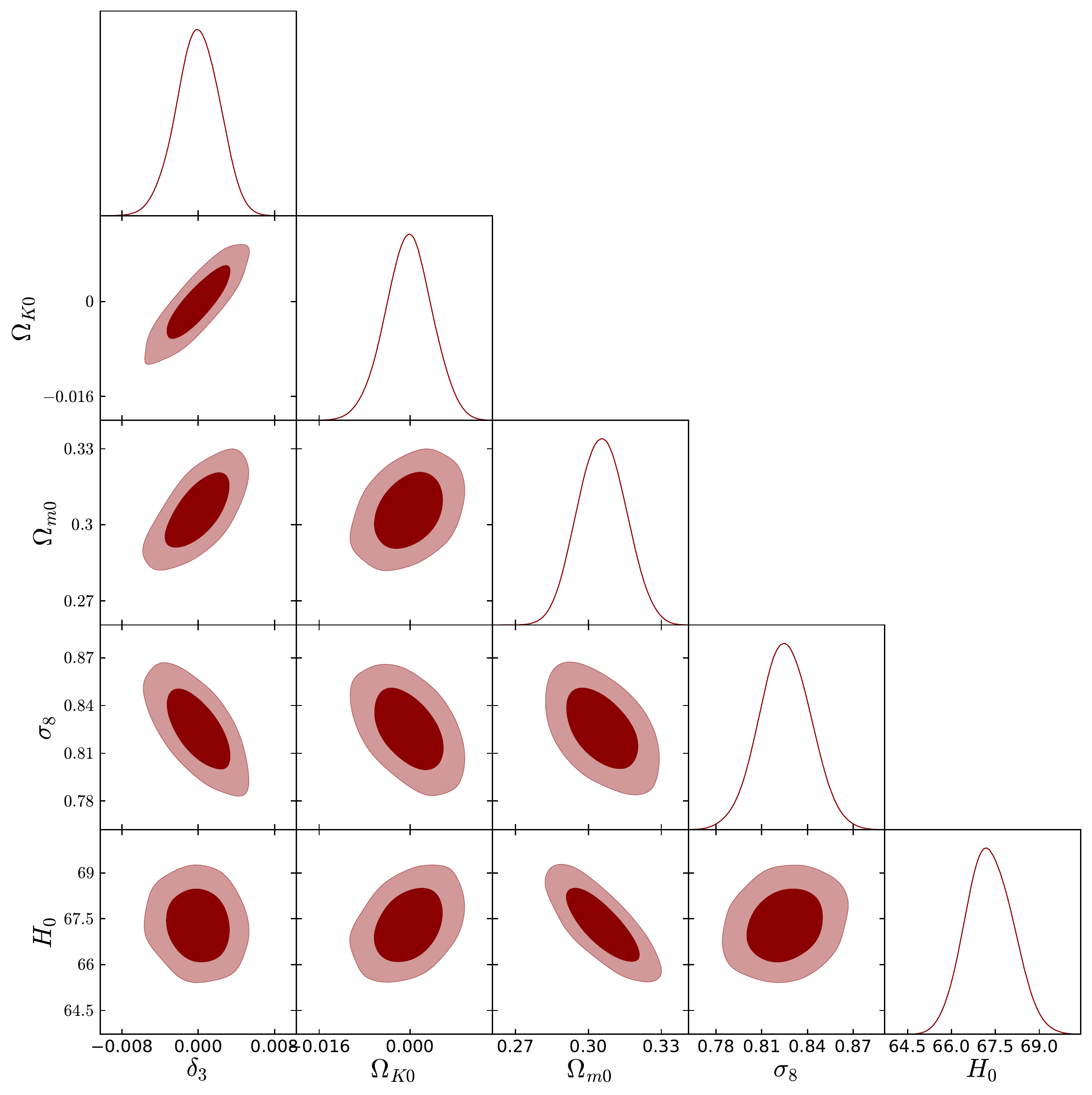}
\caption{68\% and 95\% CL joint contours and one dimensional posterior distributions for the interacting vacuum scenario (nonflat case)
where the interaction is parametrized by $\protect\delta (a) = \protect\delta%
_0 + \protect\delta_1 (1-a) + \protect\delta_2 (1-a)^2 +\protect\delta_3
(1-a)^3$ in which we fix the values of ($\protect\delta_0$, $\protect\delta%
_1 $, $\protect\delta_2$) from the previous analysis (Table \ref{tab:Int-vacuum-nonflat}) but left $\protect\delta_3$ as a free parameter. The combined observational data we set to be CMB $+$ JLA $+$ BAO $+$ CC and the results are shown in the last two columns of Table \ref{tab:Int-vacuum-nonflat}. }
\label{fig:int-vacuum-nonflat-3}
\end{figure*}

\subsection{Reconstruction of the interaction rate in the nonflat universe}

\label{results:nonflat}

For the three particular type of the dark energy fluids, we now consider a universe with a nonzero spatial curvature. The model parameters are estimated against the same four observational dataset CMB $+$ JLA $+$ BAO $+$ CC.

\subsubsection{Interacting vacuum}

In the beginning we consider that the
interaction parameter $\delta (a)$ is characterized by its Taylor series
expansion (\ref{xi}) with two free parameters $\delta_0$ and $\delta_1$. 
We constrained this scenario with the same observational data and present
the results in the second and third column of Table \ref{tab:Int-vacuum-nonflat}. 
From the estimations of $\delta_0$ and $\delta_1$, namely, 
$\delta_0 = -0.135_{- 0.177}^{+ 0.161}$ (68\% CL) and $\delta_1 = 0.132_{- 0.158}^{+ 0.174}$ (68\% CL), one can see that the mean values of $\delta_0$ and $\delta_1$ are nonzero, but, 
$(\delta_0, \delta_1) = (0, 0)$ are allowed by the data within 68\% CL,  which means that according to
the observational data, the model hardly distinguishes between an interacting and non-interacting scenario.   
We also estimated the curvature of the 
universe, $\Omega_{K}= - K/(aH)^2$, at present time,
i.e., $\Omega_{K0}$, for this interaction scenario.  
From Table \ref{tab:Int-vacuum-nonflat}, we find that, 
$\Omega_{K0}= -0.00016_{- 0.00585}^{+ 0.00549}$ at 68\% CL
(i.e., $-0.00601 < \Omega_{K0} < 0.00533 $), that means, statistically, all the possibilities of the universe (open, flat, closed) are almost equally allowed. In Fig. \ref{fig:int-vacuum-nonflat-1}  we have clearly displayed the graphical variations for this interaction scenario.  
We now point out one of the interesting observations from Fig. \ref{fig:int-vacuum-nonflat-1}. From this figure, one can see that the parameters $(\delta_0, \delta_1)$ are (strongly-) negatively correlated to each other, very similar to the interacting scenarios without the presence of curvature. So, one can realize that the presence of curvature cannot alter the behaviour of these two parameters.

We now consider the extended 
parametrization of $\delta (a)$ given in eqn. (\ref{xi-2}) that includes one more free parameter $\delta_2$. 
We fix the values of $(\delta_0, \delta_1)$ to
their corresponding mean values obtained from the analysis with eqn. (\ref{xi}) and constrain the model with one parameter left, i.e., $\delta_2$. The
observational summary has been given in the fourth and fifth columns of
Table \ref{tab:Int-vacuum-nonflat}.  Fig. \ref{fig:int-vacuum-nonflat-2} 
displays the corresponding graphical variations for this model.
From the analysis, we see that, $\delta_2$ is very very small ($\delta_2= -0.000066_{- 0.002243}^{+ 0.002264}$ at 68\% CL) and within 68\% CL, $\delta_2 = 0$ is allowed. 
From the estimation of the curvature scalar, we find that $\Omega_{K0}$ is sufficiently small, $\Omega_{K0} = -0.00032_{- 0.00424}^{+ 0.00390}$ (at 68\% CL),  but higher 
than the radiation density at present ($\mathcal{O} (10^{-4})$). Since $\Omega_{K0}$ is found to allow both positive and negative values including $\Omega_{K0} = 0$, thus, all the possibilities of the universe (open, flat and closed) are statistically allowed. 

We  discuss the last parametrization of $\delta(a)$ in 
eqn. (\ref{xi-3}) and fix the parameters $\delta_i$'s ($i = 0, 1, 2$) to their estimated
mean values, see Table \ref{tab:Int-vacuum-nonflat} and wish to
constrain the remaining free interaction parameter $\delta_3$. The results
of the analysis have been summarized in the last two columns of Table \ref{tab:Int-vacuum-nonflat} 
and in Fig. \ref{fig:int-vacuum-nonflat-3} we present the graphical behaviour of the free parameters of the model.  
From the analysis one can see that the interaction
parameter $\delta_3$ is very small with $\delta_3 =
-0.000041_{- 0.002167}^{+ 0.002268}$ (68\% CL) where $\delta_3 = 0$ is also allowed. 
The curvature parameter is constrained to be, $\Omega_{K0} = -0.00024_{- 0.00404}^{+ 0.00412}$ (68\% CL). Thus, we see that curvature density parameter is higher than that of the radiation density at present ($\mathcal{O} (10^{-4})$) and follows similar conclusion as found in the previous cases.

\begin{figure*}
\includegraphics[width=0.6\textwidth]{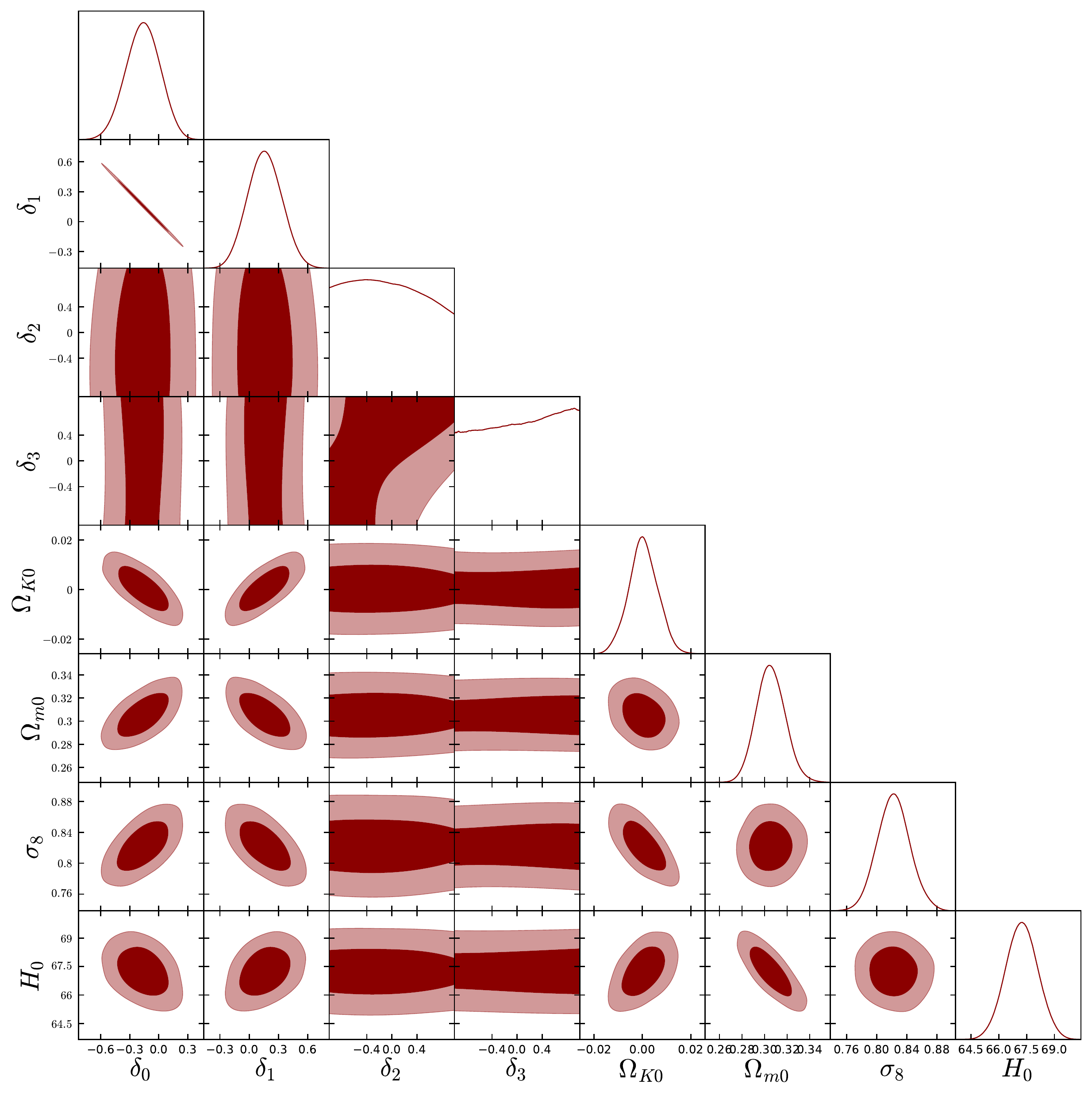}
\caption{68\% and 95\% CL joint contours and one dimensional posterior distributions for the interacting vacuum scenario (nonflat case) using the most 
general parametrization $\protect\delta (a) = \protect\delta%
_0 + \protect\delta_1 (1-a) + \protect\delta_2 (1-a)^2 +\protect\delta_3
(1-a)^3$. The combined observational data we set to be CMB $+$
JLA $+$ BAO $+$ CC and the results of the analysis are shown in 
Table \ref{tab:Int-vacuum-nonflat-general}. }
\label{fig:int-vacuum-nonflat-4}
\end{figure*}

 We now consider the general parametrization for $\delta (a)$ given 
in equation (\ref{xi-3}) and constrain the interaction model using the same datasets. See Table \ref{tab:Int-vacuum-nonflat-general} and Fig. \ref{fig:int-vacuum-nonflat-4} for summary. Including the expected degeneracies in the values of $\delta_2$ and $\delta_3$, all conclusions remain the same. Concerning the $\chi^2$ values (shown in Tables \ref{tab:Int-vacuum-nonflat} and \ref{tab:Int-vacuum-nonflat-general}), we have similar observation as already discussed in section \ref{sec-int-vacuum-flat}.

\begingroup                                                                                                                     
\squeezetable                                                                                                                   

\begin{center}
\begin{table*}
\begin{tabular}{|c|c|c|c|c|c|c|c}
\hline
Parameters & Mean with errors & Best fit & Mean with errors & Best fit & 
Mean with errors & Best fit   \\ \hline
$\Omega_c h^2$ & $0.1156_{- 0.0041- 0.0070}^{+ 0.0035+ 0.0077}$ & $%
0.1160$ & $0.1155_{- 0.0028- 0.0056}^{+ 0.0029+ 0.0058}$ & $%
0.1163$ & $0.1154_{- 0.0028- 0.0057}^{+ 0.0028+ 0.0055}$ & $%
0.1149$   \\ 

$\Omega_b h^2$ & $0.02224_{- 0.00017- 0.00033}^{+ 0.00016+ 0.00035}$ & $%
0.02225$ & $0.02224_{- 0.00016- 0.00031}^{+ 0.00016+ 0.00033}$ & $%
0.02223$ & $0.02224_{- 0.00017- 0.00034}^{+ 0.00017+ 0.00033}$ & $%
0.02240$   \\ 

$100\theta_{MC}$ & $1.04106_{- 0.00042- 0.00086}^{+ 0.00043+ 0.00085}$
& $1.04115$ & $1.04107_{- 0.00039- 0.00077}^{+ 0.00039+ 0.00074}$ & $%
1.04087$ & $1.04107_{- 0.00039- 0.00073}^{+ 0.00038+ 0.00078}$ & $%
1.04129$   \\ 

$\tau$ & $0.083_{- 0.017- 0.034}^{+ 0.017+ 0.035}$ & $%
0.075 $ & $0.083_{- 0.017- 0.033}^{+ 0.017+ 0.033}$ & $%
0.088$ & $0.083_{- 0.016- 0.031}^{+ 0.016+ 0.031}$ & $%
0.093$   \\ 

$n_s$ & $0.9709_{- 0.0060- 0.0120}^{+ 0.0060+ 0.0114}$ & $0.9710$
& $0.9706_{- 0.0060- 0.0108}^{+ 0.0055+ 0.0111}$ & $0.9689$ & $%
0.9710_{- 0.0061- 0.0109}^{+ 0.0054+ 0.0110}$ & $0.9742$   \\ 

$\mathrm{ln}(10^{10} A_s)$ & $3.098_{- 0.033- 0.066}^{+ 0.034+
0.068}$ & $3.081$ & $3.099_{- 0.034- 0.064}^{+ 0.033+
0.064}$ & $3.108$ & $3.098_{- 0.031- 0.060}^{+ 0.032+
0.060}$ & $3.121$  \\ \hline

$\delta_0$ & $-0.135_{- 0.177- 0.326}^{+ 0.161+ 0.349}$ & $%
-0.066$ & $-$ & $-$ & $-$ & $-$  \\ 

$\delta_1$ &  $0.132_{- 0.158- 0.345}^{+ 0.174 + 0.322}$ & $%
0.063$ & $-$ & $-$ & $-$ & $-$  \\ 

$\delta_2$ & $\times$ & $\times$ & $-0.000066_{- 0.002243- 0.004542}^{+ 0.002264+
0.004479}$ & $0.000585$ & $-$ & $-$  \\ 

$\delta_3$ & $\times$ & $\times$ & $\times$ & $\times$ & $-0.000041_{- 0.002167- 0.004485}^{+
0.002268+ 0.004200}$ & $-0.000739$   \\ \hline 

$\Omega_{K0}$ & $-0.00016_{- 0.00585- 0.01068}^{+ 0.00549+ 0.01107}$ & $%
-0.00182$ & $-0.00032_{- 0.00424- 0.00807}^{+ 0.00390+ 0.00820}$ & $%
0.00077$ & $-0.00024_{- 0.00404- 0.00829}^{+ 0.00412+ 0.00791}$ & $%
-0.00305$   \\ 

$\Omega_{m0}$ & $0.307_{- 0.014- 0.024}^{+ 0.012+ 0.026}$ & $%
0.303$ & $0.307_{- 0.010- 0.020}^{+ 0.010+ 0.021}$ & $%
0.309$ & $0.306_{- 0.010- 0.019}^{+ 0.010+ 0.019}$ & $%
0.311$  \\ 

$\sigma_8$ & $0.825_{- 0.020- 0.039}^{+ 0.020+ 0.040}$ & $%
0.830$ & $0.825_{- 0.018- 0.034}^{+ 0.017 + 0.035}$ & $%
0.826$ & $0.825_{- 0.016- 0.033}^{+ 0.017 + 0.033}$ & $%
0.838$   \\ 

$H_0$ & $67.23_{- 0.85- 1.69}^{+ 0.85+ 1.67}$ & $%
67.66$ & $67.21_{- 0.88- 1.66}^{+ 0.82+ 1.64}$ & $%
67.16$ & $67.30_{- 0.80- 1.52}^{+ 0.82+ 1.59}$ & $%
66.64$   \\ \hline
$\chi^2$ && 13676.08 &&  13673.516 && 13675.156 \\
\hline 
\end{tabular}%
\caption{Reconstruction of the dark matter-dark energy interaction scenario
for the interacting vacuum universe in the nonflat universe using the
combined analysis CMB $+$ JLA $+$ BAO $+$ CC.  In the columns `$\times$' means that the corresponding parameter has not been considered into the analyses while the sign `$-$' present against any parameter means its mean value has been fixed for the subsequent analyses.  }
\label{tab:Int-vacuum-nonflat}
\end{table*}
\end{center}
\endgroup                                                                                                                       
\begingroup                                                                                                                     
\squeezetable                                                                                                                   
\begin{center}                                                                                                                  
\begin{table}                                                                                                                   
\begin{tabular}{|c|c|c|c|c|}                                                                                                            
\hline                                                                                                                    
Parameters & Mean with errors & Best fit \\ \hline
$\Omega_c h^2$ & $    0.1151_{-    0.0037-    0.0073}^{+    0.0036+    0.0073}$ & $    0.1114$\\

$\Omega_b h^2$ & $    0.02224_{-    0.00016-    0.00032}^{+    0.00017+    0.00033}$ & $    0.02215$\\

$100\theta_{MC}$ & $    1.04109_{-    0.00041-    0.00082}^{+    0.00041+    0.00082}$ & $    1.04135$\\

$\tau$ & $    0.084_{-    0.017-    0.034}^{+    0.017+    0.034}$ & $    0.100$\\

$n_s$ & $    0.9713_{-    0.0059-    0.0119}^{+    0.0059+    0.0118}$ & $    0.9740$\\

${\rm{ln}}(10^{10} A_s)$ & $    3.098_{-    0.034-    0.067}^{+    0.034+    0.066}$ & $    3.130$\\

$\delta_0$ & $   -0.163_{-    0.177-    0.345}^{+    0.177+    0.334}$ & $   -0.380$\\

$\delta_1$ & $    0.160_{-    0.175-    0.331}^{+    0.176+    0.341}$ & $    0.371$\\

$\delta_2$ & $   -0.042_{-    0.958-    0.958}^{+    1.042+    1.042}$ & $   -0.016$\\

$\delta_3$ & $    0.040_{-    1.040-    1.040}^{+    0.960+    0.960}$ & $    0.888$\\

$\Omega_{K0}$ & $    0.00036_{-    0.00573-    0.01187}^{+    0.00612+    0.01168}$ & $    0.0050$\\

$\Omega_{m0}$ & $    0.305_{-    0.013-    0.024}^{+    0.013+    0.026}$ & $    0.288$\\

$\sigma_8$ & $    0.822_{-    0.022-    0.041}^{+    0.021+    0.044}$ & $    0.831$\\

$H_0$ & $   67.24_{-    0.86-    1.69}^{+    0.86+    1.69}$ & $   68.28$\\
\hline
$\chi^2$ && 13673.582 \\
\hline                                                                                                                     
\end{tabular}                                                                                                                   
\caption{Reconstruction of the dark matter-dark energy interaction scenario
for the interacting vacuum universe for the nonflat universe using the 
general parametrization $\protect\delta (a) = \protect\delta%
_0 + \protect\delta_1 (1-a) + \protect\delta_2 (1-a)^2 +\protect\delta_3
(1-a)^3$. The combined analysis has been set to be CMB $+$ JLA $+$ BAO $+$ CC.}\label{tab:Int-vacuum-nonflat-general}                                                                                                   
\end{table}                                                                                                                     
\end{center}                                                                                                                    
\endgroup      
\begin{figure*}
\includegraphics[width=0.5\textwidth]{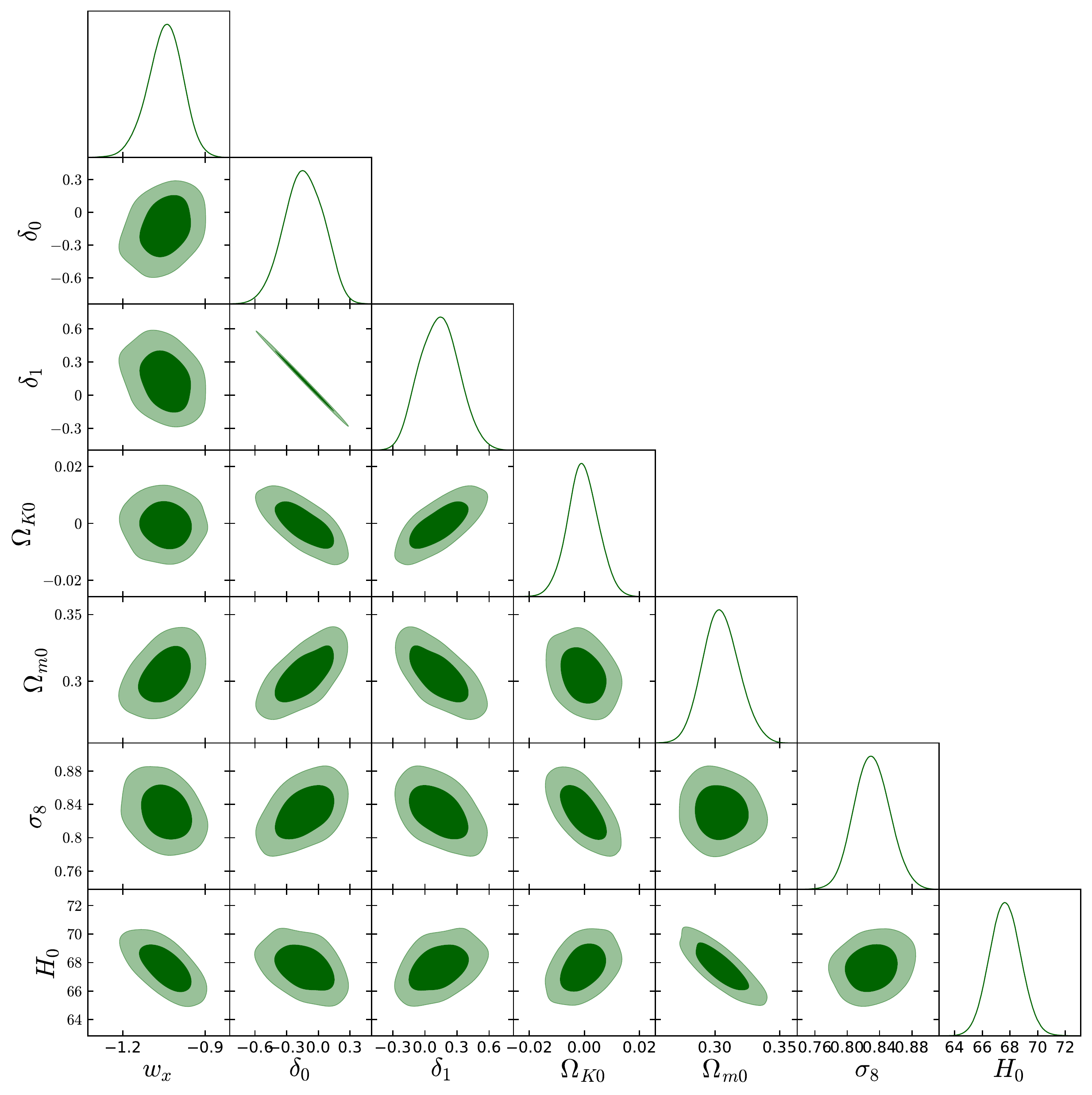}
\caption{68\% and 95\% CL joint contours and one dimensional posterior distributions for the interacting DE scenario (nonflat case) with
constant state parameter in DE, $w_x$, where the interaction is parametrized
by $\protect\delta (a) = \protect\delta_0 + \protect\delta_1 (1-a)$. The
combined data for this analysis have been set to be CMB $+$ JLA $+$ BAO $+$
CC and the results are shown in the second and third columns of 
Table \ref{tab:Int-wCDM-nonflat}.}
\label{fig:int-wCDM-nonflat-1}
\end{figure*}
\begin{figure*}
\includegraphics[width=0.5\textwidth]{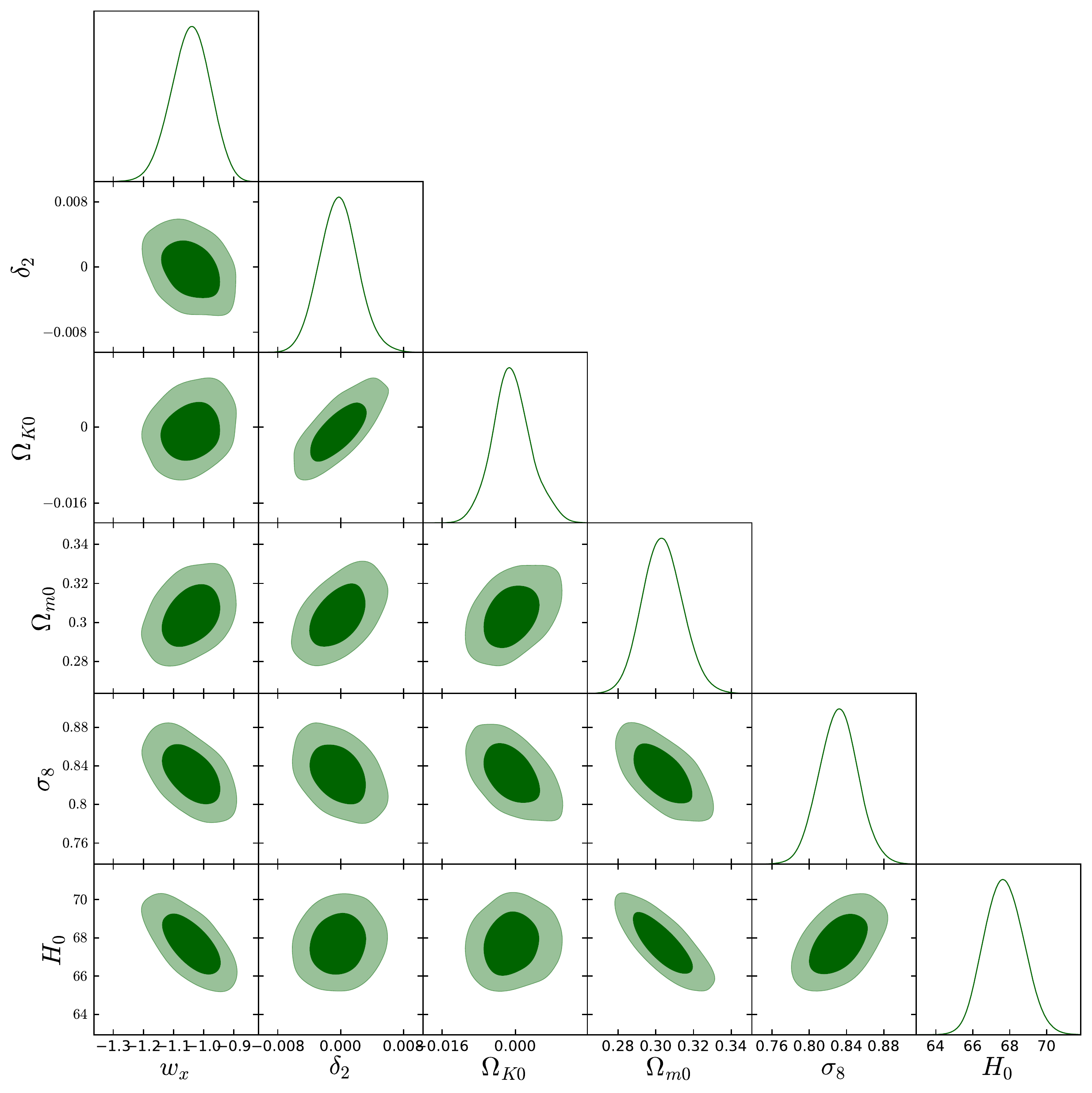}
\caption{68\% and 95\% CL joint contours and one dimensional posterior 
distributions for the interacting DE scenario (nonflat case) with
constant state parameter in DE, $w_x$, where the interaction is parametrized
by $\protect\delta (a) = \protect\delta_0 + \protect\delta_1 (1-a) + \protect%
\delta_2 (1-a)^2$, and in which we fix the first two parameters $(\delta_0, \delta_1)$ at their mean values from Table \ref{tab:Int-wCDM-nonflat}. The combined data for this analysis have been set to be CMB $+$ JLA $+$ BAO $+$ CC and the results are shown in the fourth and fifth columns of  Table \ref{tab:Int-wCDM-nonflat}. }
\label{fig:int-wCDM-nonflat-2}
\end{figure*}
\begin{figure*}
\includegraphics[width=0.5\textwidth]{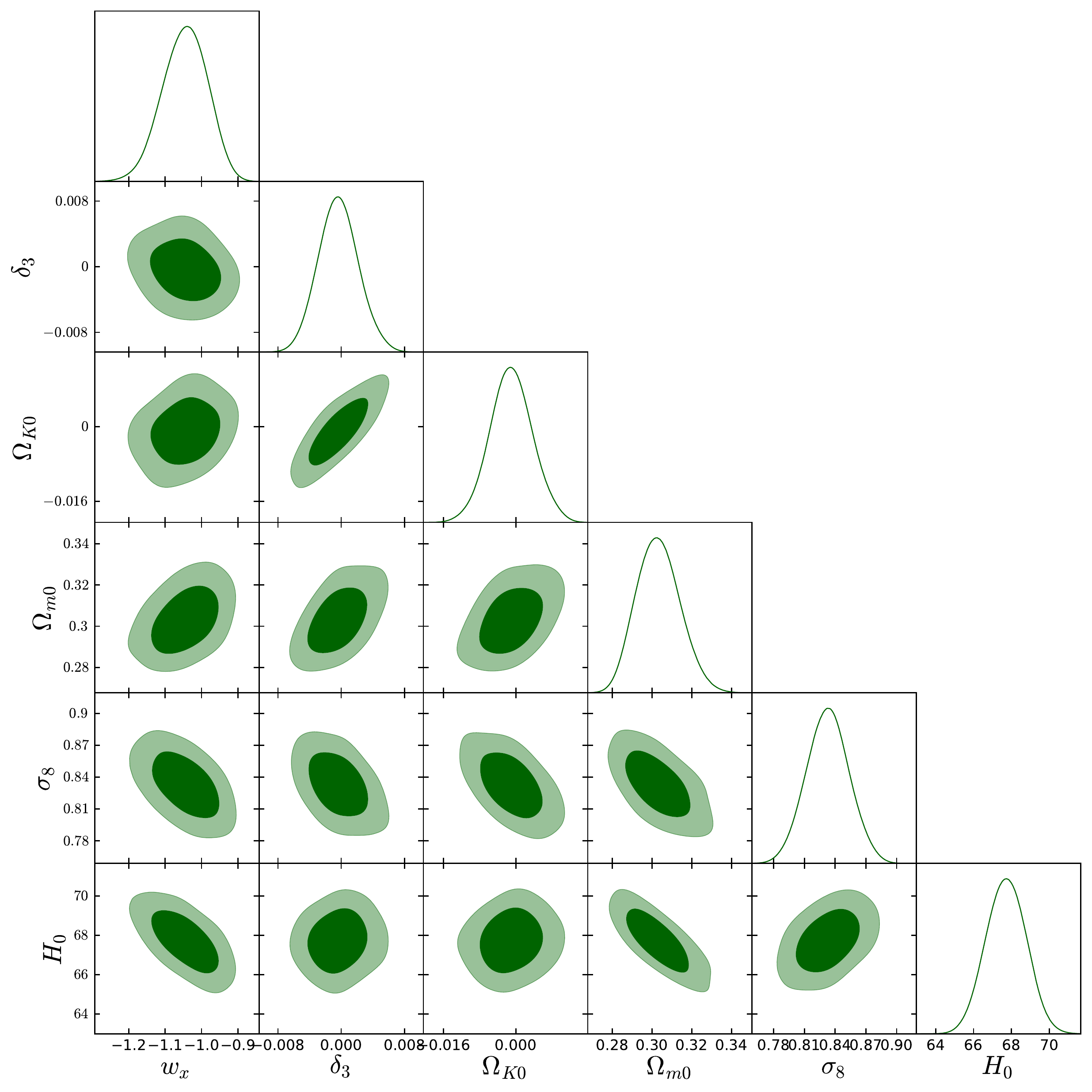}
\caption{68\% and 95\% CL joint contours and one dimensional posterior 
distributions for the interacting DE scenario (nonflat case) with
constant state parameter in DE, $w_x$, where the interaction is parametrized
by $\protect\delta (a) = \protect\delta_0 + \protect\delta_1 (1-a) + \protect%
\delta_2 (1-a)^2 + \protect\delta_3 (1-a)^3$, in which we fix the first three parameters $(\delta_0, \delta_1, \delta_2)$ from the previous analyses shown in Table \ref{tab:Int-wCDM-nonflat}. The combined data for this
analysis have been set to be CMB $+$ JLA $+$ BAO $+$ CC and the results are shown in the last two columns of Table \ref{tab:Int-wCDM-nonflat}. }
\label{fig:int-wCDM-nonflat-3}
\end{figure*}

\subsubsection{Interacting DE with constant EoS other than vacuum}

We now focus on the observational constraints on the interacting scenario when
the universe has a nonzero curvature and the dark energy has a constant
state parameter $w_x$.

We start with the parametrization $\delta (a) = \delta_0 + \delta_1 (1-a)$
of eqn. (\ref{xi}) and constrain all the model parameters with the
combined observational data that we have already mentioned. The summary of
the observational constraints has been given in the second and third column
of the Table \ref{tab:Int-wCDM-nonflat} and the graphical behaviour between various parameters have been shown in Fig. \ref{fig:int-wCDM-nonflat-1}. 
From the analysis we see that both 
$\delta_0$ and $\delta_1$ predict nonzero {\bf mean} values. More specifically, 
the estimated values are as follows: $\delta_0 = 0.134_{- 0.208}^{+ 0.175}$ at 68\% CL ($0.134_{- 0.352}^{+ 0.359}$ at 95\% CL), 
$\delta_1 = 0.134_{- 0.208}^{+ 0.175}$ at 68\% CL 
($0.1342_{-0.352}^{+ 0.359}$ at 95\% CL). However, one can clearly notice that within 68\% CL, $\delta_0 = 0 $ and $\delta_1 = 0$ are also allowed by the data.  
Regarding the dark energy state parameter, we see that, $w_x = -1.046_{- 0.058}^{+ 0.070}$ (68\% CL) which 
also includes its quintessential nature and it is 
close to the cosmological constant boundary. Furthermore, from the curvature parameter, $\Omega_{K0} = -0.00053_{- 0.00534}^{+ 0.00557}$ (at 68\% CL) one may see that since its error bars are bigger compared to its mean value, thus, the possibilities of open, closed and flat universe are all allowed. About the nature of the parameters $\delta_0$ and $\delta_1$, we have exactly similar conclusion (see Fig. \ref{fig:int-wCDM-nonflat-1}) as mentioned in other interacting scenarios. 

We now consider the second parametrization, i.e., eqn. (\ref{xi-2}) 
and fix the first two free parameters $\delta_0$
and $\delta_1$ to their corresponding mean values obtained from the 
previous analysis (column second of  
Table \ref{tab:Int-wCDM-nonflat}) and constrain the scenario. The reuslts are shown in the fourth and fifth columns of Table \ref{tab:Int-wCDM-nonflat} and in Fig. \ref{fig:int-wCDM-nonflat-2} we present the corresponding graphics. From the results, we find the parameter $\delta_2$ is very small taking $\delta_2 = -0.000249_{-0.002470}^{+ 0.002384}$ (68\% CL) and 
hence no effective contribution to $\delta (a)$ as in the previous cases.
The dark energy state parameter exhibits its phantom character, $w_x =
-1.043_{- 0.063}^{+ 0.069}$ (68\% CL). Although from the
estimation of $w_x$, we find its quintessential character is 
not ruled out either. As for the contribution of the spatial curvature, we have exactly similar conclusion as in the earlier sections. 

 After this we take the last parametrization of this
series, i.e., equation (\ref{xi-3}) where similar to the previous cases, 
we fix the first three parameters $\delta_0$, $\delta_1$ and $\delta_2$ and 
constrain $\delta_3$. The results are summarized in the last two columns of 
Table \ref{tab:Int-wCDM-nonflat-general} and the 
corresponding graphical variations are shown in Fig. \ref{fig:int-wCDM-nonflat-3}. 
We found that $\delta_3$ is very small and the dark energy state parameter $w_x$, 
represents a universe where both a phantom or a quintessence behaviour is allowed since $w_x =-1.045_{- 0.061}^{+ 0.067}$ (68\% CL).  
Finally, we find that $|\Omega_{K0}|$ is higher in this case
compared to previous two analyses with equations (\ref{xi-2}) and (\ref{xi}).

\begin{figure*}
\includegraphics[width=0.6\textwidth]{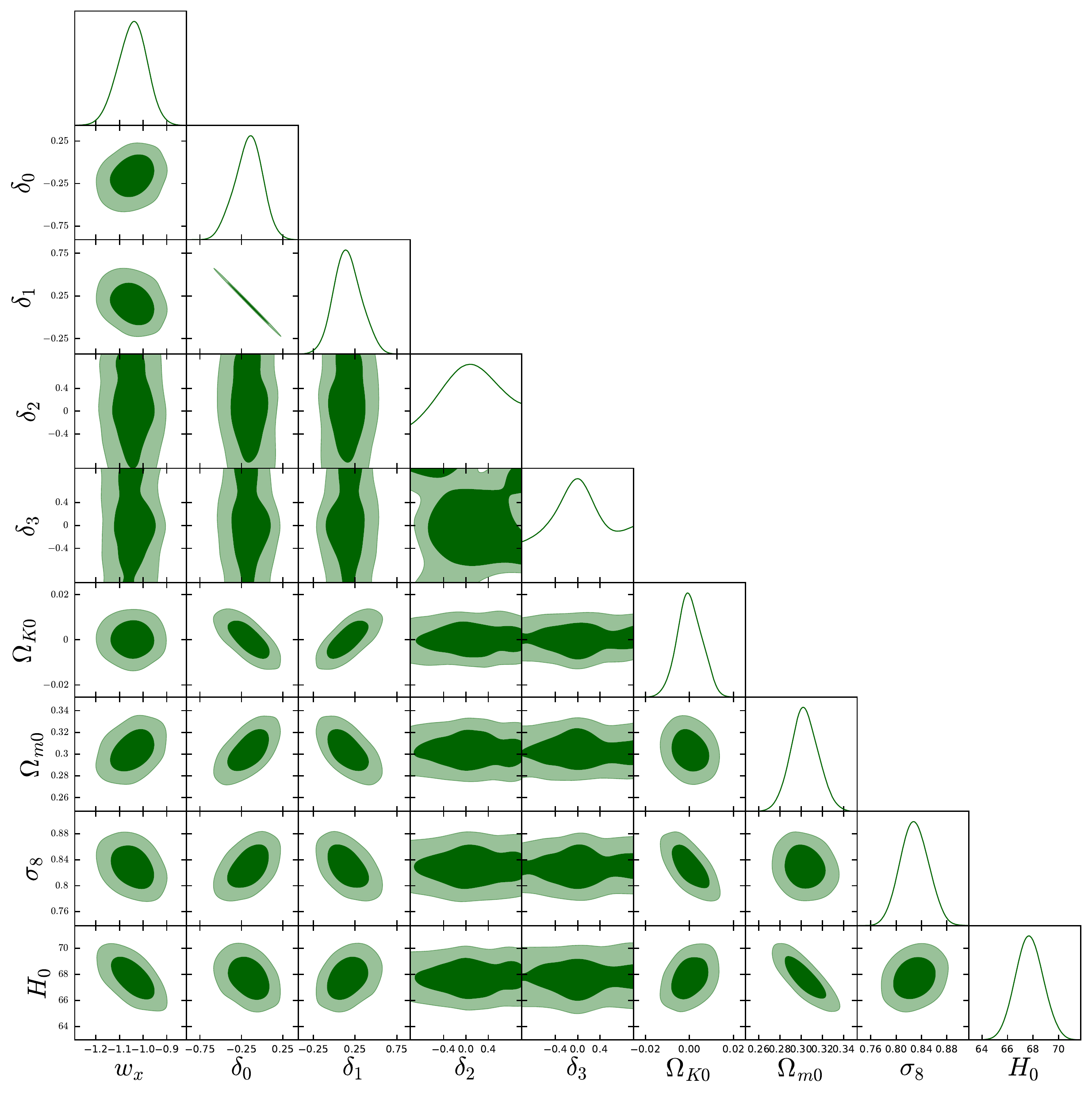}
\caption{68\% and 95\% CL joint contours and one dimensional posterior 
distributions for the interacting DE scenario (nonflat case)  with
constant state parameter in DE, $w_x$, where the interaction is parametrized
by the most general choice 
$\protect\delta (a) = \protect\delta_0 + \protect\delta_1 (1-a) + \protect%
\delta_2 (1-a)^2 + \protect\delta_3 (1-a)^3$ and $\delta_i$'s ($i= 1, 2, 3$) are kept free. The combined data for this analysis have been set to be CMB $+$ JLA $+$ BAO $+$ CC and the results of this analysis are shown in Table \ref{tab:Int-wCDM-nonflat-general}.}
\label{fig:int-wCDM-nonflat-4}
\end{figure*}

Now, as usual, we constrain this interacting  
scenario for the parametrization 
$\protect\delta (a) = \protect\delta_0 + 
\protect\delta_1 (1-a) + \protect\delta_2 (1-a)^2 + \protect\delta_3 (1-a)^3$ where 
all the interaction parameters $\delta_i$'s are kept free. The results of the analysis have been presented in Table \ref{tab:Int-wCDM-nonflat-general}  and the corresponding graphical variations are displayed in Fig. \ref{fig:int-wCDM-nonflat-4}. From this figure, one can clearly see that although $\delta_i$'s for $i=2, 3$ are not well constrained by the present data, but, the behaviour of $\delta_0$ and $\delta_1$ are again not altered as significantly. Finally, in Table \ref{tab:Int-wCDM-nonflat} and Table \ref{tab:Int-wCDM-nonflat-general}, we have shown the $\chi^2$ values for the best-fit values obtained using the combined analysis.  We have similar comment already notified in section \ref{sec-int-cons-eos-flat}.

\begingroup                                                                                                                     
\squeezetable                                                                                                                   

\begin{center}
\begin{table*}
\begin{tabular}{|c|c|c|c|c|c|c|c}
\hline
Parameters & Mean with errors & Best fit & Mean with errors & Best fit & 
Mean with errors & Best fit  \\ \hline
$\Omega_c h^2$ & $0.1164_{- 0.0038- 0.0076}^{+ 0.0039+ 0.0074}$ & $%
0.1195$ & $0.1162_{- 0.0032- 0.0064}^{+ 0.0030+ 0.0063}$ & $%
0.1207$ & $0.1160_{- 0.0036- 0.0061}^{+ 0.0032+ 0.0065}$ & $%
0.1126$   \\
 
$\Omega_b h^2$ & $0.02226_{- 0.00017- 0.00033}^{+ 0.00017+ 0.00032}$ & $%
0.02238$ & $0.02226_{- 0.00016- 0.00033}^{+ 0.00016+ 0.00032}$ & $%
0.02219$ & $0.02225_{- 0.00015- 0.00031}^{+ 0.00015+ 0.00031}$ & $%
0.02220$   \\
 
$100\theta_{MC}$ & $1.04101_{- 0.00043- 0.00088}^{+ 0.00044+ 0.00087}$
& $1.04067$ & $1.04104_{- 0.00039- 0.00078}^{+ 0.00039+ 0.00076}$ & $%
1.04087$ & $1.04101_{- 0.00038- 0.00082}^{+ 0.00043+ 0.00077}$ & $%
1.04111$   \\ 

$\tau$ & $0.082_{- 0.018- 0.035}^{+ 0.018 + 0.034}$ & $%
0.070 $ & $0.082_{- 0.018- 0.033}^{+ 0.017+ 0.035}$ & $%
0.079$ & $0.081_{- 0.017- 0.033}^{+ 0.017+ 0.031}$ & $%
0.084$   \\ 

$n_s$ & $0.9697_{- 0.0063- 0.0117}^{+ 0.0063+ 0.0126}$ & $0.9658$
& $0.9698_{- 0.0058- 0.0111}^{+ 0.0058+ 0.0111}$ & $0.9645$ & $%
0.9698_{- 0.0058- 0.0113}^{+ 0.0059+ 0.0112}$ & $0.9747$   \\ 

$\mathrm{ln}(10^{10} A_s)$ & $3.096_{- 0.035- 0.070}^{+ 0.035+
0.069}$ & $3.066$ & $3.095_{- 0.033- 0.066}^{+ 0.034+
0.067}$ & $3.099$ & $3.095_{- 0.034- 0.065}^{+ 0.033+
0.063}$ & $3.103$   \\ 

\hline 

$\delta_0$ & $-0.137_{- 0.184- 0.364}^{+ 0.197+ 0.357}$ & $%
-0.042$ & $-$ & $-$ & $-$ & $-$  \\

$\delta_1$ & $0.134_{- 0.208- 0.352}^{+ 0.175+ 0.359}$ & $%
0.043$ & $-$ & $-$ & $-$ & $-$   \\

$\delta_2$ & $\times$ & $\times$ & $-0.000249_{- 0.002470- 0.004740}^{+ 0.002384+
0.004100}$ & $0.003250$ & $-$ & $-$  
\\ 

$\delta_3$ & $\times$ & $\times$ & $\times$ & $\times$ & $-0.000361_{- 0.002509- 0.004806}^{+
0.002445+ 0.005239}$ & $-0.002758$  \\ \hline 

$w_x$ & $-1.046_{- 0.058- 0.134}^{+ 0.070+ 0.118}$ & $%
-1.092$ & $-1.043_{- 0.063- 0.128}^{+ 0.069+ 0.122}$ & $%
-1.048$ & $-1.045_{- 0.061- 0.117}^{+ 0.067+ 0.119}$ & $%
-1.072$   \\

$\Omega_{K0}$ & $-0.00053_{- 0.00534- 0.01108}^{+ 0.00557+ 0.01106}$ & $%
-0.00144$ & $-0.00075_{- 0.00441- 0.00821}^{+ 0.00393+ 0.00910}$ & $%
0.00498$ & $-0.00101_{- 0.00454- 0.00927}^{+ 0.00452+ 0.00980}$ & $%
-0.00656$   \\ 

$\Omega_{m0}$ & $0.305_{- 0.015- 0.026}^{+ 0.013+ 0.028}$ & $%
0.306$ & $0.304_{- 0.012- 0.020}^{+ 0.010+ 0.022}$ & $%
0.314$ & $0.303_{- 0.012- 0.021}^{+ 0.010+ 0.022}$ & $%
0.294$   \\
 
$\sigma_8$ & $0.831_{- 0.023- 0.042}^{+ 0.021+ 0.044}$ & $%
0.829$ & $0.832_{- 0.021- 0.040}^{+ 0.020+ 0.042}$ & $%
0.821$ & $0.833_{- 0.021- 0.040}^{+ 0.020+ 0.040}$ & $%
0.852$   \\
 
$H_0$ & $67.66_{- 1.11- 2.13}^{+ 1.10+ 2.20}$ & $%
68.27$ & $67.69_{- 1.07- 1.95}^{+ 1.06+ 2.09}$ & $%
67.64$ & $67.72_{- 1.07- 2.06}^{+ 1.06+ 2.01}$ & $%
67.89$   \\ \hline
$\chi^2$ && 13676.896 && 13676.222 && 13675.694 \\
\hline 
\end{tabular}%
\caption{Non-flat universe: Reconstruction of the interacting DE scenario
where DE has a constant EoS in DE, $w_x$, using the combined analysis CMB $+$
JLA $+$ BAO $+$ CC. In the columns `$\times$' means that the corresponding parameter has not been considered into the analyses while the sign `$-$' present against any parameter means its mean value has been fixed for the subsequent analyses. }
\label{tab:Int-wCDM-nonflat}
\end{table*}
\end{center}
\endgroup                                                                                                                       
\begingroup                                                                                                                     
\squeezetable                                                                                                                   
\begin{center}                                                                                                                  
\begin{table}                                                                                                                   
\begin{tabular}{|c|c|c|c|}                                                                                                            
\hline                                                                                                                    
Parameters & Mean with errors & Best fit \\ \hline
$\Omega_c h^2$ & $    0.1160_{-    0.0036-    0.0074}^{+    0.0036+    0.0068}$ & $    0.1158$\\

$\Omega_b h^2$ & $    0.02224_{-    0.00017-    0.00033}^{+    0.00017+    0.00034}$ & $    0.02224$\\

$100\theta_{MC}$ & $    1.04102_{-    0.00041-    0.00082}^{+    0.00042+    0.00082}$ & $    1.04119$\\

$\tau$ & $    0.082_{-    0.017-    0.034}^{+    0.017+    0.035}$ & $    0.096$\\

$n_s$ & $    0.9698_{-    0.0059-    0.0115}^{+    0.0059+    0.0119}$ & $    0.9716$\\

${\rm{ln}}(10^{10} A_s)$ & $    3.095_{-    0.033-    0.067}^{+    0.034+    0.067}$ & $    3.122$\\

$w_x$ & $   -1.046_{-    0.058-    0.124}^{+    0.067+    0.117}$ & $   -1.006$\\

$\delta_0$ & $   -0.165_{-    0.146-    0.344}^{+    0.192+    0.313}$ & $   -0.082$\\

$\delta_1$ & $    0.162_{-    0.190-    0.308}^{+    0.145+    0.340}$ & $    0.078$\\

$\delta_2$ & $    0.049_{-    0.580-    1.049}^{+    0.598+    0.951}$ & $   -0.183$\\

$\delta_3$ & $    0.013_{-    0.232-    1.013}^{+    0.987+    0.987}$ & $   -0.175$\\

$\Omega_{K0}$ & $    0.00013_{-    0.00591-    0.01091}^{+    0.00562+    0.01082}$ & $   -0.00448$\\

$\Omega_{m0}$ & $    0.303_{-    0.014-    0.025}^{+    0.013+    0.026}$ & $ 0.314$\\

$\sigma_8$ & $    0.829_{-    0.022-    0.043}^{+    0.022+    0.044}$ & $    0.847$\\

$H_0$ & $   67.72_{-    1.08-    2.08}^{+    1.08+    2.16}$ & $   66.44$\\
\hline
$\chi^2$ && 13676.21 \\
\hline                                                                                                                     
\end{tabular}                                                                                                                   
\caption{Reconstruction of the interacting DE scenario
where DE has a constant EoS in DE, $w_x$, in the spatially nonflat universe for the general parametrization $\protect\delta (a) = \protect\delta_0 + \protect\delta_1 (1-a) + \protect\delta_2 (1-a)^2 + \protect\delta_3 (1-a)^3$ 
using the combined analysis CMB $+$ JLA $+$ BAO $+$ CC.}
\label{tab:Int-wCDM-nonflat-general}                                                                                                   
\end{table}                                                                                                                     
\end{center}                                                                                                                    
\endgroup                                                                                                                       
\begingroup                                                                                                                     
\squeezetable                                                                                                                   
\begin{center}
\begin{table*}
\begin{tabular}{|c|c|c|c|c|c|c|c}
\hline
Parameters & Mean with errors & Best fit & Mean with errors & Best fit & 
Mean with errors & Best fit \\ \hline
$\Omega_c h^2$ & $0.1163_{- 0.0035 - 0.0075}^{+ 0.0033+ 0.0071}$ & 
$0.1142$ & $0.1167_{- 0.0029- 0.0058}^{+ 0.0031+ 0.0057}$ & $%
0.1172$ & $0.1165_{- 0.0026- 0.0063}^{+ 0.0035+ 0.0057}$ & $%
0.1127$   \\ 

$\Omega_b h^2$ & $0.02227_{- 0.00017- 0.00034}^{+ 0.00018+ 0.00035}$ & $%
0.02226$ & $0.02228_{- 0.00017- 0.00032}^{+ 0.00017+ 0.00032}$ & $%
0.02217$ & $0.02225_{- 0.00016- 0.00033}^{+ 0.00016+ 0.00031}$ & $%
0.02230$   \\ 

$100\theta_{MC}$ & $1.04104_{- 0.00041 - 0.00080}^{+ 0.00041 + 0.00080}$
& $1.04145$ & $1.04101_{- 0.00041- 0.00078}^{+ 0.00038+ 0.00079}$ & $%
1.04089$ & $1.04098_{- 0.00043- 0.00071}^{+ 0.00037+ 0.00079}$ & $%
1.04145$   \\ 

$\tau$ & $0.078_{- 0.017- 0.036}^{+ 0.017+ 0.033}$ & $%
0.078 $ & $0.078_{- 0.016- 0.032}^{+ 0.017+ 0.033}$ & $%
0.089$ & $0.079_{- 0.016- 0.035}^{+ 0.017+ 0.033}$ & $%
0.102$   \\ 

$n_s$ & $0.9701_{- 0.0056- 0.0112}^{+ 0.0057+ 0.0108}$ & $0.9714$
& $0.9694_{- 0.0056- 0.0108}^{+ 0.0055+ 0.0112}$ & $0.9684$ & $%
0.9694_{- 0.0065- 0.0105}^{+ 0.0053+ 0.0117}$ & $0.9777$  \\ 

$\mathrm{ln}(10^{10} A_s)$ & $3.088_{- 0.034- 0.071}^{+ 0.037+
0.067}$ & $3.091$ & $3.088_{- 0.032- 0.064}^{+ 0.033+
0.064}$ & $3.115$ & $3.091_{- 0.032- 0.067}^{+ 0.033+
0.065}$ & $3.131$   \\  \hline 

$\delta_0$ & $-0.139_{- 0.137- 0.354}^{+ 0.209+ 0.313}$ & $%
-0.222$ & $-$ & $-$ & $-$ & $-$  \\

$\delta_1$ & $0.137_{- 0.208- 0.308}^{+ 0.134+ 0.350}$ & $%
0.218$ & $-$ & $-$ & $-$ & $-$   \\

$\delta_2$ & $\times$ & $\times$ & $0.000350_{- 0.002261- 0.004713}^{+ 0.002540+
0.004543}$ & $-0.000517$  \\

$\delta_3$ & $\times$ & $\times$ & $\times$ & $\times$ & $-0.000004_{- 0.002056- 0.005068}^{+
0.002698+ 0.004179}$ & $-0.001751$   \\  \hline

$w_0$ & $-0.910_{- 0.184- 0.322}^{+ 0.152+ 0.348}$ & $%
-0.873$ & $-0.932_{- 0.170- 0.254}^{+ 0.104+ 0.299}$ & $%
-0.989$ & $-1.005_{- 0.109- 0.181}^{+ 0.081+ 0.215}$ & $%
-0.982$   \\ 

$w_a$ & $-0.454_{- 0.393- 1.162}^{+ 0.630+ 1.005}$ & $%
-0.531$ & $-0.383_{- 0.250- 0.988}^{+ 0.546+ 0.748}$ & $%
-0.180$ & $-0.133_{- 0.149- 0.590}^{+ 0.313+ 0.412}$ & $%
-0.140$   \\

$\Omega_{K0}$ & $-0.00329_{- 0.00652- 0.01188}^{+ 0.005668+ 0.012586}$ & $%
-0.00619$ & $-0.00262_{- 0.00438- 0.00885}^{+ 0.00433+ 0.00939}$ & $%
-0.00003$ & $-0.00121_{- 0.00459- 0.00810}^{+ 0.00489+ 0.00912}$ & $%
-0.00435$   \\ 

$\Omega_{m0}$ & $0.306_{- 0.015- 0.029}^{+ 0.013+ 0.028}$ & $%
0.313$ & $0.306_{- 0.012- 0.021}^{+ 0.010+ 0.022}$ & $%
0.299$ & $0.305_{- 0.011- 0.021}^{+ 0.011+ 0.021}$ & $%
0.293$   \\ 

$\sigma_8$ & $0.831_{- 0.023- 0.047}^{+ 0.023+ 0.048}$ & $%
0.825$ & $0.830_{- 0.018- 0.037}^{+ 0.019+ 0.037}$ & $%
0.852$ & $0.831_{- 0.019- 0.039}^{+ 0.019+ 0.038}$ & $%
0.847$   \\
 
$H_0$ & $67.44_{- 1.12- 2.38}^{+ 1.17+ 2.20}$ & $%
66.21$ & $67.53_{- 1.05- 2.00}^{+ 1.01+ 1.98}$ & $%
68.42$ & $67.67_{- 0.96- 1.88}^{+ 0.95+ 1.98}$ & $%
68.04$   \\ \hline
$\chi^2$ && 13676.868 && 13675.118 && 13677.000\\
\hline 
\end{tabular}%
\caption{Non-flat universe: Reconstruction of the interacting DE scenario
where DE has a dynamical EoS in DE, $w_x (a) = w_0 + w_a (1-a)$, using the
combined analysis CMB $+$ JLA $+$ BAO $+$ CC. In the columns `$\times$' means that the corresponding parameter has not been considered into the analyses while the sign `$-$' present against any parameter means its mean value has been fixed for the subsequent analyses.  }
\label{tab:Int-dyn-nonflat}
\end{table*}
\end{center}
\endgroup                                                                                                                       
\begingroup                                                                                                                     
\squeezetable                                                                                                                   
\begin{center}                                                                                                                  
\begin{table}                                                                                                                   
\begin{tabular}{|c|c|c|c|c}                                                                                                            
\hline                                                                                                                    
Parameters & Mean with errors & Best fit \\ \hline
$\Omega_c h^2$ & $    0.1161_{-    0.0037-    0.0072}^{+    0.0036+    0.0076}$ & $    0.1200$\\

$\Omega_b h^2$ & $    0.02228_{-    0.00017-    0.00035}^{+    0.00017+    0.00035}$ & $    0.02219$\\

$100\theta_{MC}$ & $    1.04103_{-    0.00042-    0.00085}^{+    0.00043+    0.00084}$ & $    1.04079$\\

$\tau$ & $    0.078_{-    0.017-    0.035}^{+    0.018+    0.034}$ & $    0.069$\\

$n_s$ & $    0.9702_{-    0.0059-    0.0117}^{+    0.0059+    0.0119}$ & $    0.9623$\\

${\rm{ln}}(10^{10} A_s)$ & $    3.088_{-    0.034-    0.068}^{+    0.034+    0.066}$ & $    3.066$\\

$w_0$ & $   -0.912_{-    0.177-    0.345}^{+    0.162+    0.350}$ & $   -0.919$\\

$w_a$ & $   -0.471_{-    0.418-    1.154}^{+    0.631+    1.035}$ & $   -0.519$\\

$\delta_0$ & $   -0.162_{-    0.183-    0.364}^{+    0.179+    0.373}$ & $   -0.169$\\

$\delta_1$ & $    0.159_{-    0.178-    0.368}^{+    0.182+    0.361}$ & $    0.169$\\

$\delta_2$ & $   -0.042_{-    0.535-    0.959}^{+    0.609+    1.042}$ & $    0.384$\\

$\delta_3$ & $    0.036_{-    0.483-    1.036}^{+    0.764+    0.965}$ & $    0.983$\\

$\Omega_{K0}$ & $   -0.00282_{-    0.00630-    0.01239}^{+    0.00682+    0.01177}$ & $    0.00171$\\

$\Omega_{m0}$ & $    0.306_{-    0.015-    0.026}^{+    0.013+    0.029}$ & $    0.308$\\

$\sigma_8$ & $    0.828_{-    0.023-    0.044}^{+    0.023+    0.047}$ & $    0.818$\\

$H_0$ & $   67.49_{-    1.18-    2.28}^{+    1.18+    2.33}$ & $   68.12$\\
\hline
$\chi^2$ && 13673.444 \\
\hline                                                                                                                     
\end{tabular}                                                                                                                   
\caption{Reconstruction of the interacting DE scenario
for the dynamical EoS in DE, $w_x (a) = w_0 + w_a (1-a)$, 
where the interaction is parametrized by 
$\protect\delta (a) = \protect\delta_0 + 
\protect\delta_1 (1-a) + \protect\delta_2 (1-a)^2 + \protect\delta_3 (1-a)^3$. The
combined analysis is CMB $+$ JLA $+$ BAO $+$ CC. }\label{tab:Int-dyn-nonflat-general}                                                                                                   
\end{table}                                                                                                                     
\end{center}                                                                                                                    
\endgroup                                                                                                                       
\begin{figure*}
\includegraphics[width=0.6\textwidth]{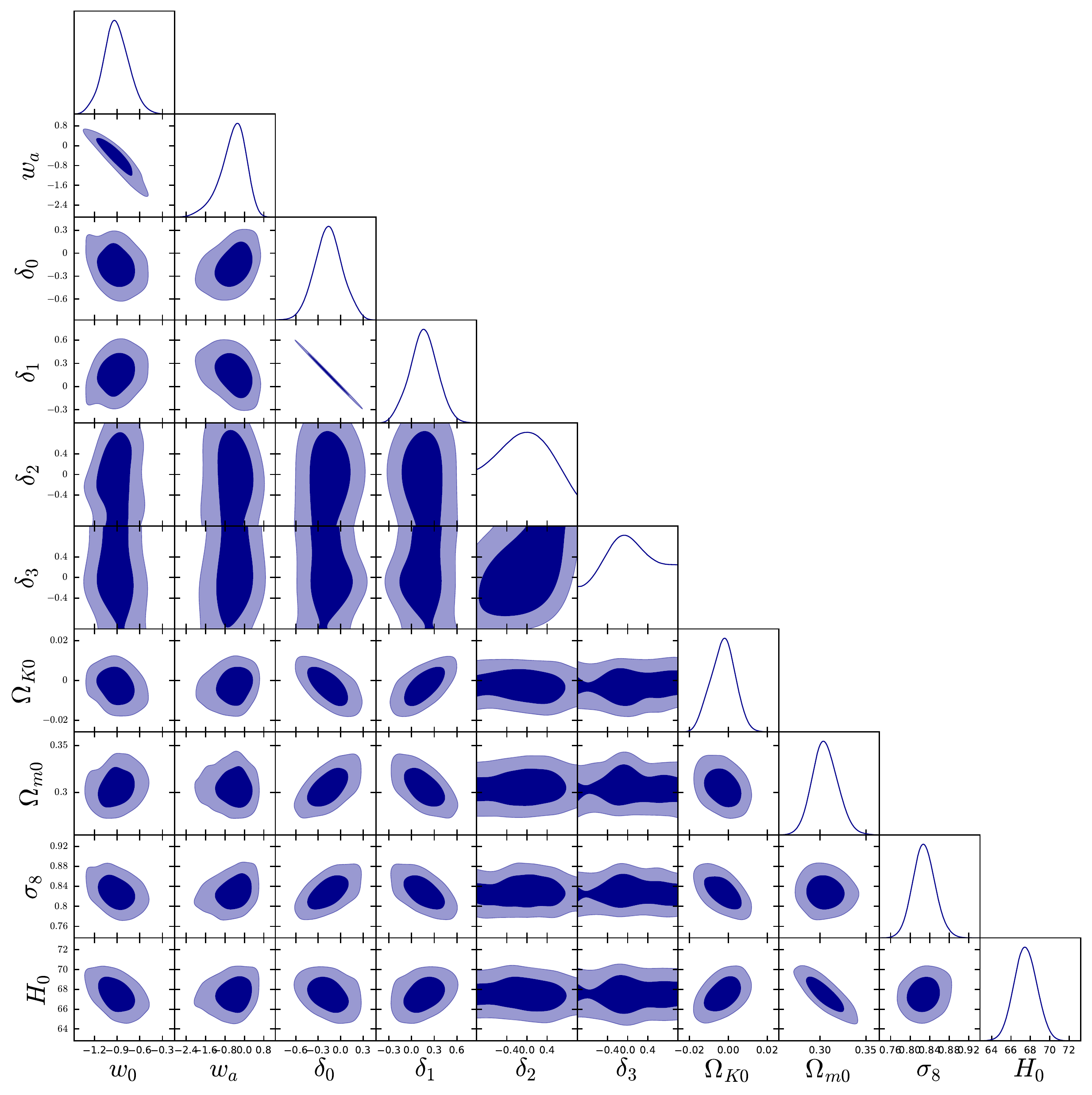}
\caption{68\% and 95\% CL joint contours and one dimensional posterior 
distributions for the interacting DE scenario (nonflat case)  with
dynamical state parameter in DE, $w_x (a) = w_0 + w_a (1-a)$, where the
interaction is parametrized by $\protect\delta (a) = \protect\delta_0 + 
\protect\delta_1 (1-a) + \protect\delta_2 (1-a)^2 + \protect\delta_3 (1-a)^3$ and all
$\delta_i$'s ($i=1, 2, 3$) are kept free. The combined data for this analysis have been set to be CMB $+$ JLA $+$
BAO $+$ CC and the results are shown in Table \ref{tab:Int-dyn-nonflat-general}.}
\label{fig:int-dyn-nonflat-4}
\end{figure*}
\begin{figure*}
\includegraphics[width=0.5\textwidth]{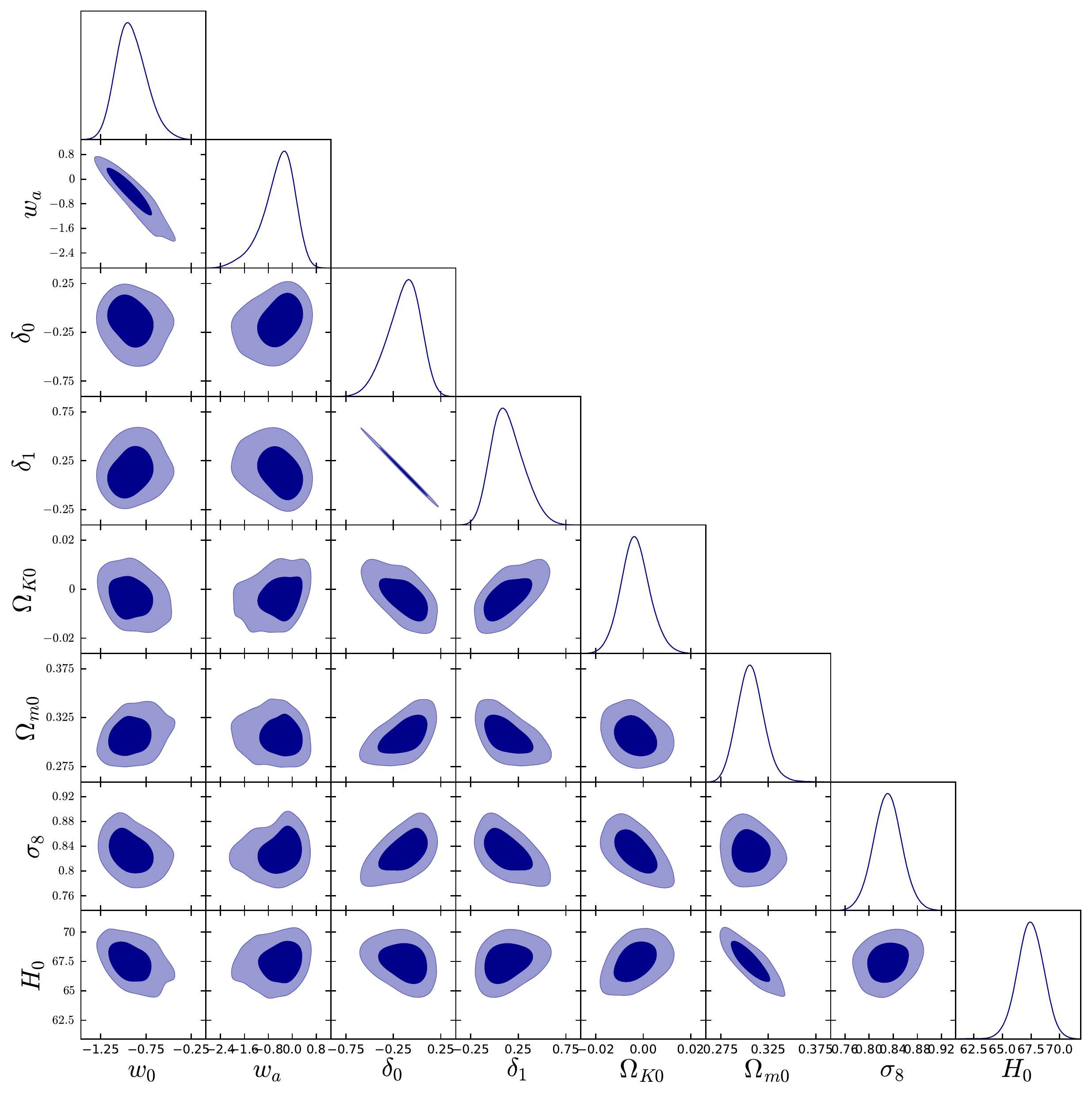}
\caption{68\% and 95\% CL joint contours and one dimensional posterior 
distributions for the interacting DE scenario (nonflat case)  with
dynamical state parameter in DE, $w_x (a) = w_0 + w_a (1-a)$, where the
interaction is parametrized by $\protect\delta (a) = \protect\delta_0 + 
\protect\delta_1 (1-a)$. The combined data for this analysis have been set
to be CMB $+$ JLA $+$ BAO $+$ CC and the results are shown in the second and third columns of Table \ref{tab:Int-dyn-nonflat}.}
\label{fig:int-dyn-nonflat-1}
\end{figure*}
\begin{figure*}
\includegraphics[width=0.5\textwidth]{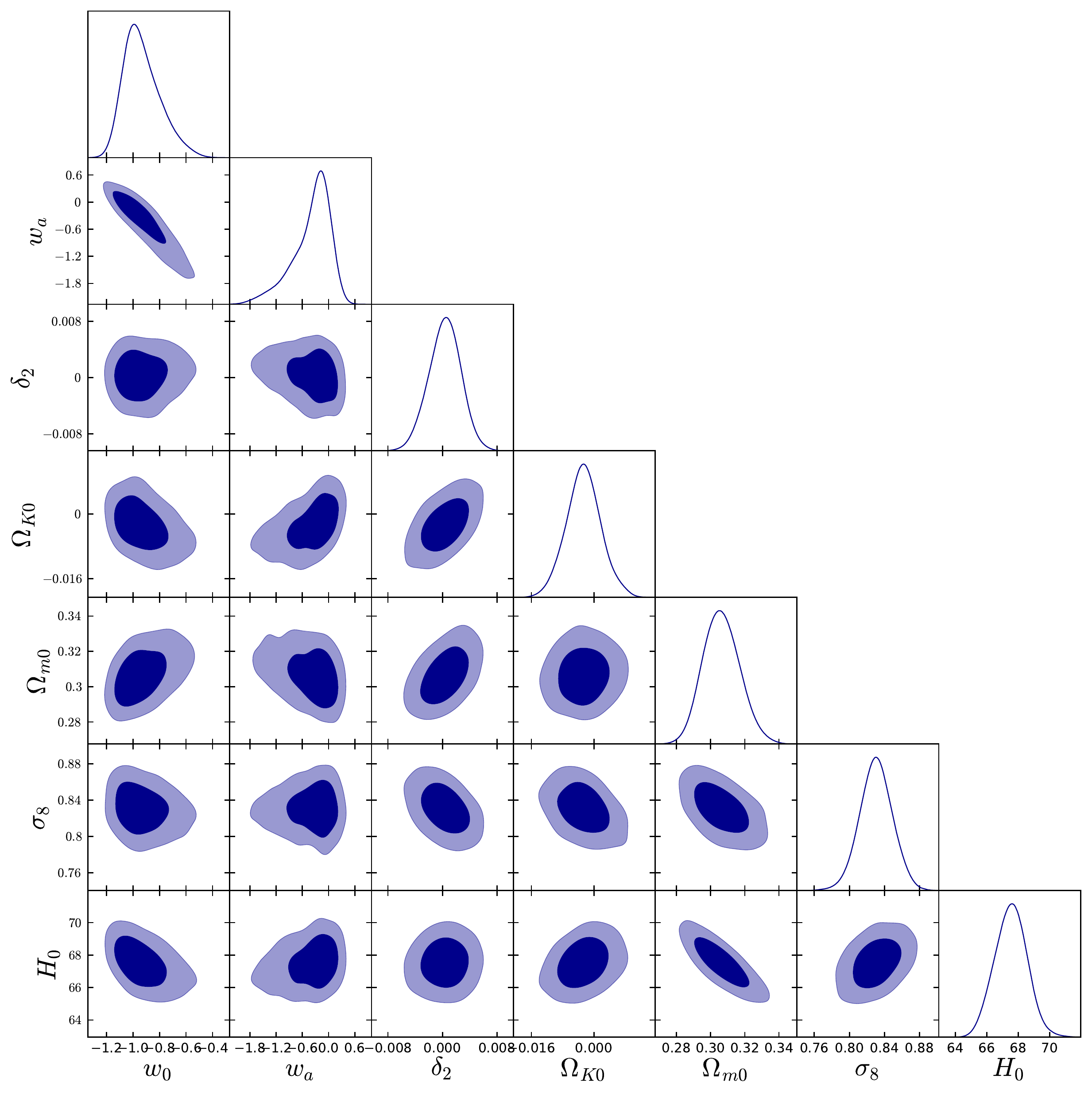}
\caption{68\% and 95\% CL joint contours and one dimensional posterior 
distributions for the interacting DE scenario (nonflat case) with
dynamical state parameter in DE, $w_x (a) = w_0 + w_a (1-a)$, where the
interaction is parametrized by $\protect\delta (a) = \protect\delta_0 + 
\protect\delta_1 (1-a)+ \protect\delta_2 (1-a)^2$ in which ($\delta_0$, $\delta_1$) are fixed at their mean values from Table \ref{tab:Int-dyn-nonflat}. The combined data for
this analysis have been set to be CMB $+$ JLA $+$ BAO $+$ CC and the results are shown in the fourth and fifth columns of Table \ref{tab:Int-dyn-nonflat}.  }
\label{fig:int-dyn-nonflat-2}
\end{figure*}
\begin{figure*}
\includegraphics[width=0.5\textwidth]{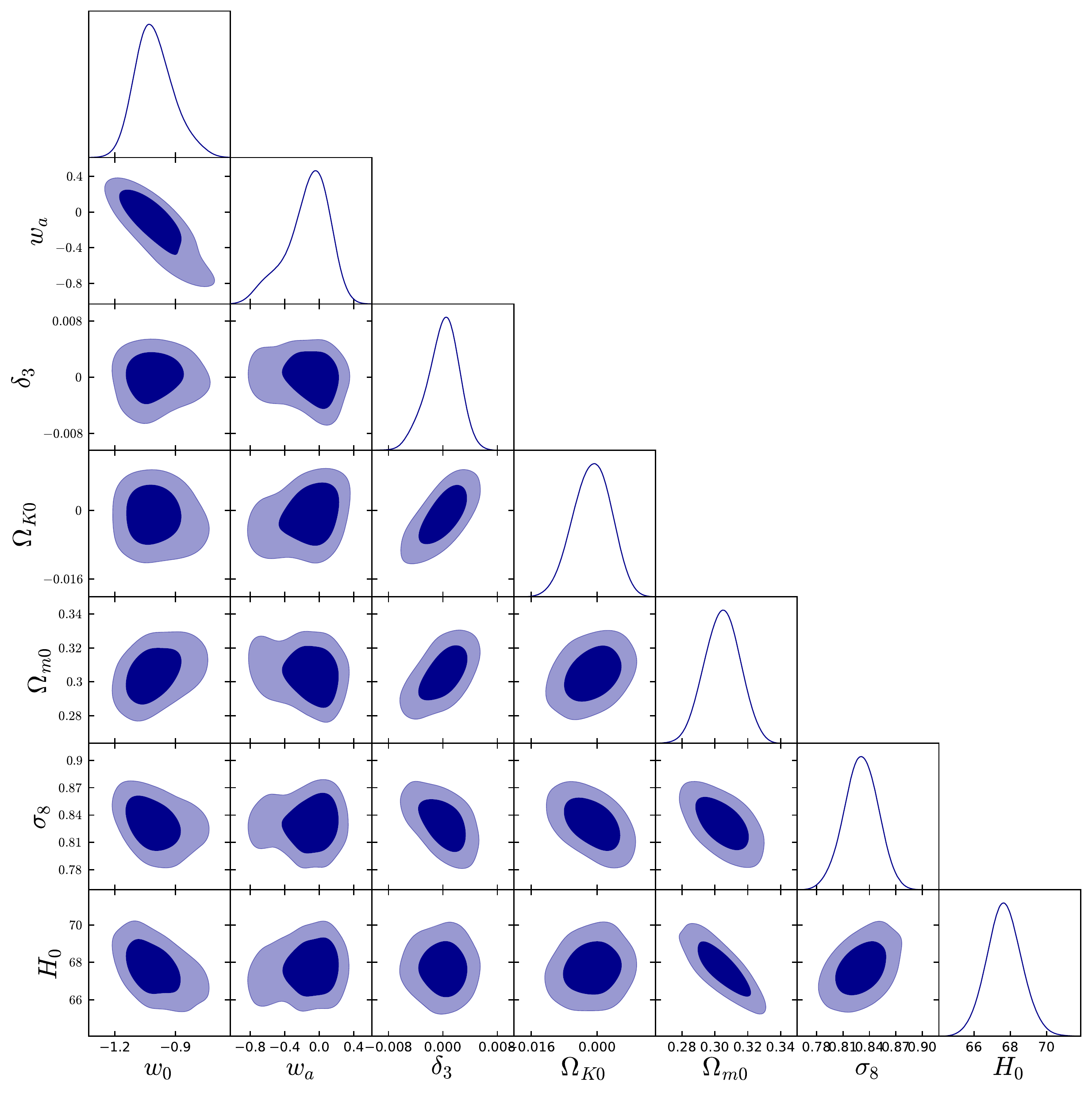}
\caption{68\% and 95\% CL joint contours and one dimensional posterior 
distributions for the interacting DE scenario (nonflat case) with
dynamical state parameter in DE, $w_x (a) = w_0 + w_a (1-a)$, where the
interaction is parametrized by $\protect\delta (a) = \protect\delta_0 + 
\protect\delta_1 (1-a) + \protect\delta_2 (1-a)^2 + \protect\delta_3 (1-a)^3$, and in which ($\delta_0$, $\delta_1$, $\delta_2$) are fixed at their mean values from Table \ref{tab:Int-dyn-nonflat}. The combined data for this analysis have been set to be CMB $+$ JLA $+$ BAO $+$ CC and the results are shown in the last two columns of Table \ref{tab:Int-dyn-nonflat}. }
\label{fig:int-dyn-nonflat-3}
\end{figure*}
\subsubsection{Interacting DE with dynamical EoS}

Finally, we discuss the interacting model in the nonflat FLRW universe when
the dark energy equation of state is dynamical, given by the CPL
parametrization, eqn. (\ref{cpl}).

We begin with the 
first parametrization $\delta (a) = \delta_0
+ \delta_1 (1-a)$ of eqn. (\ref{xi}).  The observational
constraints on the model are shown in the second and third
columns of Table \ref{tab:Int-dyn-nonflat}. In Fig. \ref{fig:int-dyn-nonflat-1}
we present the corresponding graphical analysis. 
Our analysis reveals as usual 
that the  mean values of
both $\delta_0$ and $\delta_1$ are non-null but within 68\% CL, 
both of them allow zero values. 
The dark energy state parameter shows its quintessential behaviour 
as, $w_0$ is $w_0= -0.910_{- 0.184}^{+ 0.152}$ (at 68\% CL). 
However, the phantom character of $w_0$ is still allowed within 
the 68\% CL where $-1.095< w_0 < -0.758$. 
Regarding the spatial curvature,
we notice that within 68\% CL, $-0.00981 <
\Omega_{K0} < 0.00238$, which is similar to findings of this work in other scenarios. 
From Fig. \ref{fig:int-dyn-nonflat-1} we again conclude that, irrespective of the dynamical nature in the DE state parameter, the negative correlation between $\delta_0$ and $\delta_1$ exists and such correlation is very strong.

We then start with the  second parametrization of $\delta(a)$, i.e., 
equation (\ref{xi-2}) where we fix the two parameters $\delta_0$ and $\delta_1$ to their obtained values, and constrain the interacting scenario with the same astronomical data. The results are summarized in the fourth and fifth columns of Table \ref{tab:Int-dyn-nonflat} and in Fig. \ref{fig:int-dyn-nonflat-2} we show the corresponding graphics. Our results show that $\delta_2$ assumes a very small value with $\delta_2= 0.000350_{- 0.002261}^{+ 0.002540}$ (68\% CL). Thus, the contribution of $\delta_2$ towards the interaction parameter $\delta (a)$ is insignificant. While on the other hand, the present value of the dark energy equation of state keeps its quintessential character (see Table \ref{tab:Int-dyn-nonflat}), although the observational data allow it to cross the phantom divide line, since within 68\% CL, $-1.102< w_0 < -0.828$. For the curvature parameter, our conclusion follows the similar statements made in earlier sections.

After the fittings with first two parametrizations, namely, (\ref{xi}) and (\ref{xi-2}), we start with the next parametrization (\ref{xi-3}) and fit the interacting scenario where we fix the first parameters $\delta_0$, $\delta_1$, 
$\delta_2$ and constrain the last parameter $\delta_3$ along with 
other free parameters. The summary is given in the last two columns of Table \ref{tab:Int-dyn-nonflat} and in Fig. \ref{fig:int-dyn-nonflat-3} we display the corresponding graphics. The conclusion is that $\delta_3$ is very small and hence there is no effective contribution to $\delta (a)$; the mean value of the dark energy state parameter crosses the `$-1$' boundary although it may be quintessential in 68\% CL; and the curvature parameter assumes the negative value (indication for an open universe) while the positive and zero values are allowed by the data.

As the last example, we take the  general parametrization 
$\protect\delta (a) = \protect\delta_0 + 
\protect\delta_1 (1-a) + \protect\delta_2 (1-a)^2 + 
\protect\delta_3 (1-a)^3$ and constrain the interacting scenario using the same combined analysis employed throughout this work. The results have been 
summarized in Table \ref{tab:Int-dyn-nonflat-general} and the graphical dependence between several free parameters including the one dimensional posterior distributions 
and the two dimensional contour plots are shown in Fig. \ref{fig:int-dyn-nonflat-4}.   
From Fig. \ref{fig:int-dyn-nonflat-4}, one can see that the parameters $\delta_2$ and $\delta_3$ are not properly constrained since from the analysis we cannot put any upper or lower bounds on them. This is not surprising because we already mentioned that the parameters space for this interaction scenario increases considerably compared to the previous interaction models and  one may expect such degeneracy in the parameters. From Fig. \ref{fig:int-dyn-nonflat-4}, one can see that the strong negative correlation between $\delta_0$ and $\delta_1$ does not change. This is probably the most important finding of this work because such nature is independent of the curvature of the universe and with the dimension of the parameters space under consideration.  We conclude this section with the following comment on $\chi^2$. Following the similar fashion, in Tables \ref{tab:Int-dyn-nonflat} and \ref{tab:Int-dyn-nonflat-general}, we show the $\chi^2$ values obtained from the best-fit values of the reconstructed scenarios. Leaving the $\chi^2$ value of the general scenario as commented earlier, we focus on the $\chi^2$ values of Table \ref{tab:Int-dyn-nonflat} and we draw similar conclusion as described in section \ref{sec-int-DE-Dyn-flat}.

\section{Summary and Discussions}
\label{discussion}

In the present work we focus on a specific class of interacting models where
we denote the presence of interaction through the deviation in the
evolution law of the standard cold dark matter sector; we assume
that the evolution of CDM is governed by the law $\rho_c \propto a^{-3
+\delta (a)}$, where $\delta (a)$ is assumed to evolve and any $\delta (a)$ apart from
its null values, reflects the interaction in the dark sector. 
The flow of energy can be characterized with the sign of $\delta (a)$. For instance, 
$\delta (a) < 0$ implies an energy flow from CDM to DE while its positive value
indicates the flow of energy from DE to CDM. As the functional
form of $\delta (a)$ is not known, we take a standard approach $-$
the Taylor series expansion of $\delta (a)$ around the present scale factor $%
a_0 = 1$ as follows

\begin{eqnarray}
\delta \left( a\right) =\delta _{0}+\delta _{1}\left( 1-a\right) +\delta
_{2}\left( 1-a\right) ^{2}+ \delta_3 (1-a)^3+...  \notag
\end{eqnarray}
where $\delta_i$'s, $i= 0, 1, 2, 3,..,$ are constants with $\delta_0$ as
the current value of the parameter $\delta (a)$. 
 However, the consideration of a large
number of free parameters in a cosmological model generally increases the
degeneracy amongst the parameters and  it is naturally expected that 
the observational analysis with the entire parametrization (\ref{xi-3}) might 
be plagued with this issue. In the present work we have performed two separate 
reconstructions of the interaction scenarios. To start with, 
we consider the Taylor series expansion up to its second term, i.e. $\delta
(a) = \delta _{0}+\delta _{1}\left( 1-a\right)$ and constrain the
interaction parameters, $\delta_0$ and $\delta_1$. Next we increase one more
term in the Taylor expansion as $\delta \left( a\right) =\delta _{0}+\delta
_{1}\left( 1-a\right) +\delta _{2}\left( 1-a\right) ^{2}$, but this time, we
fix the free parameters ($\delta_0$, $\delta_1$) of this Taylor expansion to
their corresponding mean values obtained in the previous analysis with the
parametrization $\delta (a) = \delta _{0}+\delta _{1}\left( 1-a\right)$, and
constrain the free parameter $\delta_2$. In a similar fashion, we consider
the final parametrization in this series, $\delta \left( a\right) =\delta
_{0}+\delta _{1}\left( 1-a\right) +\delta _{2}\left( 1-a\right) ^{2}+
\delta_3 (1-a)^3$, and similarly we fix the values of $(\delta_0, \delta_1,
\delta_2)$ from the previous analyses and constrain the last interaction
parameter $\delta_3$. We note that, in a similar way, one can continue with
more free parametrizations. But from the present analysis we find that the parameters $\delta_2$ and $%
\delta_3$ are in fact very small so that effectively the terms containing 
$\delta_2$, $\delta_3$ might be neglected, and any further generalization is hardly expected to improve the analysis. 

Then we consider the full parametriation up to third order, 
$\delta \left( a\right) =\delta_{0}+\delta _{1}\left( 1-a\right) +\delta _{2}\left( 1-a\right) ^{2}+
\delta_3 (1-a)^3$ where we consider the parameters $\delta_i$'s to be free
and constrained all the interaction scenarios. The analyses performed with this 
entire parametrization now reflect that
the parameters $\delta_2$ and $\delta_3$ are indeed degenerate as expected.

We have considered three dark energy models for both spatially flat and curved universe with various parametrization of the interaction. In whatever way we proceed, and whatever kind of a dark energy we choose, $\delta_3$, $\delta_4$ are always found to be small, and $\delta_1$, $\delta_2$ are strongly negatively correlated, and they actually compensate the contribution from each other, particularly for small values of $z$. So there is a very strong indication that there is hardly any interaction in the dark sector of the universe, at least for the present interaction function $Q$ given in eqn. (\ref{Q}). Even if there is any, that should have been in a distant past. The present dark matter and dark energy hardly infringe upon the independent evolution of each other.

We have also considered models with a nonzero spatial curvature. It is found that these models do not rule out such a possibility. However, the fate of the possibility of an interaction between DE and CDM hardly improves in the presence of a spatial curvature. In all such scenarios we note that the estimated values of $H_0$ are very similar to Planck \cite{Ade:2015xua}, and hence, all of them are many sigmas apart from the local estimation of $H_0$ \cite{Riess:2016jrr}. This perhaps might be an issue for further investigations. 
Finally, we display the $\chi^2$ values obtained for the best-fit scenarios in the last row of Tables \ref{tab:Int-vacuum-flat} -- \ref{tab:Int-dyn-nonflat-general}. We note that from the $\chi^2$ values of all the interacting scenarios (including both spatially flat and nonflat cases), no new physics comes out.

\begin{acknowledgments}
The authors gratefully acknowledge the referee for his/her important suggestions and comments to improve the quality of the work and its presentation. 
W. Yang's work is supported by the National
Natural Science Foundation of China under Grants No.  11705079 and No.  11647153. SP acknowledges the Faculty Research and Professional Development Fund (FRPDF) Scheme of Presidency University, Kolkata, India. 
AP was supported by FONDECYT postdoctoral grant no. 3160121.
\end{acknowledgments}


\end{document}